\documentclass[a4paper,fleqn]{cas-sc}

\usepackage{siunitx}
\usepackage{chemformula}
\usepackage{upgreek}

\usepackage{caption}
\usepackage{subcaption}

\usepackage[dvipsnames]{xcolor}	
\usepackage[authoryear]{natbib}
\usepackage{mathtools}

\providecommand{\citet}[1]{\cite{#1}}
\providecommand{\citep}[1]{\cite{#1}}

\newcommand*{\myexternaldocument}[1]{%
    \externaldocument{#1}%
    \addFileDependency{#1.tex}%
    \addFileDependency{#1.aux}%
}
\makeatletter
\newcommand*{\addFileDependency}[1]{
  \typeout{(#1)}
  \@addtofilelist{#1}
  \IfFileExists{#1}{}{\typeout{No file #1.}}
}
\makeatother

\newcommand{\ie}{\textit{i.e.}}
\newcommand{\eg}{\textit{e.g.}}
\newcommand{\etc}{\textit{etc.}}
\renewcommand{\S}{Section~}
\newcommand{\Eq}{Eq.~}
\newcommand{\Eqs}{Eqs.~}
\newcommand{\F}{Figure~}
\newcommand{\T}{Table~}
\newcommand{\Fs}{Figures~}
\newcommand{\siu}[1]{\,\si{#1}}

\newcommand{\sn}{\sin\theta}
\newcommand{\cn}{\cos\theta}

\newcommand{\eiwt}{e^{-i\omega t}}
\newcommand{\eimp}{e^{im\phi}}

\newcommand{\B}{\mathbf{B}}
\newcommand{\Bnet}{\B_\mathrm{net}}

\newcommand{\Binm}{B^{i}_{nm}}
\newcommand{\Benm}{B^{e}_{nm}}

\newcommand{\Benmp}{B^{e}_{n'm'}}
\newcommand{\Ynm}{Y_{nm}}

\newcommand{\YnmM}{Y_{n,-m}}
\newcommand{\dYnmM}{Y^\star_{n,-m}}
\newcommand{\Ypq}{Y_{\shapen\shapem}}
\newcommand{\dYnm}{Y^\star_{nm}}

\newcommand{\dYnmConj}{Y^{\star*}_{nm}}

\newcommand{\bP}{\Lambda}

\newcommand{\Pnmu}{P_n^m}

\newcommand{\shapemax}{\shapen_\mathrm{max}}

\newcommand{\dbdt}{\frac{\partial\B}{\partial t}}
\newcommand{\lapl}{\nabla^2}

\newcommand{\ddr}{\frac{d}{dr}}

\newcommand{\dydth}[1]{\frac{\partial Y_{#1}}{\partial\theta}}

\newcommand{\aovr}{\left(\frac{\roB}{r}\right)}
\newcommand{\rova}{\left(\frac{r}{\roB}\right)}

\newcommand{\jn}{j_{n}}
\newcommand{\yn}{y_{n}}
\newcommand{\jd}{\jn^{\star}}
\newcommand{\yd}{\yn^{\star}}

\newcommand{\jnP}{j_{n+1}}
\newcommand{\ynP}{y_{n+1}}

\newcommand{\inkr}{i_n(kr)}
\newcommand{\knkr}{k_n(kr)}

\newcommand{\mix}{w}
\newcommand{\g}{\chi}
\newcommand{\mixm}{\mix_{nm}^-}
\newcommand{\mixp}{\mix_{nm}^+}

\newcommand{\shapen}{p}
\newcommand{\shapem}{q}
\newcommand{\sumpq}{\sum_{\shapen,\shapem}}
\newcommand{\sumnmpq}{\sum_{n',m',\shapen,\shapem}}
\newcommand{\gpq}{\g^\bdy_{\shapen\shapem}}
\newcommand{\gpqo}{\g^\Nl_{\shapen\shapem}}

\newcommand{\low}{l}

\newcommand{\bdy}{\low}
\newcommand{\Nl}{N}
\newcommand{\il}{\mathrm{i}}

\newcommand{\bcdev}{\varepsilon}
\newcommand{\bdb}{\bcdev_\bdy}
\newcommand{\bdo}{\bcdev_\Nl}

\newcommand{\rb}{r_\low}
\newcommand{\rbavg}{\overline{r}_\bdy}

\newcommand{\roB}{R}
\newcommand{\ko}{k_\Nl}

\newcommand{\abari}{\overline{a}_{nm}^{\il}}

\newcommand{\Amp}{\mathcal{A}}
\newcommand{\Ampne}{\Amp_n^e}
\newcommand{\rbtp}{r_\bdy(\theta,\phi)}

\newcommand{\SYchar}{\Xi}

\newcommand{\dSYrev}{\SYchar_{n'm'\shapen\shapem}^{\star\,nm}}

\newcommand{\Ampnti}{\Amp_{n}^{t,\il}}
\newcommand{\Ampnpd}{\Amp_{n'}^\star}
\newcommand{\Ampnto}{\Amp_{n}^{t,\Nl}}
\newcommand{\K}{K}

\newcommand{\Kni}{\K_n^\il}
\newcommand{\ki}{k_{\il}}
\newcommand{\kiP}{k_{\il+1}}

\newcommand{\Dnmi}{\Delta_{nm}^{\il}}
\newcommand{\bdi}{\bcdev_\il}
\newcommand{\rbi}{\overline{r}_\il}

\newcommand{\jniip}{j_{n'}^{\il,\il}}
\newcommand{\yniip}{y_{n'}^{\il,\il}}

\newcommand{\bPnip}{\bP^\il_{n'}}

\newcommand{\gpqi}{\g^\il_{\shapen\shapem}}

\myexternaldocument{aux_all}

\renewcommand*{\thefootnote}{\fnsymbol{footnote}}

\date{May 17, 2021}

\begin{document}
	
	\title[mode=title]{An analytic solution for evaluating the magnetic field induced from an arbitrary, asymmetric ocean world}
	\shorttitle{Magnetic induction from asymmetric ocean worlds}
	\shortauthors{M.J.~Styczinski et al.}
	
	\author[1,2]{Marshall J. Styczinski}[orcid=0000-0003-4048-5914]
	\cormark[1]
	\ead{mjstyczi@uw.edu}
	\credit{Conceptualization, calculations, software, writing}
	\cortext[cor1]{Corresponding author}
	\address[1]{Department of Physics, University of Washington, Box 351560, 3910 15th Ave NE, Seattle, WA 98195-1560, USA.}
	\address[2]{UW Astrobiology Program, University of Washington, Box 351580, 3910 15th Ave NE, Seattle, WA 98195-1580, USA.}
	
	\author[3]{Steven D. Vance}[orcid=0000-0002-4242-3293]
	\ead{svance@jpl.caltech.edu}
	\credit{Geophysical models, applications, mentoring}
	\address[3]{Jet Propulsion Laboratory, California Institute of Technology, 4800 Oak Grove Dr, Pasadena, CA 91109-8001, USA.}
	
	\author[2,4]{Erika M. Harnett}[orcid=0000-0002-0025-7274]
	\ead{eharnett@uw.edu}
	\credit{Mentoring, editing}
	\address[4]{Department of Earth and Space Sciences, University of Washington, Box 351310, 4000 15th Ave NE, Seattle, WA 98195-1310, USA.}
	
	\author[3]{Corey J. Cochrane}[orcid=0000-0002-4935-1472]
	\ead{corey.j.cochrane@jpl.nasa.gov}
	\credit{External field models}
	
	\begin{abstract}
		Magnetic investigations of icy moons have provided some of the most compelling evidence available confirming the presence of subsurface, liquid water oceans.
		In the exploration of ocean moons, especially Europa, there is a need for mathematical models capable of predicting the magnetic fields induced under a variety of conditions, including in the case of asymmetric oceans.
		Existing models are limited to either spherical symmetry or assume an ocean with infinite conductivity.
		In this work, we derive an analytic result capable of determining the induced magnetic moments for an arbitrary, layered body.
		Crucially, we find that degree-2 tidal deformation results in changes to the induced dipole moments.
		We demonstrate application of our results to models of plausible asymmetry from the literature within the oceans of Europa and Miranda and the ionospheres of Callisto and Triton.
		For the models we consider, we find that in the asymmetric case, the induced magnetic field differs by more than $2\siu{nT}$ near the surface of Europa, $0.25$--$0.5\siu{nT}$ at $1\,R$ above Miranda and Triton, and is essentially unchanged for Callisto.
		For Miranda and Triton, this difference is as much as 20--30\% of the induced field magnitude.
		If measurements near the moons can be made precisely to better than a few tenths of a nT, these values may be used by future spacecraft investigations to characterize asymmetry within the interior of icy moons.
	\end{abstract}
	
	
	\maketitle
	
	\section{Introduction}
	
	Any moon subjected to an oscillating magnetic field is a candidate for magnetic sounding---the determination of subsurface properties from magnetic measurements.
	Liquid water oceans containing dissolved salts conduct electricity, causing them to respond to oscillating fields that may be applied by their parent planet.
	This concept has proven especially useful in the study of icy moons in the outer solar system, as the induced magnetic fields that result from this interaction have been measured by orbiting spacecraft.
	
	Induced magnetic fields observed by analysis of \textit{Galileo} magnetometer data have provided the most compelling direct detection of oceans within Jupiter's moons Europa, Ganymede, and Callisto \citep{kivelson2000galileo,zimmer2000subsurface,kivelson2002permanent}.
	This demonstrated methodology will be deployed by the planned \textit{Europa Clipper} and \textit{JUICE} missions to characterize the oceans of these bodies.
	The same technique also shows promise for the investigation of the large moons of Uranus and Neptune \citep{khurana2019single,weiss2020agu,cochrane2021search,arridge2021electromagnetic}; magnetometer investigations are likely to play a central role in future missions to these moons, too.
	
	Magnetic sounding of moons requires a method for calculating the \textit{expected} induced field that would result from a particular model of the interior conductivity structure, based on the measured oscillation of the parent planet's magnetosphere far from the moon.
	Forward models of planetary magnetic induction typically assume a spherically symmetric body for simplicity, as in the oft-cited \citet[Ch. 5]{parkinson1983introduction}.
	Various interior models can then be supposed, the results calculated, and induced fields compared against spacecraft measurements near the moon to determine which model best fits the measurements.
	However, to date it has not been possible to analytically determine the induced magnetic field resulting from excitation of a planetary ocean that is not spherically symmetric.
	
	The ability to suppose an asymmetric model for the conductivity structure of a moon is critical to an interpretation of measurements from upcoming missions, especially \textit{Europa Clipper,} which plans flybys of Europa just $25\siu{km}$ above the surface---$0.02R_E$ in altitude \citep{campagnola2019tour}.
	Close to the moon, induced magnetic moments of quadrupole and higher order will have their largest effect relative to the dominant dipole moments.
	However, a spherically symmetric solution would predict only dipole moments are induced.
	
	Further, asymmetries in the shape of subsurface oceans are fully expected, especially in the degree-2 shapes ({\eg}\ $J_2$ and $C_{22}$ in the gravitational moments).
	The moons are expected to be oblate spheroids because they spin as they orbit, and the asymmetric gravity field applied by their parent planet will introduce a bulging pattern.
	These shapes can even impact the induced dipole moments, by affecting where electric currents can flow within the body.
	Interpretation of magnetic measurements near icy moons will require an understanding of the effects asymmetry may have on the induced field.
	Sufficient measurements may even permit a detailed characterization of the asymmetric interior shape.
	
	In this work, we refine our previously derived solution for highly conducting oceans \citep{styczinski2021induced}.
	Here, we generalize the approach to apply to cases of arbitrary conductivity, described by uniformly conducting layers with boundaries of arbitrary, but near-spherical shape\footnote{
		The meaning of ``near-spherical'' is clarified in \S\ref{sec:nonspherical}---in short, this assumption should hold well in considering asymmetric oceans for most solar system moons.
		The consequences of our assumptions are elaborated in \S\ref{sec:limitations}.
	}.
	Then, we apply our mathematical result to plausible models of asymmetry within the conductivity structure of the moons Europa, Miranda, Callisto, and Triton, informed by the literature and calculated using the geophysical modeling framework \textit{PlanetProfile} \citep{vance2018geophysical,vance2021magnetic}.
	A summary of the results from the example cases studied is in \S\ref{sec:application}; a full collection of the model results is contained in the Supplemental Text, \S\ref{supp:fullResults}.
	
	The excitation field applied by the parent planet typically has complicated interactions with the asymmetric boundary shape, resulting in changes to the magnetic field that vary with latitude, longitude, and time.
	To simplify the analysis of the examples we study, we choose to evaluate the magnetic fields at the J2000 reference epoch, approximately noon UTC on January 1, 2000.
	We include animations of the differences in induced magnetic field throughout the synodic period for some components, as well as plots showing the differences in all vector components of the induced field, in the Supplemental Text (\Fs\ref{fig:suppEuropaHigh}--\ref{fig:suppTriton}).
	The Supplemental Material also contains Python code\footnote{Also available as a Zenodo archive: \url{https://zenodo.org/record/5002956}.} for evaluating our model, including creation of all plots in this manuscript.
	
	\section{Methods}\label{sec:methods}
	At core, we apply Maxwell's laws at boundaries between regions and find the induced magnetic moments that satisfy the resulting equations.
	The same is true of the spherically symmetric, recursive solution developed by \citet{srivastava1966theory} and prominently applied by \citet{zimmer2000subsurface}, \citet{seufert2011multi}, and others.
	However, in our application, we describe the boundaries between regions of varying electrical conductivity by an arbitrary shape, rather than by a constant radius.
	The shape of the boundary has important consequences, and determines which magnetic moments are induced by a given excitation field applied by the parent planet.
	Here, we summarize the key points in our derivation and the final result for determining the induced magnetic field for an asymmetric conducting ocean.
	
	A derivation of the Srivastava recursive method is recounted in more modern notation by the oft-cited \citet{parkinson1983introduction}.
	The Parkinson derivation has formed the basis for several influential studies of magnetic sounding of Europa, notably \citet{zimmer2000subsurface}, \citet{khurana2002searching}, and \citet{hand2007empirical}.
	The Parkinson derivation contains several errors and inconsistencies that were repeated by the aforementioned studies, so we cover each step in our derivation from first principles and in fine detail in the Supplemental Text (\S\ref{supp:derivation}).
	Despite the errors, the Parkinson derivation still gives the physically correct result for a spherically symmetric conductor.
	The consequences on the comparison to prior work that result from these errors are described in \S\ref{sec:parkinsonErrors}.
	
	The key steps in our approach are:
	\begin{enumerate}
		\item Define a radial electrical conductivity structure within the body.
		\item Apply Maxwell's laws at the boundaries between uniformly conducting regions.
		\item Expand the boundary radii in a first-order Taylor expansion in terms of spherical harmonics.
		\item Solve the resulting linear system of equations for the induced magnetic moments.
	\end{enumerate}
	Two main approximations are required for this approach to adequately describe a physical system.
	
	First, we must assume that the electrical conductivity is uniform within each layer within the planetary body.
	This assumption holds approximately for any planetary body with layers that are global in extent, although geophysical conditions changing with depth require a greater number of radial layers in order to be adequately represented \citep{vance2021magnetic}.
	The consequence of this assumption is that our model is not expected to be capable of predicting the induced field of a body containing localized melt lenses \citep{schmidt2011active}, sills \citep{michaut2014domes}, {\etc}\ within an ice shell.
	
	Second, we assume that the boundaries are \textit{near-spherical,} in that we may retain only the first term in a Taylor expansion about the boundary radius.
	This approximation is valid if the differences expected from the second-order terms are negligible, {\ie}\ smaller than the expected measurement precision.
	Second-order terms are expected to be negligible for all icy moons in the solar system.
	The validity of this approximation is discussed further in \S\ref{sec:taylor}.
	
	\subsection{Basic physics of magnetic induction}\label{sec:equations}
	\nocite{dennery2012mathematics,arfken2012mathematical}
	In the presence of an electrical conductor, an oscillating magnetic field will induce a secondary, induced magnetic field in accordance with Maxwell's laws.
	The strength of the induced field depends on the angular frequency $\omega$ of the oscillating applied field (called the \textit{excitation} field) and the material properties everywhere within the conducting body: the electrical conductivity $\sigma$ and the magnetic permittivity $\mu$.
	It is customary to assume $\mu=\mu_o$, as this approximation holds very well on planetary scales \citep{saur2010induced}.
	The faster the oscillation and the greater the electrical conductivity, the stronger the induced magnetic field, up to a maximum such that the radial component of the excitation field is exactly cancelled at the outer boundary of a perfect conductor.
	
	For simplicity in deriving our model, we neglect movement of conducting material within the body (as in the case of ocean currents), which can itself induce secondary fields \citep{saur2010induced,vance2021magnetic}.
	We also neglect currents outside the body.
	Although currents in the plasma environment around the body are not generally negligible, the principle of superposition permits us to consider each contribution to the net electromagnetic response independently---the net magnetic field is then the sum from each individual contribution.
	In this work, we consider only the induced magnetic moments generated by the interaction of the primary excitation field with the conducting body, as this is the dominant interaction that induces magnetic fields from within solar system moons.
	
	Inside the conducting body, the net magnetic field $\B$ obeys a diffusion equation:
	\begin{equation}
		\lapl\B = \mu\sigma\dbdt\label{eqm:diffusion},
	\end{equation}
	where $\mu$ is magnetic permeability and $\sigma$ is electrical conductivity, both functions of position.
	For a sinusoidal time dependence $\eiwt$ with angular frequency $\omega$,
	\begin{equation}
		\lapl\B = -i\omega\mu\sigma\B\label{eqm:diffusion2},
	\end{equation}
	or alternatively
	\begin{equation}
		\lapl\B = -k^2\B\label{eqm:helmholtz},
	\end{equation}
	where $k$ is known as the wavenumber.
	\Eq\ref{eqm:helmholtz} is a vector Helmholtz equation, with
	\begin{equation}
		k = \sqrt{i\omega\mu\sigma}.\label{eqm:kdef}
	\end{equation}
	Outside the body, both the excitation field and the induced field, as well as their sum, obey the Laplace equation:
	\begin{equation}
		\lapl\B = 0\label{eqm:laplace}.
	\end{equation}
	In this outer region, the magnetic field can be determined from \citep{jackson1999classical}
	\begin{equation}
		\Bnet = -\nabla\left[\sum_{n,m}\roB\left(\Benm\rova^n + \Binm\aovr^{n+1}\right)\Ynm(\theta,\phi)\right]\label{eqm:multipole},
	\end{equation}
	where the excitation moments $\Benm$ are determined by measuring the oscillating field far from the body, $\Binm$ are the induced magnetic moments, $\roB$ is a unit of radial distance (typically the outer radius of the conducting body), and $\Ynm$ are the spherical harmonics.
	
	As the induced field can only be as strong as the excitation field, and that only happens in the limit of high conductivity, it is often convenient to consider the ratio of the induced and excitation moments, $\Binm/\Benm$, sometimes called the ``$Q$-response.''
	In the case of spherical symmetry and in the high-conductivity limit, this ratio approaches $n/(n+1)$.
	Because $\Binm/\Benm$ can be complex, we define the complex amplitude $\Ampne$ such that
	\begin{equation}
		\Binm = \frac{n}{n+1}\Ampne\Benm\label{eqm:ampdef}.
	\end{equation}
	
	\subsubsection{Comparison to derivations used in prior work}\label{sec:parkinsonErrors}
	In the commonly studied case of a uniform excitation field, with $n=1$, $\Ampne$ may be expressed in terms of the real amplitude $A$ and phase delay $\upphi$ \citep[{\eg}\ as defined by][]{zimmer2000subsurface}:
	\begin{equation}
		\Amp^e_1 = Ae^{-i\upphi}, \label{eqm:aeip} 
	\end{equation}
	allowing for ready comparison with prior work.
	
	Note that the negative exponent in \Eq\ref{eqm:aeip} results from our definition of $k$ in \Eq\ref{eqm:kdef}, with $+i$ under the square root, and the contrasting choice of \citet{parkinson1983introduction}---and several authors of prior work who have relied on the Parkinson derivation, notably \citet{zimmer2000subsurface}, \citet{khurana2002searching}, \citet{hand2007empirical}, and \citet{arridge2021electromagnetic}.
	The \citet{parkinson1983introduction} derivation contains an error, in that choosing $-i$ in the definition of $k$ results in the \textit{modified} spherical Bessel equation, with the \textit{modified} spherical Bessel functions $\inkr$ and $\knkr$ as solutions, as detailed in \citet{schilling2007time} and \citet{seufert2011multi}.
	\citet{parkinson1983introduction} incorrectly arrives at the standard spherical Bessel equation (\Eq\ref{eq:bessel}).
	In effect, both our input wavenumber $k$ and our result for the complex response amplitude $\Amp$ are equal to the complex conjugate of the analogous quantities from prior work, hence the negative exponent in \Eq\ref{eqm:aeip}.
	See \S\ref{supp:errors} for more information.
	We define the relationship between $\Amp^e_1$, $A$, and $\upphi$ as in \Eq\ref{eqm:aeip} to facilitate comparison with the rich set of prior work related to this topic.
	
	\subsection{Radial conductivity structure}\label{sec:structure}
	Within the planetary body, we must suppose a radial conductivity structure $\sigma(r)$.
	Solution of the vector Helmholtz equation (\Eq\ref{eqm:helmholtz}) in spherical coordinates is made a tractable problem by our assumption that each layer has uniform conductivity.
	In this work, we use the \textit{PlanetProfile} framework \citep{vance2018geophysical,vance2021magnetic} to model the electrical conductivity of ocean waters under geophysical conditions likely found within icy moons.
	The conductivity within ice and rock are typically small ($<10^{-2}\siu{S/m}$, \citeauthor{glover2015geophysical}, \citeyear{glover2015geophysical}).
	The conductivity of an iron core may be much higher ($>10^6\siu{S/m}$, \citeauthor{khurana2002searching}, \citeyear{khurana2002searching}), but its contribution to the induction response is likely a few percent at most \citep{seufert2011multi} due to its smaller size and screening of the excitation field by the ocean.
	We neglect the conductivity of ice, rock, and a possible iron core, but these are easily inserted into our model.
	
	We require a spherically symmetric conductivity model as a starting point for our derivation because we assume only the boundaries between the conducting regions are asymmetric.
	This is a requirement for an analytic solution of the boundary condition equations.
	Although the conductivity within each asymmetric layer is likely to differ with a change in depth, our near-spherical approximation ensures that this difference is small compared to the global effect of the asymmetric boundary.
	This approximation is most valid when there are sharp contrasts in conductivity between the layers, {\eg}\ between the ocean and nonconducting ice shell, because the depth-dependence of the conductivity in the ocean has a much smaller effect on conductivity than the layer transition.
	
	\subsection{Applying Maxwell's laws at the boundaries}\label{sec:maxwell}
	In spherical coordinates, the vector components of general solutions to \Eqs\ref{eqm:diffusion} and \ref{eqm:laplace} are spherical harmonics and their $\theta$ derivatives, multiplied by either spherical Bessel functions and their $r$ derivatives (inside) or a power series in $r$ (outside) \citep{moffatt1978field,jackson1999classical}.
	As we have assumed $\mu=\mu_o$ everywhere, and currents are not confined to the boundary surfaces, Maxwell's laws dictate that the vector components of $\B$ must be continuous at every boundary.
	This continuity gives us equalities that relate coefficients for the general solutions that describe each region, and ultimately will yield a solvable set of linear equations.
	
	In this work, we use the fully normalized, complex spherical harmonics:
	\begin{align}
		\Ynm &= \sqrt{\frac{2n+1}{4\pi}\frac{(n-m)!}{(n+m)!}}\,\Pnmu(\cn)\,\eimp, \\
		\YnmM &= (-1)^m\Ynm^* \label{eqm:YnmConj}
	\end{align}
	for degree $n$ and order $m$, where $\Pnmu$ are the associated Legendre functions with the Condon--Shortley phase \citep{condonshortley}.
	In contrast, much of the geomagnetism and geodetics communities use real-valued spherical harmonics, in terms of $\sin\phi$ and $\cos\phi$, in the Schmidt normalization, and omit the Condon--Shortley phase.
	We choose our convention out of necessity to match that common in quantum physics, so that we may use results from historic works \citep{wigner1931gruppen,brink1968angular,edmonds1996angular} involving the products of spherical harmonics.
	See \S\ref{supp:derivation} for more details.
	
	For the spherical harmonics, we define
	\begin{align}
		\dYnm \equiv \dydth{nm} &= \frac{1}{\sn}\left( -\mixm Y_{n-1,m} + \mixp Y_{n+1,m} \right) \quad \mathrm{for}~ m \ge 0, \label{eqm:dYdef}\\
		\mixm &= (n+1)\sqrt{\frac{n^2 - m^2}{(2n-1)(2n+1)}}, \quad \mixp = n\sqrt{\frac{(n+1)^2 - m^2}{(2n+1)(2n+3)}}, \label{eqm:wdef}\\
		\dYnmM &= (-1)^m\dYnmConj \label{eqm:dYconj},
	\end{align}
	where $*$ denotes a complex conjugate.
	For spherical Bessel functions of the first and second kind $\jn(kr)$ and $\yn(kr)$, we define
	\begin{equation}
		\jd \equiv \ddr(r\jn) = (n+1)\jn - kr\,\jnP, \quad
		\yd \equiv \ddr(r\yn) = (n+1)\yn - kr\,\ynP \label{eqm:jderiv}.
	\end{equation}
	The starred functions appear in the tangential boundary condition equations and the overall solutions.
	
	\subsection{Near-spherical boundary shapes}\label{sec:nonspherical}
	We must define surfaces $\rbtp$ for arbitrary boundaries that we will insert into the internal and external boundary condition equations.
	We choose boundary surfaces of the form
	\begin{equation}
		\rbtp = \rbavg + \bdb\sumpq\gpq\Ypq(\theta,\phi) \label{eqm:rio}
	\end{equation}
	because expanding in spherical harmonics will allow us to make use of well-known relations from the mathematics of quantum mechanics.
	Here, $\rbavg$ is the mean radius of boundary $\bdy$, $\bdb$ is the amplitude of deviation from spherical symmetry, $\gpq$ is a dimensionless constant that indicates the relative amount of each harmonic represented in the boundary surface, and $\Ypq$ are fully normalized spherical harmonics of degree $\shapen$ and order $\shapem$.
	We use the index $\low$ to indicate that this surface describes the outer boundary of the lower region.
	Care must be taken to select values for $\g$ that prevent boundary surfaces from overlapping, so as to describe physically possible interior structures.
	Contour maps plotting the difference between layers are instructive for this purpose, as in \F\ref{fig:mirandaContour}, which was created with the analysis software we provide as Supplemental Material.
	
	As we wish to describe physical surfaces, \Eq\ref{eqm:rio} must be real-valued.
	The spherical harmonics we use are complex in general; so too must $\gpq$ be.
	Topographical descriptions in the literature against which we wish to compare are often described in terms of coefficients for real-valued, Schmidt semi-normalized spherical harmonics.
	The provided software includes a tool for converting this format into the coefficients $\gpq$, as we have used in our analysis.
	In \Eq\ref{eqm:rio}, for $\rbavg$ in meters, $\bdb$ is also in meters.
	For a perturbation represented by a pure harmonic of degree $\shapen$ and order $\shapem$, $\bdb$ is the maximum radial displacement from a perfect sphere, and $\gpq$ is 1/$\max(\Ypq)$.
	
	Surfaces described by \Eq\ref{eqm:rio} are near-spherical in that we make the approximation $\bdb \ll \rbavg$ for all $\bdy$.
	Equivalently, $\bdb/\rbavg \ll 1$, and we retain terms up to first order in $\bdb/\rbavg$ only.
	This enables us to use a Taylor expansion in the boundary conditions that truncates quickly, adding only one term containing a product of two spherical harmonics.
	Products of spherical harmonics may be expressed as a linear combination of different harmonics \citep{wigner1931gruppen,condonshortley}.
	This results in ``mixing'' of harmonics in the excitation field from $n=1$ into other $n$, so a uniform excitation field induces magnetic moments of quadrupole order or higher for this shape, in addition to altering the original dipole moments.
	
	\subsubsection{Taylor expansion of boundary shapes}\label{sec:taylor}
	We account for asymmetry in the conductivity structure of the body by expanding the boundary surfaces about their average radii.
	To first order, a Taylor expansion of a function $f(r)$ about $\rb$ has terms
	\begin{equation}
		f(\rb) \approx f(\rbavg) + (\rb-\rbavg)\frac{\partial f(r)}{\partial r}\Bigr|_{r=\rbavg} = f(\rbavg) + \bdb\Big[\sumpq\gpq\Ypq(\theta,\phi)\Big]\frac{\partial f(r)}{\partial r}\Bigr|_{r=\rbavg}.\label{eqm:Taylor}
	\end{equation}
	We evaluate these terms, then insert the result in place of the boundary radius in the general expressions for the field within each region (\Eqs\ref{eq:brext}--\ref{eq:bpint}) in order to match the components of the field at the boundaries.
	
	The next term in the expansion (the second-order term) will be proportional to $\bdb^2$ and the second derivative of $f(r)$.
	Each derivative will contribute a division by $r$ \citep[from the Bessel functions;][]{marion1980classical}, so each successive term gains another power of $\bdb/\rbavg$, the ratio of the maximum deviation from spherical symmetry to the average radius of the boundary surface.
	We assume the boundaries are near-spherical, as $\bdb/\rbavg$ is small---for Enceladus, for example, the maximum possible value is about $21\siu{km}/231\siu{km}\approx 0.09$ \citep{hemingway2019enceladus}.
	For Europa, the maximum ratio is even smaller, at about $30\siu{km}/1530\siu{km}\approx 0.02$ \citep{billings2005great}.
	The excitation moments experienced by icy moons in the solar system range from $7\siu{nT}$ for Triton up to about $300\siu{nT}$ for Miranda \citep{cochrane2021search}.
	The largest differences in magnetic field at the surface of these bodies from the first-order term can be about 10\% for highly conducting, maximally asymmetric oceans.
	The second-order term can then only be about 1\% of the maximum, up to a few $\si{nT}$ on the surface.
	A lander on a highly asymmetric moon may need to account for second-order effects.
	However, at orbital/flyby altitudes and with realistic ocean shapes and conditions, we find that second-order effects are likely to be under $1\siu{nT}$, and thus insignificant in a dynamic space environment.
	
	\section{Results}\label{sec:results}
	
	\subsection{Analytic solution for the induced magnetic moments}\label{sec:solution}
	Through application of the methods described above, for $\Binm$ of a body containing $\Nl$ conducting layers and an average outer radius $\roB$, we obtain a solution
	\begin{equation}
		\Binm = \frac{n}{n+1}\Ampne\Benm + n\sum_{\il=1}^{\Nl}\Ampnti\Kni\Dnmi, \label{eqm:Binm}
	\end{equation}
	with
	\begin{equation}
		\Dnmi \equiv
		\begin{dcases}
			\sumnmpq\frac{\bdi\gpqi}{\rbi}\dSYrev\, \frac{\abari}{\roB}\left(\jniip + \bPnip\yniip\right)\left(\ki^2\rbi^2 - \kiP^2\rbi^2\right) & \mathrm{for}~\il<\Nl, \\
			\sumnmpq\frac{\bdo\gpqo}{\roB}\dSYrev\, \frac{2n'+1}{n'+1}\Ampnpd\Benmp & \mathrm{for}~\il=\Nl.
		\end{dcases} \label{eqm:Dnmi}
	\end{equation}
	Each quantity is fully determined by the interior structure model and the excitation moments, so the induced magnetic field can be directly calculated.
	$\Ampne$ is the complex amplitude for the zeroth-order (spherically symmetric) excitation, $\Ampnti$ is the complex amplitude from the ``output'' harmonic $n,m$ for the tangential component from the $\il^\text{th}$ layer, and $\Ampnpd$ is the complex amplitude from the ``input'' harmonic that mixes into the output harmonic.
	As with $\Ampne$, the product $\Ampnto\Ampnpd$ is asymptotic to $(1+0i)$ in the limit $|\ko\roB|\rightarrow\infty$.
	The radial first-order term that would multiply a quantity $\Amp^r$ analogous to $\Amp^t$ is identically zero, so it does not appear.
	$\Dnmi$ represents the strength of the induced magnetic field from each asymmetric boundary that ``mixes'' into the output harmonic $n,m$ from the known excitation field.
	$\Kni$ is a propagation factor that determines how the induced magnetic field resulting from a buried asymmetric boundary gets attenuated by the conducting layers above it.
	$\dSYrev$ is a complicated function of Clebsch--Gordan coefficients that determines the strength of coupling between the given excitation moment $n',m'$, a given boundary shape harmonic $\shapen,\shapem$, and the induced moment $n,m$.
	Please refer to the full derivation in \S\ref{supp:derivation} for more details, including definitions of every quantity.
	
	\Eq\ref{eqm:Binm} is our final result.
	Along with a suitable conductivity model and excitation moments from a magnetospheric model of the parent planet, this result may be applied to determine the induced magnetic field for a non-spherical ocean or ionosphere or both.
	
	\subsection{Application to ocean worlds in our solar system}\label{sec:application}
	\nocite{anderson1998europa,anderson2001shape}
	
	We now apply our model to several examples of icy moons in our solar system that are expected to contain asymmetry in their conductivity structure.
	We provide computer programs written in Python for evaluating our model as Supplemental Material, including the code for generating the plots in this Section.
	A full detailing of the results for each example, including contour maps of the asymmetry model and differences in the induced magnetic field arising from asymmetry, are contained in the Supplemental Text (\S\ref{supp:fullResults}).
	Here, we focus on the reasoning for the modeled asymmetry and the key findings from each application.
	
	Conductivities within the ocean layers are determined using the \textit{PlanetProfile} geophysical modeling framework \citep{vance2018geophysical,vance2021magnetic}.
	Several bulk parameters are used as inputs, such as surface temperature and moment of inertia, which are typically determined from spacecraft measurements; \T\ref{table:suppPlanetProps} lists each of the parameters we used.
	Ice shell thicknesses are assumed for each body.
	The interior structure is then determined self-consistently, accounting for solid-state convection in ice and adiabatic convection in the ocean, and using recently developed Gibbs-energy-based thermodynamics of solids and liquids.
	The ocean is assumed to be well-mixed, with a specified salt composition and concentration that is constant throughout the ocean.
	Depth-dependent pressures and temperatures are then combined with the salinity to determine ocean conductivities using the Gibbs Seawater model \citep{mcdougall2011getting} for Seawater or the model of \citet{larionov1984conductivity} for \ch{MgSO4} oceans.
	Oceans are then reduced to three radial conducting layers before inserting them into our model, to reduce computational load.
	A greater number of layers may impact detailed analyses of spacecraft data, but the impact is expected to be small compared to other effects if ocean composition is uniform with depth \citep{vance2021magnetic}.
	Asymmetry is then applied by describing the boundaries between radial layers as in \Eq\ref{eqm:rio}.
	Ocean layers are assumed to be concentric, in that they each have the same asymmetric shape but scaled proportionally to the radius of the boundary.
	
	Contour maps for Europa and Miranda show the asymmetry we model in the outermost ocean boundary, the ice--ocean interface for these bodies.
	The interiors for Callisto and Triton are assumed spherically symmetric; asymmetry is applied to their ionospheres based on a day--night dichotomy in ionization rate.
	Interior contour maps may appear similar to those seen in the literature showing tidal flexure, but the similarity is only because they are each global maps portraying low-degree spherical harmonics. The difference from diurnal tides is relatively small \citep[about 30\,m for Europa;][]{moore2000tidal} so we assume static asymmetry in the ice--ocean interface.
	As the ionospheric asymmetry we assume for Callisto and Triton is due to day--night differences, the asymmetric shapes for these bodies will oscillate throughout their orbital periods (all the moons we study rotate synchronously).
	However, the strongest excitations applied to Callisto and Triton occur at their synodic periods with the parent planet's rotation, 10--40 times more rapid than their orbital periods.
	For simplicity, we assume a fixed asymmetry in their ionospheres and consider our results an order-of-magnitude estimate for these bodies.
	
	In all cases, the excitation moments applied to the moon are determined by taking a Fourier transform of the magnetic field at the position of the moon, evaluated over a 10-year time series.
	Body positions are determined using SPICE kernels available from NAIF\footnote{Generic SPICE kernels are hosted by NAIF at \url{https://naif.jpl.nasa.gov/naif/data\_generic.html}.}$^,$\footnote{The specific kernels we used are \texttt{pck00010.tpc} and \texttt{gm\_de431.tpc} for planetary constants, \texttt{naif0012.tls} for leap seconds, and \texttt{jup310.bsp}, \texttt{ura111.bsp}, and
		\texttt{nep081.bsp} for ephemerides.}.
	The magnetic field of the parent planet is calculated using a magnetosphere model from the literature: JRM09 + the Connerney current sheet model for Jupiter \citep{connerney2018new,connerney1981modeling}, AH$_5$ for Uranus \citep{herbert2009aurora}, and that of \citet{connerney1991magnetic} for Neptune.
	Although we determine excitation moments for all significant periods, in this work we consider only the largest excitation, at the synodic period.
	
	\subsubsection{Europa}\label{sec:resultsEuropa}
	
	\begin{figure}
		\centering
		\includegraphics[width=\textwidth]{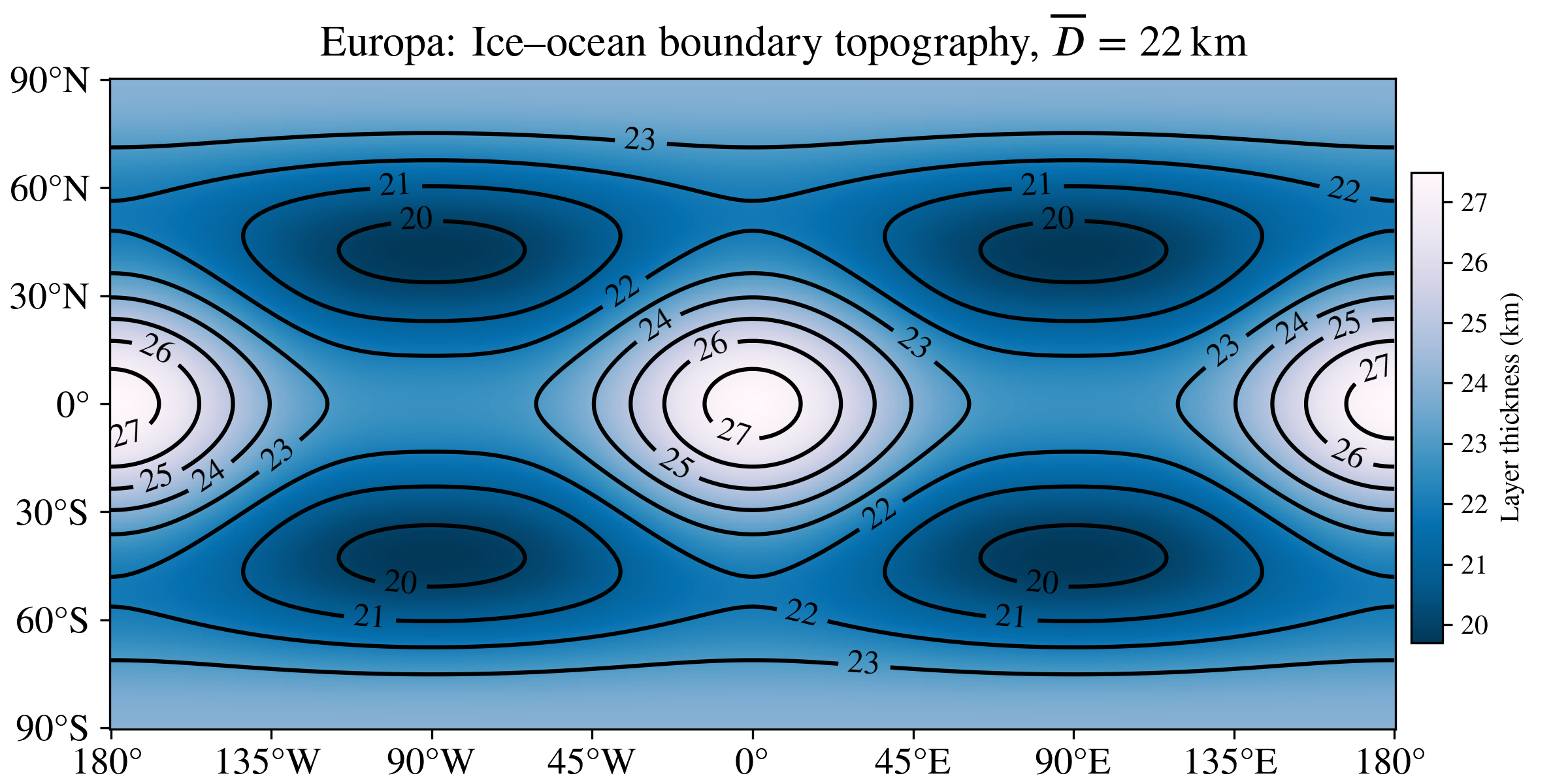}
		\caption{
			Asymmetry model for Europa, showing the thickness of the ice shell as a function of latitude and longitude in IAU coordinates, where $(0^\circ,0^\circ)$ is the sub-jovian point.
			The outer surface of Europa is assumed to be a perfect sphere and the ice--ocean boundary is perturbed.
			Then, both surfaces have tidal deformation added in accordance with $J_2$ and $C_{22}$ values reported by \citet{anderson1998europa}.
			We have supposed an ice shell asymmetry model approximating the results of \citet{tobie2003tidally}; compare to \F12a of that work.
			Ice thickness is $22.5\siu{km}$ on average and ranges from $20$--$27\siu{km}$.
		}
		\label{fig:europaContour}
	\end{figure}
	
	To model plausible asymmetry in the interior of Europa, we suppose a shape approximating the results of \citet{tobie2003tidally} for the ice shell thickness.
	In addition, we suppose a tidally deformed shape consistent with the static gravity coefficients $J_2$ and $C_{22}$ inferred from radio Doppler shifts due to acceleration of the \textit{Galileo} spacecraft \citep{anderson1998europa}---see \S\ref{supp:gravity} for more details.
	\citet{tobie2003tidally} studied the dynamics of tidal heating and convection in Europa's ice shell with numerical simulations, concluding that the shell is likely thickest at the sub- and anti-jovian points and thinnest around $60^\circ$ latitude on the leading and trailing hemispheres.
	\F\ref{fig:europaContour} shows the shape model we use to approximate this ice shell, which we have selected based on \F12a of \citet{tobie2003tidally}.
	We model the ocean using 3 concentric layers with the same boundary shape between layers, scaled to the radius of each boundary.
	A Seawater ocean is assumed; interior structure, including electrical conductivity, is evaluated using \textit{PlanetProfile} as described above (\S\ref{sec:application}).
	Conduction in the core is ignored for simplicity.
	
	\begin{figure}
		\centering
		\includegraphics[width=\textwidth]{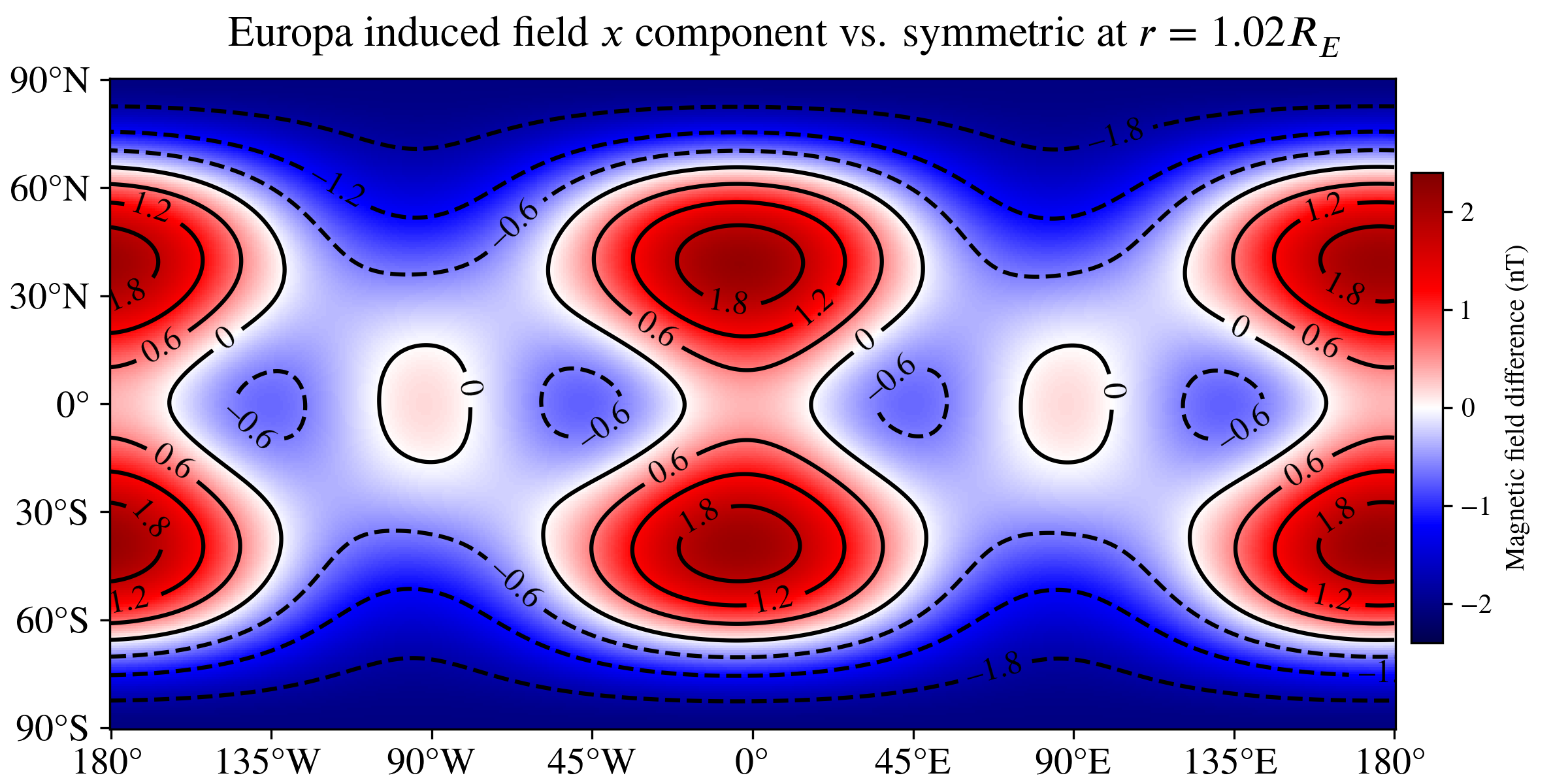}
		\caption{
			Difference in the $B_x$ component (in IAU coordinates) of the induced magnetic field of Europa resulting from asymmetry in the ice--ocean boundary corresponding to the shape displayed in \F\ref{fig:europaContour} and static gravity inferred by \citet{anderson1998europa}.
			A Seawater ocean composition is assumed.
			The magnetic fields are evaluated at the J2000 reference epoch and at $25\siu{km}$ altitude, consistent with the closest approach of several flybys planned by the \textit{Europa Clipper} mission.
			This component of the induced field is changed by more than $1\siu{nT}$ in many locations, demonstrating that expected asymmetry in Europa's ice shell is likely to have measurable effects in the nearest flybys by \textit{Europa Clipper.}
			These closest flybys may help constrain the shape of Europa's ice shell from magnetic measurements if precision reaches a few tenths of a $\siu{nT}$.
			Tidal deformation is responsible for about 1/3 of the difference in this component.
			This global map of differences resulting from asymmetry changes throughout the synodic period; an animation showing the same map as it varies during the 11.2-hour period is included in the Supplemental Material.
		}
		\label{fig:europaHighXdiff}
	\end{figure}\nocite{rambaux2013tides,iess2014gravity}
	
	\F\ref{fig:europaHighXdiff} shows the difference in the $B_x$ component (in IAU coordinates) of the induced magnetic field that results from the asymmetric structure at an altitude of $25\siu{km}$.
	This altitude is chosen to assess the possible impact of our results on the \textit{Europa Clipper} mission, which plans several flybys with closest approaches at this altitude or less \citep{campagnola2019tour}.
	The displayed vector component is aligned with the strongest excitation applied by Jupiter; analogous maps for all components are shown in the Supplemental Material (\F\ref{fig:suppEuropaHigh}).
	Differences of over $1.8\siu{nT}$ are observed in several locations at this time, the J2000 reference epoch.
	Tidal deformation is responsible for about 1/3 of the difference in the $B_x$ component (\F\ref{fig:europaHighXdiff}) and about half of the difference in the $B_z$ component (\F\ref{fig:suppEuropaHighZdiff}), which can be as much as $2.4\siu{nT}$.
	Although the background field of Jupiter is around $500\siu{nT}$ \citep{connerney2018new}, features of order $1\siu{nT}$ are routinely considered measurable in the analysis of spacecraft data \citep{horbury2020solar}.

	\begin{figure}
		\centering
		\includegraphics[width=0.75\textwidth]{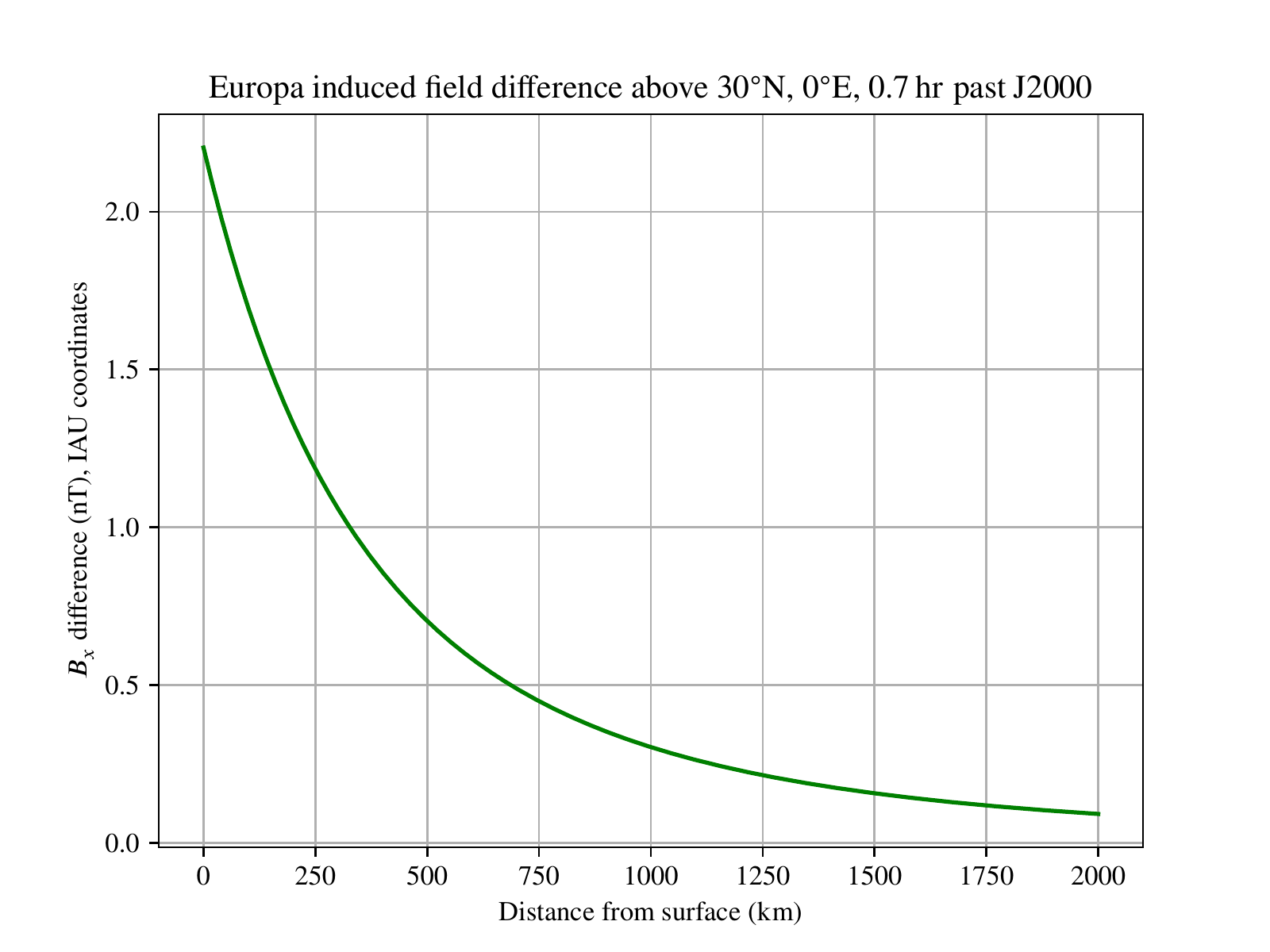}
		\caption{
			Difference in induced field $B_x$ component for our Europa Seawater model as a function of altitude for a fixed point in time.
			The selected surface location $(30^\circ\mathrm{N},0^\circ\mathrm{E})$ and time ($0.7\siu{hr}$ past J2000) maximize the observed difference relative to the spherically symmetric case for this interior model and component---see \F\ref{fig:europaHighXdiff}. 
			Beyond about $1500\siu{km}$ altitude, the difference is around $0.2\siu{nT}$ and likely negligible.
		}
		\label{fig:europaFalloff}
	\end{figure}
	
	\begin{figure}
		\centering
		\includegraphics[width=\textwidth]{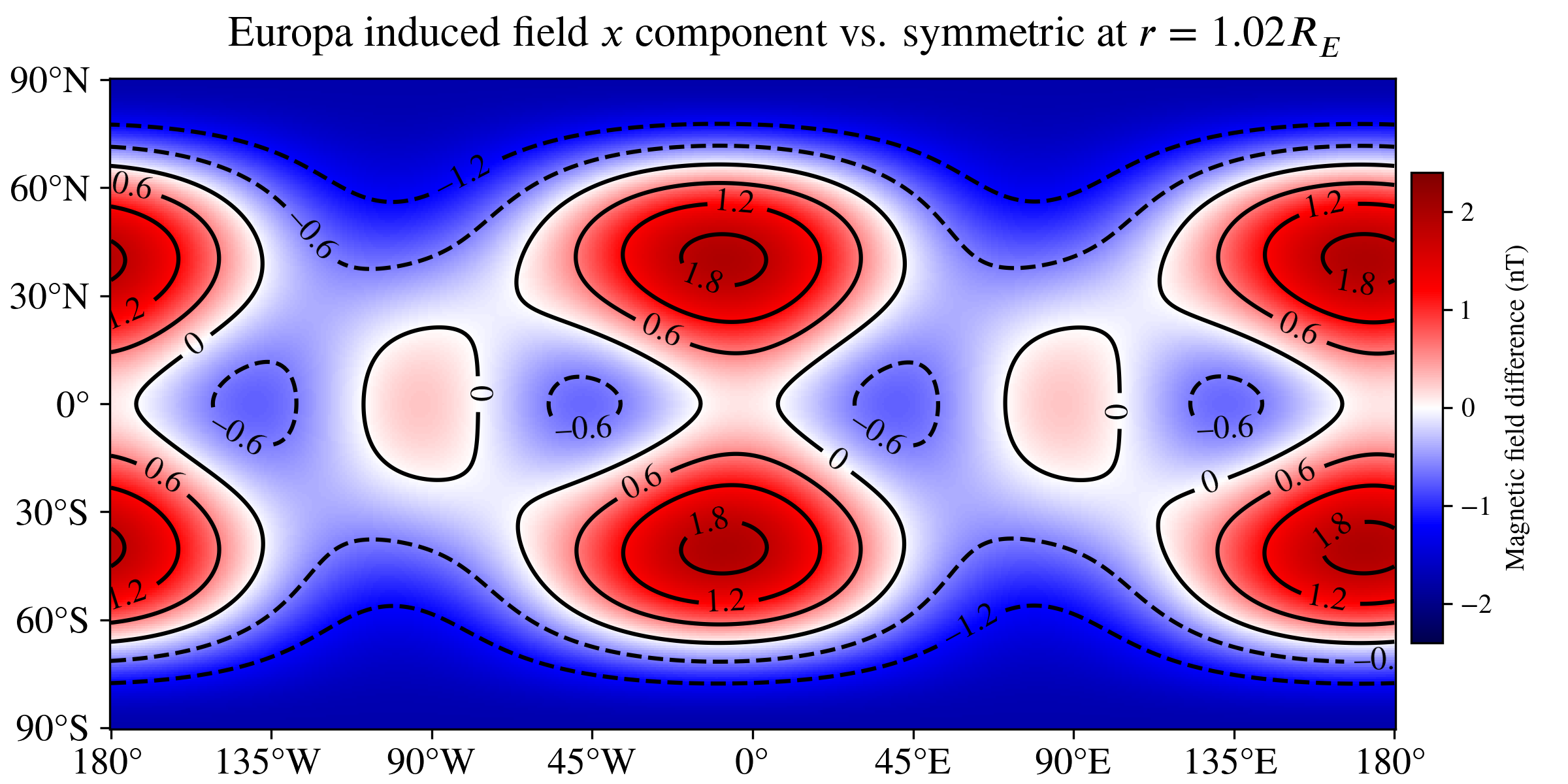}
		\caption{
			Difference in the $B_x$ component of the induced magnetic field of Europa resulting from the asymmetry model shown in \F\ref{fig:europaContour}, as in \F\ref{fig:europaHighXdiff}, but with a salinity 10\% that of Seawater.
			Comparison between the two cases shows these differences to be smaller for this component (60--90\% as great), as well as having a slight phase shift relative to the Seawater ocean.
			The colormap and contours for this plot are fixed to match those of \F\ref{fig:europaHighXdiff}.
		}
		\label{fig:europaLowXdiff}
	\end{figure}
	
	The J2000 epoch is at an arbitrary point in time during Europa's synodic period.
	At about $0.7\siu{hr}$ after J2000, the difference in $B_x$ resulting from asymmetry is maximized.
	As seen in \F\ref{fig:europaHighXdiff}, the difference is also maximized near $30^\circ\mathrm{N}$, $0^\circ\mathrm{E}$.
	To form an initial estimate of how this asymmetric ocean model will impact spacecraft measurements at a variety of altitudes, we evaluated the induced field in a straight line normal to the surface above this location.
	\F\ref{fig:europaFalloff} shows the difference resulting from asymmetry along this line.
	Near the surface, the difference is about $2.2\siu{nT}$.
	The difference drops to $0.2\siu{nT}$, likely small enough to ignore, only above an altitude of about $1500\siu{km}$.
	Note also that the differences are sometimes small nearly everywhere---for example, the differences are below $0.6\siu{nT}$ at all points $25\siu{km}$ above the surface around $t=3.4\siu{hr}$ past J2000 (see \F\ref{fig:suppSmallDiff}).
	
	We also evaluated a near-copy of this model, only with a salinity 10\% that of Seawater, to examine the influence of asymmetry at other concentrations considered in the literature \citep{vance2021magnetic}.
	The difference in the $B_x$ component of the induced magnetic field resulting from asymmetry for this lower-salinity model is shown in \F\ref{fig:europaLowXdiff}.
	In this case, the differences are smaller by 10--40\% and a slight phase shift is observed.
	However, compared to the spherically symmetric analog, the differences are still as much as $2\siu{nT}$.
	Europa's relatively large size causes it to behave as a strong conductor and reject penetration by the time-varying excitation field with a strong induced field; this effect has a much stronger dependence on size than on conductivity or oscillation frequency (through an exponential dependence on $(r\sqrt{i\omega\mu\sigma})$ in Bessel functions $\jn(kr)$, see \Eq\ref{eqm:kdef}).
	
	\subsubsection{Comparison with prior work}\label{sec:resultsPrev}
	
	\begin{figure}
		\centering
		\includegraphics[width=0.75\textwidth]{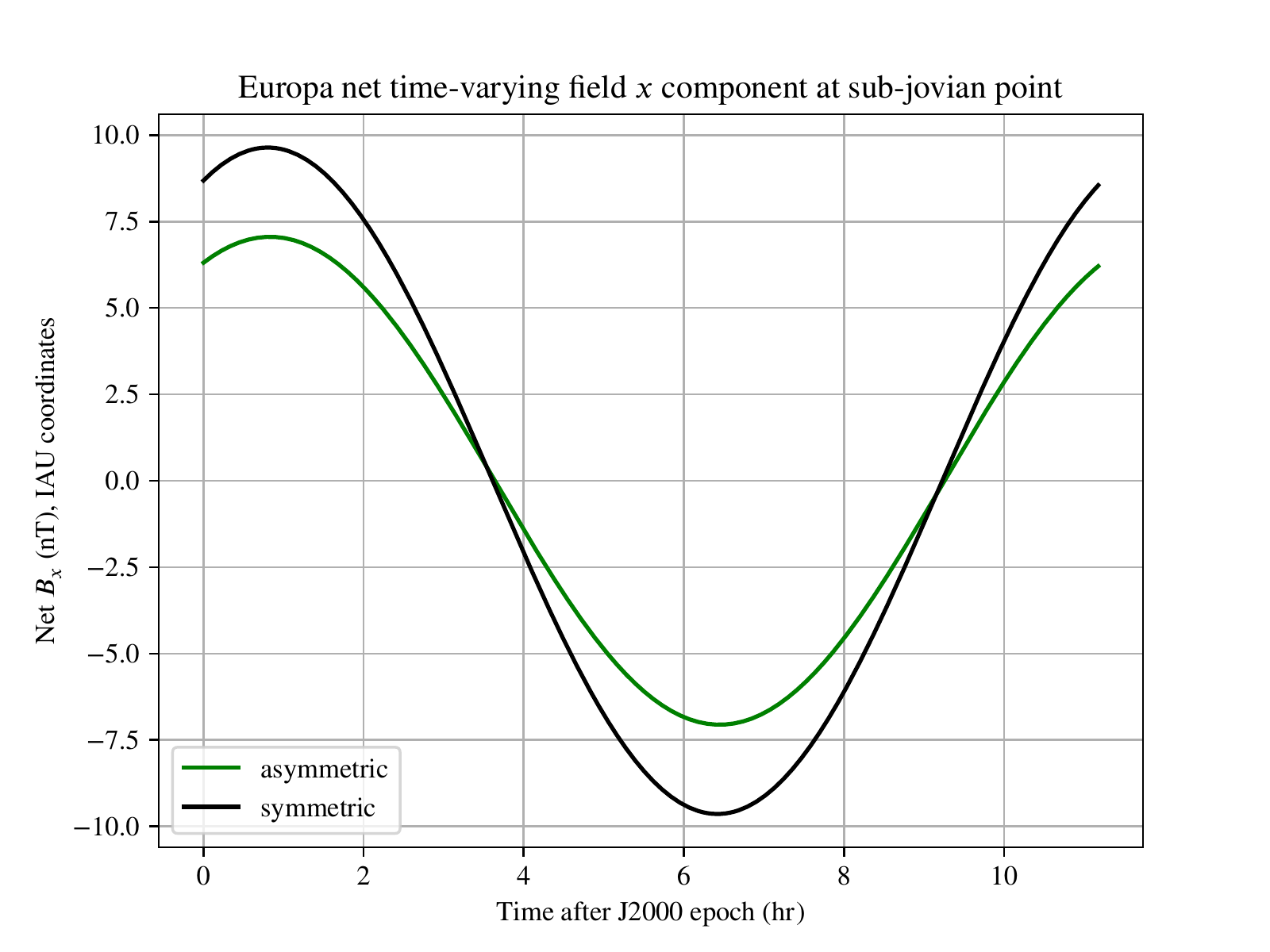}
		\caption{
			Vertical component of the net magnetic field a lander located at the sub-jovian point would measure on Europa for the asymmetric ice--ocean boundary studied in our previous work \citep{styczinski2021induced}, along with that of an analogous, spherically symmetric model.
			The ocean is treated as having effectively infinite conductivity, as in prior work.
			The effects of an ionosphere are neglected.
			The difference between the asymmetric and symmetric cases predicted by our complete model in this work is several times larger than the estimate from our previous work, about $2.5\,\siu{nT}$ at most.
			This model does not include tidal deformation.
		}
		\label{fig:tSeriesLander}
	\end{figure}
	
	In prior work \citep{styczinski2021induced}, we derived an approximation for evaluating the effects of asymmetry on induced fields, but in the limit of high conductivity in the conducting region ($|kr|\rightarrow\infty$).
	In this approximation, only the shape of the outermost boundary matters, because the time-varying excitation field is rapidly attenuated within the conducting material by the induced currents that generate the induced field.
	For comparison with that work, we now consider the exact same boundary shape studied in that work for Europa, {\ie}\ by setting
	\begin{equation}
		\bcdev_\low=2.5\siu{km}, \quad r_\low=1537.5\siu{km}, \quad \g^\low_{2,-2}=\g^\low_{2,2}=\frac{1}{2}\cdot 4\sqrt{\frac{2\pi}{15}}    
	\end{equation}
	in \Eq\ref{eqm:rio} (note the difference in normalization of $\gpq$ in the earlier work).
	\F\ref{fig:tSeriesLander} shows the vertical component of the net magnetic field a lander at the sub-jovian point would measure throughout a synodic period, considering only that excitation period, analogous to \F2 of \citet{styczinski2021induced}.
	In that work, the greatest difference was near the same points in time, though only about $0.5\siu{nT}$.
	
	Armed now with a rigorously derived, explicit formula for evaluating the mixing coefficients $\dSYrev$ (\S\ref{supp:wignerSY}), we find that the reason for the greater effect size compared to our previous work is that we had underestimated these coefficients.
	Our previous method for determining the values of $\dSYrev$ did not account for non-orthogonality of the $\dYnm$ (see \S\ref{supp:nonortho}), and so scaled incorrectly when multiple excitation harmonics or multiple shape harmonics are represented.
	We now estimate a maximum difference of about $2.5\siu{nT}$ for this extreme limiting case, absent tidal deformation.
	
	\subsubsection{Miranda}\label{sec:resultsMiranda}
	
	Miranda is the innermost large moon of Uranus.
	The orbital configuration of the large uranian moons implies past orbital resonances that likely caused tidal heating \citep{cuk2020dynamical}, and Miranda's surface shows possible signs of past geologic activity \citep{beddingfield2020hidden} that might have been provoked by this heating.
	The presence of ammonia-bearing compounds on the surface of nearby Ariel \citep{cartwright2020evidence} implies lowered melting temperatures for the oceans on these moons \citep{croft1988equation}, so liquid oceans may persist to the present even in this far-out system.
	With a radius of $235.8\siu{km}$, Miranda is also comparable in size to Enceladus (radius $252.1\siu{km}$).
	Enceladus has been found to have a marked asymmetry in its ice shell, with a notably thinner portion around the south pole \citep[where plume activity is concentrated;][]{iess2014gravity,hemingway2019enceladus}.
	Together, these factors encourage a comparison between Miranda and Enceladus.
	
	\begin{figure}
		\centering
		\includegraphics[width=\textwidth]{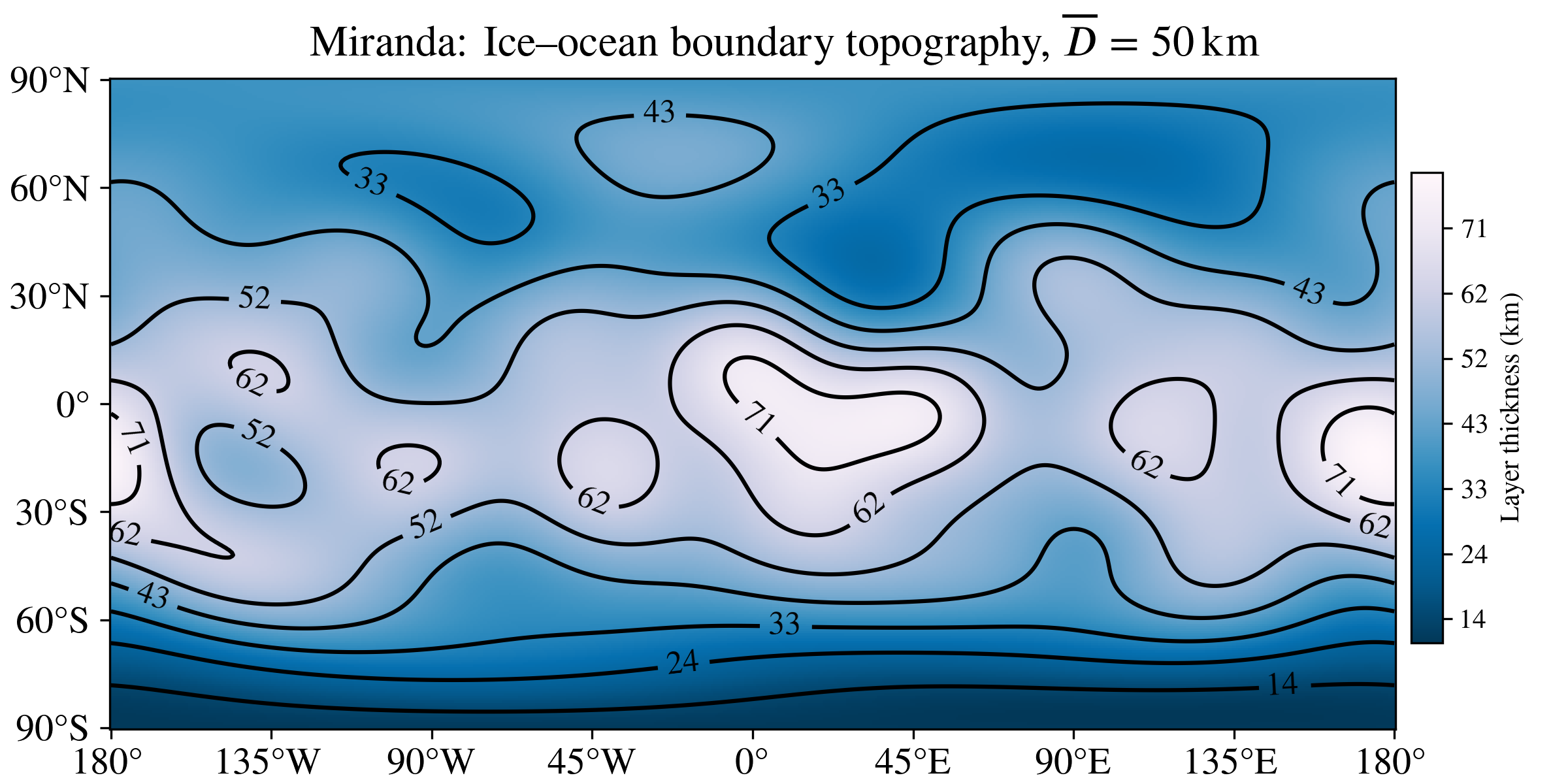}
		\caption{
			Possible ``Enceladus-like'' asymmetry in the ice shell of Miranda used in this study.
			The topography of the ice--ocean boundary for this body is scaled up from a model for Enceladus by \citet{hemingway2019enceladus} inferred from isostatic compensation and spacecraft measurements.
			Compare to \F11d from that work.
		}
		\label{fig:mirandaContour}
	\end{figure}
	
	To model possible Enceladus-like asymmetry in the interior of Miranda, we scale the asymmetric topography of the ice--ocean boundary for Enceladus favored by \citet{hemingway2019enceladus} based on isostatic compensation combined with gravity measurements and models of tidal heating in the ice shell.
	To account for the colder surface temperature of Miranda, which receives about 10\% the insolation as Enceladus, we assume a $50\siu{km}$ thick ice shell, and scale the asymmetry model of \citet{hemingway2019enceladus} from a $20\siu{km}$ average thickness to match.
	We consider this an upper estimate of asymmetry that may be possible within Miranda's interior, as viscous relaxation is likely to smooth out the exaggerated features somewhat.
	The shape of the ice--ocean boundary model we have supposed for Miranda is shown in \F\ref{fig:mirandaContour}---compare to \F11d of \citet{hemingway2019enceladus}.
	We assume an ocean composition of Seawater with 10\% the salinity of Earth's oceans, approximating the salinity of Enceladus inferred from \textit{Cassini} Dust Analyzer sampling of the plumes \citep{postberg2009sodium}.
	
	Miranda's small size and proximity to Uranus prevent it from retaining a significant ionosphere.
	Comparison of the plasma environment \citep{mauk1987hot} to that near Callisto has led us to suppose a simple, uniformly conducting ionosphere that extends from the surface to $100\siu{km}$ altitude and has a total height-integrated conductivity of $800\siu{km}$ \citep{hartkorn2017structure} to serve as an upper limit.
	We assume a spherically symmetric ionosphere for Miranda; conduction in the ionosphere will serve only to attenuate the differences resulting from asymmetry as measured outside the body.
	
	\begin{figure}
		\centering
		\includegraphics[width=\textwidth]{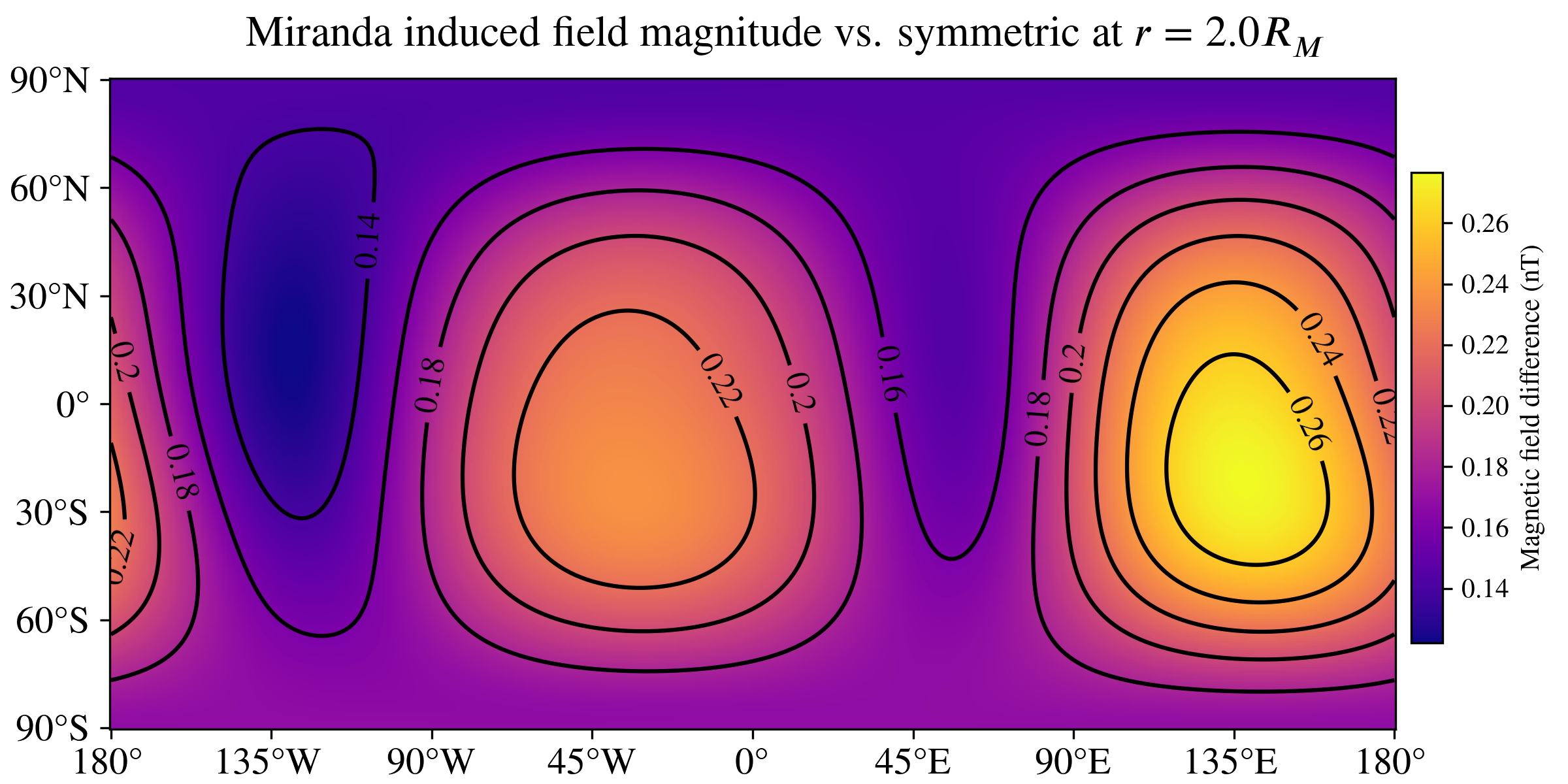}
		\caption{
			Difference in the magnitude of the induced magnetic field of Miranda resulting from the interior model shown in \F\ref{fig:mirandaContour}.
			Magnetic fields are evaluated at the J2000 reference epoch and at $r=2R_M$, a plausible flyby distance for a future mission.
			The simple geometry of the differences shown here reflects the fact that the change is dominated by increases in the dipole moments.
			This happens because the asymmetry model results in more conducting material residing closer to the surface, increasing the overall response to the excitation field.
			Unlike for Europa, the signal from asymmetry is over 20\% of the total induced field in most places (compare to \F\ref{fig:suppMirandaMag}).
			An animation showing the same map as it varies throughout the 35-hour synodic period is included in the Supplemental Material.
		}
		\label{fig:mirandaMagDiff}
	\end{figure}
	
	\F\ref{fig:mirandaMagDiff} shows the difference in the magnitude of the induced magnetic field resulting from the asymmetry model we have applied.
	The magnetic fields are again evaluated at the J2000 reference epoch, but this time at a distance $r=2R_M$, a plausible range around closest approach for flybys by a future Uranus orbiter.
	In this case, the differences are a significant fraction of the total induced field---over 20\% in most places.
	As in other examples, all vector components are shown in the Supplemental Material (\F\ref{fig:suppMiranda}).
	Greater salinity in the ocean will result in larger differences resulting from asymmetry, though typically a smaller fraction of the total induced field.
	A closer approach will also increase the differences substantially, especially because of the quadrupole and octupole moments represented in the induced field, whose field strengths decrease faster than those of the dipole moments.
	
	\subsubsection{Callisto and Triton}\label{sec:resultsCallistoTriton}
	
	For both Callisto and Triton, we model spherically symmetric interiors and suppose asymmetry in their ionospheric structure based on a day--night dichotomy.
	This day--night difference occurs when the ionization rate is heavily influenced by EUV flux from the Sun, as is the expected case for Callisto \citep{hartkorn2017structure} and Triton \citep{krasnopolsky1995photochemistry}.
	To approximate the asymmetric conductivity structure introduced by the day--night dichotomy, we apply a single degree-1 real harmonic for each moon, such that the uniformly conducting ionosphere bulges outward at local noon and inward at local midnight.
	Contour maps for these simple shapes are included in the Supplemental Text (\Fs\ref{fig:suppCallistoContour} and \ref{fig:suppTritonContour}).
	Note that the asymmetric shape of the ionosphere will rotate in the frame of reference of the satellite on the same time scale as the orbital period.
	However, we aim to estimate the order of magnitude for the signal from asymmetry and demonstrate application of our model, so for simplicity we assume the asymmetric structure is fixed.
	
	The ionosphere of Callisto is assumed to extend from the surface to $100\siu{km}$ altitude and to have a total height-integrated conductivity of $800\siu{S}$, consistent with the estimate of Pedersen conductivity concluded by \citet{hartkorn2017induction}.
	The upper boundary of this shape is then perturbed as described above.
	\textit{Voyager 2} radio measurements revealed Triton to have a highly conductive ionosphere, with an ion density peaking well above the surface \citep{tyler1989voyager}.
	To model Triton's ionosphere, we assume a uniformly conducting shell beginning at $250\siu{km}$ altitude and extending upward for $200\siu{km}$.
	The upper boundary of this shape is then perturbed as described above, and the neutral atmosphere below is assumed to be nonconducting.
	In the highly conducting ionosphere of Triton, convection electric fields will tend to transport plasma from areas of high ion density to areas with lower density, smoothing out the asymmetry somewhat \citep{schunk2009ionospheres}.
	To account for this, we decrease the amplitude of asymmetry compared to Callisto (to $\pm60\siu{km}$ rather than $\pm100\siu{km}$).
	
	Our conductivity models for both moons also contain conducting oceans in their interiors.
	For the interior of Callisto, we have assumed parameters consistent with previous studies of magnetic induction at this moon \citep[\eg\ ][]{vance2021magnetic}, namely a $100\siu{km}$-thick ice shell atop an ocean of $10\siu{wt\%}$ \ch{MgSO4 (aq)}.
	Magnetic induction at Triton has not yet been studied in detail; we have chosen the same composition for Triton's ocean as for Callisto, and a slightly thicker ice shell of $112\siu{km}$.
	Our Triton interior model is based on the moment of inertia and surface temperatures assumed for Pluto by \citet{hussmann2006subsurface}.
	
	\begin{figure}
		\centering
		\includegraphics[width=\textwidth]{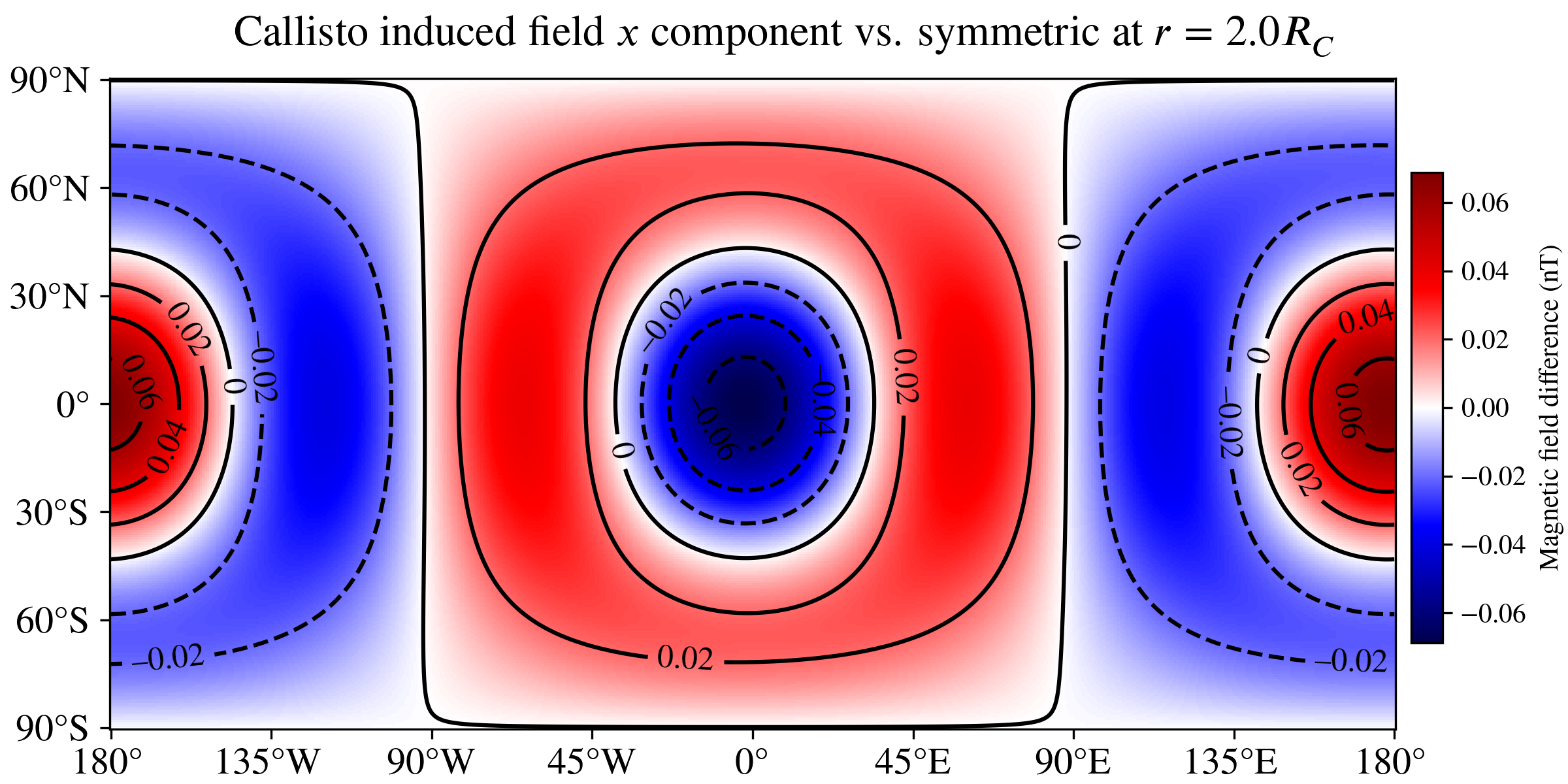}
		\caption{
			Difference in the $B_x$ component of the induced magnetic field of Callisto resulting from asymmetry in the ionosphere representing a day--night dichotomy.
			Magnetic fields are evaluated at $r=2R_C$, consistent with a spacecraft flyby.
			The induced field is nearly unaffected by the large asymmetry in the ionosphere because of the small ionospheric conductivity \citep{hartkorn2017induction} and because the odd-degree shape harmonic we used does not cause any change to the dominant induced dipole moment.
		}
		\label{fig:callistoXdiff}
	\end{figure}
	
	\begin{figure}
		\centering
		\includegraphics[width=\textwidth]{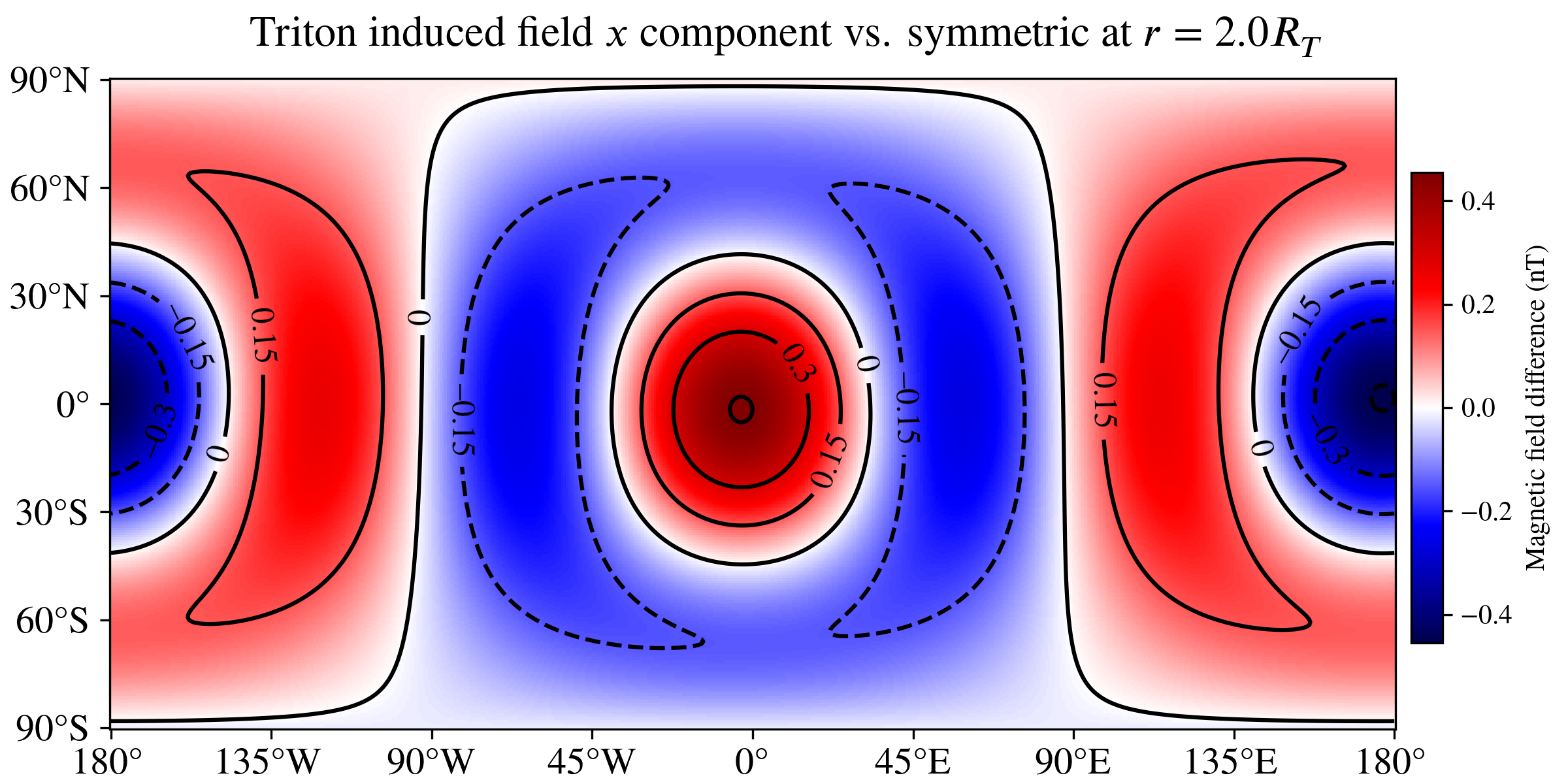}
		\caption{
			Difference in the $B_x$ component of the induced magnetic field of Triton resulting from asymmetry in the ionosphere representing a day--night dichotomy.
			Magnetic fields are evaluated at $r=2R_T$, consistent with a flyby as part of a possible future mission.
			The high-altitude, highly conducting, asymmetric ionosphere we model introduces a significant difference relative to a spherically symmetric analog (about 1/3 of the total induced field, see \F\ref{fig:suppTritonMag}).
			This suggests ionospheric asymmetry is likely to play an important role in future magnetic investigations of Triton.
			However, the observed differences are limited to induced moments of quadrupole order and larger, as the asymmetric shape we use to represent the day--night dichotomy cannot change the induced dipole moments.
		}
		\label{fig:tritonXdiff}
	\end{figure}
	
	\Fs\ref{fig:callistoXdiff} and \ref{fig:tritonXdiff} show the differences in the $B_x$ component of the induced magnetic field resulting from asymmetry in the ionospheres of Callisto and Triton, respectively.
	The magnetic fields are evaluated at the J2000 epoch and at $r=2R$, where $R$ is the body radius; as with Miranda, this distance is selected as a plausible flyby distance for past or future missions.
	Differences in the magnetic field from asymmetry in the Callisto ionosphere appear to be negligible, due primarily to the small ionospheric conductivity there.
	However, for Triton the high conductivity supports large differences, with the asymmetry accounting for as much as 1/3 of the total induced field (compare to \F\ref{fig:suppTritonMag}).
	For both moons, the degree-$1$ shape we use to perturb the upper ionospheric boundary is not capable of changing the induced dipole moment, so all differences will decay faster with distance than the main induction field from the spherically symmetric layers.
	Any asymmetry representing a day--night dichotomy will be dominated by odd-degree shapes describing the boundary due to their symmetry properties, so a more detailed asymmetric shape is unlikely to significantly change this result.
	
	\section{Discussion}\label{sec:discussion}
	
	The model we have developed is versatile in how it can be used to calculate induced magnetic fields.
	Our final result (\Eq\ref{eqm:Binm}) is a direct generalization of the recursive method presented by \citet{parkinson1983introduction} that is often used in magnetic sounding studies of icy moons \citep[\eg\ ][]{schilling2007time,hand2007empirical,seufert2011multi}.
	Describing asymmetric boundaries between regions in terms of spherical harmonics, as in \Eq\ref{eqm:rio}, permits a wide array of reasonable shapes to be supposed, allowing for forward models of the induced magnetic field of realistic, asymmetric oceans for the first time.
	
	Our model takes only seconds to run, supporting Monte Carlo/Bayesian statistical methods that require a vast array of possible configurations in order to constrain interior properties.
	Such ``big data'' approaches are simply not possible with numerical solutions that require hours or days to evaluate.
	Furthermore, as our model determines the complex induced magnetic moments at a particular epoch, explicit time dependence is added simply with a factor $e^{-i\omega t}$ for each period of excitation.
	This allows our model to immediately be applied to evaluate the magnetic field at any date and time.
	Such functionality is already a part of the Python code we used to calculate results and create all figures, included as Supplemental Material.
	
	\subsection{Limitations of the analytic model}\label{sec:limitations}
	
	The necessary assumptions we have made in order to obtain an analytic result create some limitations to the applicability of the solution.
	The primary limiting assumptions are that of uniform conductivity within each conducting layer and that asymmetric boundaries are near-spherical.
	
	\subsubsection{Uniform conductivity layers}\label{sec:uniform}
	
	Assuming uniform conductivity within each layer in the conducting body restricts our result to applications where the sequence of conducting layers along a line outward from the center is independent of latitude and longitude.
	In principle, it is possible to represent any closed surface containing the body center using a spherical harmonic expansion with high enough degree $\shapen$.
	However, the advantages of taking such an approach with our model, {\eg}\ to represent a melt lens within an ice shell, are likely outweighed by the additional complexity and computation time required.
	Higher-degree harmonics in such an expansion will have a much decreased weighting compared to low-degree harmonics, so the improved accuracy in the representation of the boundary will result in more modest gains in accuracy of representing the induced magnetic field.
	For these reasons, we recommend forming a low-degree ($\shapemax \leq 8$) approximation for non-spherical boundaries to use with our model.
	
	This condition requires that our solution is only approximate in cases where a rapid lateral change in conductivity is expected---for example, at the terminator line in Callisto's ionosphere \citep{hartkorn2017structure}.
	Far from the rapid lateral cutoff, our approach may still yield valid results, especially if the induced currents travel mostly parallel to the discontinuity.
	In the example of Callisto's ionosphere, photoionization from solar EUV photons is the primary source of plasma, so a sharp dropoff in charge carriers, and hence conductivity, is expected behind the terminator line.
	To completely account for an abrupt lateral discontinuity in conductivity, a numerical solution is required.
	However, approximating a dichotomy like that in Callisto's ionosphere with our approach still captures significant first-order effects of asymmetry and yields insights not possible to obtain with a spherically symmetric conductivity model.
	
	\subsubsection{Near-spherical boundary shape}\label{sec:nearSphericalConsqs}
	
	The second major assumption that limits applicability of our solution is that the boundaries between conducting regions are near-spherical.
	This assumption is required in order to limit our solution to a solvable set of equations describing a first-order change in the induced magnetic field.
	In other words, we are assuming that the second-order effects, {\ie}\ the induced moments that result from the interaction of asymmetric boundaries with the new moments induced by the first-order expansion, are negligible.
	Each successive term in the Taylor expansion that determines the order of the solution (\Eq\ref{eqm:Taylor}) gains a new factor $\bdb/\rbavg$ compared to the previous term.
	All other factors keep the same order of magnitude, although some mixing coefficients may grow or shrink slightly depending on the particular first-order induced moment and boundary shape considered.
	$\bdb/\rbavg$ will generally be quite small for bodies large enough to differentiate, as their self-gravity is already sufficient to pull them into a near-spherical shape.
	
	The value of the boundary deviation parameter $\bdb/\rbavg$ for the bodies we studied ranges from $0.016$ for Europa \citep[model]{tobie2003tidally} to $0.056$ for the large Triton ionosphere, with the value for Miranda an outlier at $0.210$.
	The asymmetry model for Enceladus \citep[after][]{hemingway2019enceladus} has a value of $0.071$.
	We accounted for the thicker Miranda ice shell essentially by scaling up this parameter; our Miranda model serves to quantify an upper limit to the resulting differences from the spherically symmetric case for the assumed bulk properties.
	We found in \S\ref{sec:resultsMiranda} that the differences caused by this asymmetry were a significant fraction of the total induced field, but the total induced field was only on the order of $1\siu{nT}$, so second-order effects would still be negligible in this case.
	If we had supposed a more saline ocean for Miranda, the induced field would grow substantially, and second-order effects could potentially become measurable.
	This may become a concern for future missions to the Uranus system, but we leave investigation of such second-order effects for future work.
	
	\subsection{Summary of example applications}\label{sec:examples}
	
	Applying our analytic result to example moons as in \S\ref{sec:application}, we provide evidence that the expected asymmetries are likely to result in measurable signals that will impact future magnetometer investigations.
	With the possible exception of Callisto, the conditions we studied resulted in significant differences compared to analogous spherically symmetric models: we find changes to the induced field of up to about $2.4\siu{nT}$ for Europa, $0.5\siu{nT}$ for Triton, and $0.25\siu{nT}$ for Miranda at plausible distances for orbiting spacecraft, and with marked variation in latitude, longitude, and time.
	In the case of asymmetric oceans, where degree-$2$ shape harmonics will be strongly represented, a significant amount of the change arises from alteration of the induced dipole moment.
	The size of this signal is likely to be significant for the upcoming \textit{Europa Clipper} mission, which plans several near flybys of under $25\siu{km}$.
	For the Europa Seawater model we studied, our initial estimates (\F\ref{fig:europaFalloff}) show that these effects will have measurable effects for a variety of flyby altitudes, but will also vary throughout the excitation period.
	However, we only considered this behavior for a single plausible model of asymmetry in Europa's interior.
	It is not yet clear whether the effects of plausible asymmetry may be disregarded under some general condition, {\eg}\ flybys above some threshold altitude.
	We recommend this as a topic for future work.
	
	For Europa, we studied the same asymmetric shape under two different compositions, end-member Seawater oceans with salinities previously considered in the literature \citep[\eg\ ][]{vance2021magnetic}.
	From comparing these results (\Fs\ref{fig:europaHighXdiff}, \ref{fig:europaLowXdiff}), we can conclude that lowering the ocean composition had a relatively small impact on the signal arising from asymmetry.
	However, ocean composition is likely to play a bigger role at smaller moons such as Miranda and Enceladus than at Europa because the asymmetry there may be a significant fraction of radius.
	In addition, the smaller radii of the conducting boundaries render the predicted induced moments more sensitive to the electrical conductivity for the same reason Europa is less sensitive---an exponential dependence on ($r\sqrt{i\omega\mu\sigma}$) in the Bessel functions $\jn(kr)$.
	The radii are large enough for Europa that ratios of these functions are close to unity, but the same does not apply to Miranda or Enceladus, thus causing changes in $\sigma$ to have a greater effect on the induced field.
	The strong dependence on radius in these functions also causes asymmetric oceans that rise nearer to the surface to have outsize effects on the induced moments.
	This is evident in the differences in the magnitude of the induced field for Miranda in \F\ref{fig:mirandaMagDiff}, which are universally positive for all latitudes and longitudes.
	
	We did not consider magnetic fields arising due to currents in the plasma environment outside the bodies beyond the current sheet models used to determine the excitation moments.
	Such plasma fields will be critical to any future analysis of spacecraft measurements, but to include them in induction modeling is beyond the scope of this work.
	
	\section{Conclusions}
	
	With the mathematical methods described in this work, we have derived an analytic model capable of calculating the induced magnetic field for any arbitrary planetary body to first order, provided it is near-spherical.
	This capability is an essential step forward in considering and accounting for the expected global asymmetries in ocean worlds.
	We have demonstrated that these asymmetries will produce measurable effects on the induced fields used in magnetic sounding investigations.
	These effects will introduce a source of systematic noise in measurements of order $2\siu{nT}$ near the surface of Europa, and will have a small impact on characterization studies using flyby data at all altitudes because the largest differences result from changes to the induced dipole moments.
	Future spacecraft investigations may be able to characterize asymmetry within the interiors of icy moons if they are able to achieve measurement precision on the order of a few tenths of a $\si{nT}$.
	Our analytic model and the software we provide for its evaluation support study of the interior structure of icy moons with an unprecedented level of detail.
	
	\section*{Acknowledgments}
	This work was supported by NASA Headquarters under the NASA Earth and Space Science Fellowship Program --- Grant 80NSSC18K1236.
	Part of the research was carried out at the Jet Propulsion Laboratory, California Institute of Technology, under a contract with the National Aeronautics and Space Administration (80NM0018D0004)
	The authors acknowledge that portions of this work have been carried out on the traditional lands of the Coast Salish and Tongva peoples.
	The authors are indebted to D.~Hemingway for data required to reproduce their interior model for Enceladus.
	The authors also thank H.~Hay for helpful discussion on tidal deformation and P.~Jandir for feedback regarding notation.
	
	\bibliographystyle{cas-model2-names}
	\bibliography{references}

\begin{thebibliography}{57}
\expandafter\ifx\csname natexlab\endcsname\relax\def\natexlab#1{#1}\fi
\providecommand{\url}[1]{\texttt{#1}}
\providecommand{\href}[2]{#2}
\providecommand{\path}[1]{#1}
\providecommand{\DOIprefix}{doi:}
\providecommand{\ArXivprefix}{arXiv:}
\providecommand{\URLprefix}{URL: }
\providecommand{\Pubmedprefix}{pmid:}
\providecommand{\doi}[1]{\href{http://dx.doi.org/#1}{\path{#1}}}
\providecommand{\Pubmed}[1]{\href{pmid:#1}{\path{#1}}}
\providecommand{\bibinfo}[2]{#2}
\ifx\xfnm\relax \def\xfnm[#1]{\unskip,\space#1}\fi
\bibitem[{Anderson et~al.(2001)Anderson, Jacobson, McElrath, Moore, Schubert
  and Thomas}]{anderson2001shape}
\bibinfo{author}{Anderson, J.D.}, \bibinfo{author}{Jacobson, R.A.},
  \bibinfo{author}{McElrath, T.P.}, \bibinfo{author}{Moore, W.B.},
  \bibinfo{author}{Schubert, G.}, \bibinfo{author}{Thomas, P.C.},
  \bibinfo{year}{2001}.
\newblock \bibinfo{title}{Shape, mean radius, gravity field, and interior
  structure of {C}allisto}.
\newblock \bibinfo{journal}{Icarus} \bibinfo{volume}{153},
  \bibinfo{pages}{157--161}.
\bibitem[{Anderson et~al.(1998)Anderson, Schubert, Jacobson, Lau, Moore and
  Sjogren}]{anderson1998europa}
\bibinfo{author}{Anderson, J.D.}, \bibinfo{author}{Schubert, G.},
  \bibinfo{author}{Jacobson, R.A.}, \bibinfo{author}{Lau, E.L.},
  \bibinfo{author}{Moore, W.B.}, \bibinfo{author}{Sjogren, W.L.},
  \bibinfo{year}{1998}.
\newblock \bibinfo{title}{Europa's differentiated internal structure:
  {I}nferences from four {G}alileo encounters}.
\newblock \bibinfo{journal}{Science} \bibinfo{volume}{281},
  \bibinfo{pages}{2019--2022}.
\newblock \DOIprefix\doi{10.1126/science.281.5385.2019}.
\bibitem[{Arfken et~al.(2012)Arfken, Weber and Harris}]{arfken2012mathematical}
\bibinfo{author}{Arfken, G.}, \bibinfo{author}{Weber, H.},
  \bibinfo{author}{Harris, F.}, \bibinfo{year}{2012}.
\newblock \bibinfo{title}{Mathematical Methods for Physicists}.
\newblock \bibinfo{edition}{Seventh} ed., \bibinfo{publisher}{Academic Press}.
\bibitem[{Arridge and Eggington(2021)}]{arridge2021electromagnetic}
\bibinfo{author}{Arridge, C.S.}, \bibinfo{author}{Eggington, J.W.},
  \bibinfo{year}{2021}.
\newblock \bibinfo{title}{Electromagnetic induction in the icy satellites of
  {U}ranus}.
\newblock \bibinfo{journal}{Icarus} ,
  \bibinfo{pages}{114562}\DOIprefix\doi{10.1016/j.icarus.2021.114562}.
\bibitem[{Beddingfield and Cartwright(2020)}]{beddingfield2020hidden}
\bibinfo{author}{Beddingfield, C.B.}, \bibinfo{author}{Cartwright, R.J.},
  \bibinfo{year}{2020}.
\newblock \bibinfo{title}{Hidden tectonism on {M}iranda's {E}lsinore {C}orona
  revealed by polygonal impact craters}.
\newblock \bibinfo{journal}{Icarus} \bibinfo{volume}{343},
  \bibinfo{pages}{113687}.
\bibitem[{Billings and Kattenhorn(2005)}]{billings2005great}
\bibinfo{author}{Billings, S.E.}, \bibinfo{author}{Kattenhorn, S.A.},
  \bibinfo{year}{2005}.
\newblock \bibinfo{title}{The great thickness debate: Ice shell thickness
  models for {E}uropa and comparisons with estimates based on flexure at
  ridges}.
\newblock \bibinfo{journal}{Icarus} \bibinfo{volume}{177},
  \bibinfo{pages}{397--412}.
\bibitem[{Brink and Satchler(1968)}]{brink1968angular}
\bibinfo{author}{Brink, D.M.}, \bibinfo{author}{Satchler, G.R.},
  \bibinfo{year}{1968}.
\newblock \bibinfo{title}{Angular momentum}.
\newblock \bibinfo{edition}{Second} ed., \bibinfo{publisher}{Clarendon Press}.
\bibitem[{Campagnola et~al.(2019)Campagnola, Buffington, Lam, Petropoulos and
  Pellegrini}]{campagnola2019tour}
\bibinfo{author}{Campagnola, S.}, \bibinfo{author}{Buffington, B.B.},
  \bibinfo{author}{Lam, T.}, \bibinfo{author}{Petropoulos, A.E.},
  \bibinfo{author}{Pellegrini, E.}, \bibinfo{year}{2019}.
\newblock \bibinfo{title}{Tour design techniques for the {E}uropa {C}lipper
  mission}.
\newblock \bibinfo{journal}{Journal of Guidance, Control, and Dynamics}
  \bibinfo{volume}{42}, \bibinfo{pages}{2615--2626}.
\newblock \DOIprefix\doi{10.2514/1.G004309}.
\bibitem[{Cartwright et~al.(2020)Cartwright, Beddingfield, Nordheim, Roser,
  Grundy, Hand, Emery, Cruikshank and Scipioni}]{cartwright2020evidence}
\bibinfo{author}{Cartwright, R.J.}, \bibinfo{author}{Beddingfield, C.B.},
  \bibinfo{author}{Nordheim, T.A.}, \bibinfo{author}{Roser, J.},
  \bibinfo{author}{Grundy, W.M.}, \bibinfo{author}{Hand, K.P.},
  \bibinfo{author}{Emery, J.P.}, \bibinfo{author}{Cruikshank, D.P.},
  \bibinfo{author}{Scipioni, F.}, \bibinfo{year}{2020}.
\newblock \bibinfo{title}{Evidence for ammonia-bearing species on the uranian
  satellite {A}riel supports recent geologic activity}.
\newblock \bibinfo{journal}{The Astrophysical Journal Letters}
  \bibinfo{volume}{898}, \bibinfo{pages}{L22}.
\bibitem[{Cochrane et~al.(2021)Cochrane, Vance, Nordheim, Styczinski, Masters
  and Regoli}]{cochrane2021search}
\bibinfo{author}{Cochrane, C.J.}, \bibinfo{author}{Vance, S.D.},
  \bibinfo{author}{Nordheim, T.A.}, \bibinfo{author}{Styczinski, M.},
  \bibinfo{author}{Masters, A.}, \bibinfo{author}{Regoli, L.H.},
  \bibinfo{year}{2021}.
\newblock \bibinfo{title}{In search of subsurface oceans within the uranian
  moons}.
\newblock \href{http://arxiv.org/abs/2105.06087}{\tt arXiv:2105.06087}.
  \bibinfo{note}{{S}ubmitted to Journal of Geophysical Research: Planets}.
\bibitem[{Condon and Shortley(1951)}]{condonshortley}
\bibinfo{author}{Condon, E.U.}, \bibinfo{author}{Shortley, G.H.},
  \bibinfo{year}{1951}.
\newblock \bibinfo{title}{The theory of atomic spectra}.
\newblock \bibinfo{publisher}{Cambridge University Press}.
\bibitem[{Connerney et~al.(1991)Connerney, Acu\~{n}a and
  Ness}]{connerney1991magnetic}
\bibinfo{author}{Connerney, J.}, \bibinfo{author}{Acu\~{n}a, M.H.},
  \bibinfo{author}{Ness, N.F.}, \bibinfo{year}{1991}.
\newblock \bibinfo{title}{The magnetic field of {N}eptune}.
\newblock \bibinfo{journal}{Journal of Geophysical Research: Space Physics}
  \bibinfo{volume}{96}, \bibinfo{pages}{19023--19042}.
\bibitem[{Connerney et~al.(1981)Connerney, Acu\~{n}a and
  Ness}]{connerney1981modeling}
\bibinfo{author}{Connerney, J.E.P.}, \bibinfo{author}{Acu\~{n}a, M.H.},
  \bibinfo{author}{Ness, N.F.}, \bibinfo{year}{1981}.
\newblock \bibinfo{title}{Modeling the jovian current sheet and inner
  magnetosphere}.
\newblock \bibinfo{journal}{Journal of Geophysical Research: Space Physics}
  \bibinfo{volume}{86}, \bibinfo{pages}{8370--8384}.
\bibitem[{Connerney et~al.(2018)Connerney, Kotsiaros, Oliversen, Espley,
  Joergensen, Joergensen, Merayo, Herceg, Bloxham, Moore, Bolton and
  Levin}]{connerney2018new}
\bibinfo{author}{Connerney, J.E.P.}, \bibinfo{author}{Kotsiaros, S.},
  \bibinfo{author}{Oliversen, R.J.}, \bibinfo{author}{Espley, J.R.},
  \bibinfo{author}{Joergensen, J.L.}, \bibinfo{author}{Joergensen, P.S.},
  \bibinfo{author}{Merayo, J.M.G.}, \bibinfo{author}{Herceg, M.},
  \bibinfo{author}{Bloxham, J.}, \bibinfo{author}{Moore, K.M.},
  \bibinfo{author}{Bolton, S.J.}, \bibinfo{author}{Levin, S.M.},
  \bibinfo{year}{2018}.
\newblock \bibinfo{title}{A new model of {J}upiter's magnetic field from
  {J}uno's first nine orbits}.
\newblock \bibinfo{journal}{Geophysical Research Letters} \bibinfo{volume}{45},
  \bibinfo{pages}{2590--2596}.
\bibitem[{Croft et~al.(1988)Croft, Lunine and Kargel}]{croft1988equation}
\bibinfo{author}{Croft, S.}, \bibinfo{author}{Lunine, J.},
  \bibinfo{author}{Kargel, J.}, \bibinfo{year}{1988}.
\newblock \bibinfo{title}{Equation of state of ammonia-water liquid: Derivation
  and planetological applications}.
\newblock \bibinfo{journal}{Icarus} \bibinfo{volume}{73},
  \bibinfo{pages}{279--293}.
\bibitem[{{\'C}uk et~al.(2020){\'C}uk, El~Moutamid and
  Tiscareno}]{cuk2020dynamical}
\bibinfo{author}{{\'C}uk, M.}, \bibinfo{author}{El~Moutamid, M.},
  \bibinfo{author}{Tiscareno, M.S.}, \bibinfo{year}{2020}.
\newblock \bibinfo{title}{Dynamical history of the uranian system}.
\newblock \bibinfo{journal}{The Planetary Science Journal} \bibinfo{volume}{1},
  \bibinfo{pages}{22}.
\bibitem[{Dennery and Krzywicki(2012)}]{dennery2012mathematics}
\bibinfo{author}{Dennery, P.}, \bibinfo{author}{Krzywicki, A.},
  \bibinfo{year}{2012}.
\newblock \bibinfo{title}{Mathematics for physicists}.
\newblock \bibinfo{publisher}{Dover}.
\bibitem[{Edmonds(1996)}]{edmonds1996angular}
\bibinfo{author}{Edmonds, A.R.}, \bibinfo{year}{1996}.
\newblock \bibinfo{title}{Angular momentum in quantum mechanics}.
\newblock \bibinfo{publisher}{Princeton University Press}.
\bibitem[{Glover(2015)}]{glover2015geophysical}
\bibinfo{author}{Glover, P.W.J.}, \bibinfo{year}{2015}.
\newblock \bibinfo{title}{Geophysical properties of the near surface {E}arth:
  Electrical properties}, in: \bibinfo{editor}{Schubert, G.} (Ed.),
  \bibinfo{booktitle}{Treatise on Geophysics}. \bibinfo{edition}{second} ed..
  \bibinfo{publisher}{Elsevier}. volume~\bibinfo{volume}{11}, pp.
  \bibinfo{pages}{89--137}.
\newblock \DOIprefix\doi{10.1016/B978-0-444-53802-4.00189-5}.
\bibitem[{Hand and Chyba(2007)}]{hand2007empirical}
\bibinfo{author}{Hand, K.P.}, \bibinfo{author}{Chyba, C.F.},
  \bibinfo{year}{2007}.
\newblock \bibinfo{title}{Empirical constraints on the salinity of the europan
  ocean and implications for a thin ice shell}.
\newblock \bibinfo{journal}{Icarus} \bibinfo{volume}{189},
  \bibinfo{pages}{424--438}.
\newblock \DOIprefix\doi{10.1016/j.icarus.2007.02.002}.
\bibitem[{Hartkorn and Saur(2017)}]{hartkorn2017induction}
\bibinfo{author}{Hartkorn, O.}, \bibinfo{author}{Saur, J.},
  \bibinfo{year}{2017}.
\newblock \bibinfo{title}{Induction signals from {C}allisto's ionosphere and
  their implications on a possible subsurface ocean}.
\newblock \bibinfo{journal}{Journal of Geophysical Research: Space Physics}
  \bibinfo{volume}{122}, \bibinfo{pages}{11--677}.
\newblock \DOIprefix\doi{10.1002/2017JA024269}.
\bibitem[{Hartkorn et~al.(2017)Hartkorn, Saur and
  Strobel}]{hartkorn2017structure}
\bibinfo{author}{Hartkorn, O.}, \bibinfo{author}{Saur, J.},
  \bibinfo{author}{Strobel, D.F.}, \bibinfo{year}{2017}.
\newblock \bibinfo{title}{Structure and density of {C}allisto's atmosphere from
  a fluid-kinetic model of its ionosphere: Comparison with {H}ubble {S}pace
  {T}elescope and {G}alileo observations}.
\newblock \bibinfo{journal}{Icarus} \bibinfo{volume}{282},
  \bibinfo{pages}{237--259}.
\newblock \DOIprefix\doi{10.1016/j.icarus.2016.09.020}.
\bibitem[{Hemingway and Mittal(2019)}]{hemingway2019enceladus}
\bibinfo{author}{Hemingway, D.J.}, \bibinfo{author}{Mittal, T.},
  \bibinfo{year}{2019}.
\newblock \bibinfo{title}{Enceladus's ice shell structure as a window on
  internal heat production}.
\newblock \bibinfo{journal}{Icarus} \bibinfo{volume}{332},
  \bibinfo{pages}{111--131}.
\bibitem[{Herbert(2009)}]{herbert2009aurora}
\bibinfo{author}{Herbert, F.}, \bibinfo{year}{2009}.
\newblock \bibinfo{title}{Aurora and magnetic field of {U}ranus}.
\newblock \bibinfo{journal}{Journal of Geophysical Research: Space Physics}
  \bibinfo{volume}{114}.
\bibitem[{Horbury et~al.(2020)Horbury, O'Brien, Carrasco~Blazquez, Bendyk,
  Brown, Hudson, Evans, Oddy, Carr, Beek, Cupido, Bhattacharya, Dominguez,
  Matthews, Myklebust, Whiteside, Bale, Baumjohann, Burgess, Carbone, Cargill,
  Eastwood, Erd\"{o}s, Fletcher, Forsyth, Giacalone, Glassmeier, Goldstein,
  Hoeksema, Lockwood, Magnes, Maksimovic, Marsch, Matthaeus, Murphy,
  Nakariakov, Owen, Owens, Rodriguez-Pacheco, Richter, Riley, Russell,
  Schwartz, Vainio, Velli, Vennerstrom, Walsh, Wimmer-Schweingruber, Zank,
  M\"{u}ller, Zouganelis and Walsh}]{horbury2020solar}
\bibinfo{author}{Horbury, T.S.}, \bibinfo{author}{O'Brien, H.},
  \bibinfo{author}{Carrasco~Blazquez, I.}, \bibinfo{author}{Bendyk, M.},
  \bibinfo{author}{Brown, P.}, \bibinfo{author}{Hudson, R.},
  \bibinfo{author}{Evans, V.}, \bibinfo{author}{Oddy, T.M.},
  \bibinfo{author}{Carr, C.M.}, \bibinfo{author}{Beek, T.J.},
  \bibinfo{author}{Cupido, E.}, \bibinfo{author}{Bhattacharya, S.},
  \bibinfo{author}{Dominguez, J.A.}, \bibinfo{author}{Matthews, L.},
  \bibinfo{author}{Myklebust, V.R.}, \bibinfo{author}{Whiteside, B.},
  \bibinfo{author}{Bale, S.D.}, \bibinfo{author}{Baumjohann, W.},
  \bibinfo{author}{Burgess, D.}, \bibinfo{author}{Carbone, V.},
  \bibinfo{author}{Cargill, P.}, \bibinfo{author}{Eastwood, J.},
  \bibinfo{author}{Erd\"{o}s, G.}, \bibinfo{author}{Fletcher, L.},
  \bibinfo{author}{Forsyth, R.}, \bibinfo{author}{Giacalone, J.},
  \bibinfo{author}{Glassmeier, K.H.}, \bibinfo{author}{Goldstein, M.L.},
  \bibinfo{author}{Hoeksema, T.}, \bibinfo{author}{Lockwood, M.},
  \bibinfo{author}{Magnes, W.}, \bibinfo{author}{Maksimovic, M.},
  \bibinfo{author}{Marsch, E.}, \bibinfo{author}{Matthaeus, W.H.},
  \bibinfo{author}{Murphy, N.}, \bibinfo{author}{Nakariakov, V.M.},
  \bibinfo{author}{Owen, C.J.}, \bibinfo{author}{Owens, M.},
  \bibinfo{author}{Rodriguez-Pacheco, J.}, \bibinfo{author}{Richter, I.},
  \bibinfo{author}{Riley, P.}, \bibinfo{author}{Russell, C.T.},
  \bibinfo{author}{Schwartz, S.}, \bibinfo{author}{Vainio, R.},
  \bibinfo{author}{Velli, M.}, \bibinfo{author}{Vennerstrom, S.},
  \bibinfo{author}{Walsh, R.}, \bibinfo{author}{Wimmer-Schweingruber, R.F.},
  \bibinfo{author}{Zank, G.}, \bibinfo{author}{M\"{u}ller, D.},
  \bibinfo{author}{Zouganelis, I.}, \bibinfo{author}{Walsh, A.P.},
  \bibinfo{year}{2020}.
\newblock \bibinfo{title}{The {S}olar {O}rbiter magnetometer}.
\newblock \bibinfo{journal}{Astronomy \& Astrophysics} \bibinfo{volume}{642},
  \bibinfo{pages}{A9}.
\bibitem[{Hussmann et~al.(2006)Hussmann, Sohl and
  Spohn}]{hussmann2006subsurface}
\bibinfo{author}{Hussmann, H.}, \bibinfo{author}{Sohl, F.},
  \bibinfo{author}{Spohn, T.}, \bibinfo{year}{2006}.
\newblock \bibinfo{title}{Subsurface oceans and deep interiors of medium-sized
  outer planet satellites and large trans-neptunian objects}.
\newblock \bibinfo{journal}{Icarus} \bibinfo{volume}{185},
  \bibinfo{pages}{258--273}.
\bibitem[{Iess et~al.(2014)Iess, Stevenson, Parisi, Hemingway, Jacobson,
  Lunine, Nimmo, Armstrong, Asmar, Ducci and Tortora}]{iess2014gravity}
\bibinfo{author}{Iess, L.}, \bibinfo{author}{Stevenson, D.},
  \bibinfo{author}{Parisi, M.}, \bibinfo{author}{Hemingway, D.},
  \bibinfo{author}{Jacobson, R.}, \bibinfo{author}{Lunine, J.},
  \bibinfo{author}{Nimmo, F.}, \bibinfo{author}{Armstrong, J.},
  \bibinfo{author}{Asmar, S.}, \bibinfo{author}{Ducci, M.},
  \bibinfo{author}{Tortora, P.}, \bibinfo{year}{2014}.
\newblock \bibinfo{title}{The gravity field and interior structure of
  {E}nceladus}.
\newblock \bibinfo{journal}{Science} \bibinfo{volume}{344},
  \bibinfo{pages}{78--80}.
\bibitem[{Jackson(1999)}]{jackson1999classical}
\bibinfo{author}{Jackson, J.D.}, \bibinfo{year}{1999}.
\newblock \bibinfo{title}{Classical electrodynamics}.
\newblock \bibinfo{publisher}{John Wiley \& Sons}.
\newblock \DOIprefix\doi{10.1119/1.19136}.
\bibitem[{Khurana and {the \textit{Trident} Team}(2019)}]{khurana2019single}
\bibinfo{author}{Khurana, K.}, \bibinfo{author}{{the \textit{Trident} Team}},
  \bibinfo{year}{2019}.
\newblock \bibinfo{title}{Single-pass magnetometric ocean detection at
  {T}riton}.
\newblock \bibinfo{journal}{Ocean Worlds 4} \bibinfo{volume}{2168},
  \bibinfo{pages}{6038}.
\bibitem[{Khurana et~al.(2002)Khurana, Kivelson and
  Russell}]{khurana2002searching}
\bibinfo{author}{Khurana, K.K.}, \bibinfo{author}{Kivelson, M.G.},
  \bibinfo{author}{Russell, C.T.}, \bibinfo{year}{2002}.
\newblock \bibinfo{title}{Searching for liquid water in {E}uropa by using
  surface observatories}.
\newblock \bibinfo{journal}{Astrobiology} \bibinfo{volume}{2},
  \bibinfo{pages}{93--103}.
\newblock \DOIprefix\doi{10.1089/153110702753621376}.
\bibitem[{Kivelson et~al.(2000)Kivelson, Khurana, Russell, Volwerk, Walker and
  Zimmer}]{kivelson2000galileo}
\bibinfo{author}{Kivelson, M.G.}, \bibinfo{author}{Khurana, K.K.},
  \bibinfo{author}{Russell, C.T.}, \bibinfo{author}{Volwerk, M.},
  \bibinfo{author}{Walker, R.J.}, \bibinfo{author}{Zimmer, C.},
  \bibinfo{year}{2000}.
\newblock \bibinfo{title}{Galileo magnetometer measurements: {A} stronger case
  for a subsurface ocean at {E}uropa}.
\newblock \bibinfo{journal}{Science} \bibinfo{volume}{289},
  \bibinfo{pages}{1340--1343}.
\newblock \DOIprefix\doi{10.1126/science.289.5483.1340}.
\bibitem[{Kivelson et~al.(2002)Kivelson, Khurana and
  Volwerk}]{kivelson2002permanent}
\bibinfo{author}{Kivelson, M.G.}, \bibinfo{author}{Khurana, K.K.},
  \bibinfo{author}{Volwerk, M.}, \bibinfo{year}{2002}.
\newblock \bibinfo{title}{The permanent and inductive magnetic moments of
  {G}anymede}.
\newblock \bibinfo{journal}{Icarus} \bibinfo{volume}{157},
  \bibinfo{pages}{507--522}.
\bibitem[{Krasnopolsky and Cruikshank(1995)}]{krasnopolsky1995photochemistry}
\bibinfo{author}{Krasnopolsky, V.A.}, \bibinfo{author}{Cruikshank, D.P.},
  \bibinfo{year}{1995}.
\newblock \bibinfo{title}{Photochemistry of {T}riton's atmosphere and
  ionosphere}.
\newblock \bibinfo{journal}{Journal of Geophysical Research: Planets}
  \bibinfo{volume}{100}, \bibinfo{pages}{21271--21286}.
\bibitem[{Larionov and Kryukov(1984)}]{larionov1984conductivity}
\bibinfo{author}{Larionov, E.}, \bibinfo{author}{Kryukov, P.},
  \bibinfo{year}{1984}.
\newblock \bibinfo{title}{The conductivity of {MgSO$_4$} aqueous-solutions in
  the range of temperatures 298--423 {K} and pressures 0.1--784.6 {MPa}}.
\newblock \bibinfo{journal}{Izvestiya Sibirskogo Otdeleniya Akademii Nauk SSSR
  Seriya Khimicheskikh Nauk} \bibinfo{volume}{5}, \bibinfo{pages}{20--23}.
\bibitem[{Marion and Heald(1980)}]{marion1980classical}
\bibinfo{author}{Marion, J.B.}, \bibinfo{author}{Heald, M.A.},
  \bibinfo{year}{1980}.
\newblock \bibinfo{title}{Classical Electromagnetic Radiation}.
\newblock \bibinfo{edition}{Second} ed., \bibinfo{publisher}{Academic Press},
  \bibinfo{address}{New York}.
\bibitem[{Mauk et~al.(1987)Mauk, Krimigis, Keath, Cheng, Armstrong, Lanzerotti,
  Gloeckler and Hamilton}]{mauk1987hot}
\bibinfo{author}{Mauk, B.}, \bibinfo{author}{Krimigis, S.},
  \bibinfo{author}{Keath, E.}, \bibinfo{author}{Cheng, A.},
  \bibinfo{author}{Armstrong, T.}, \bibinfo{author}{Lanzerotti, L.},
  \bibinfo{author}{Gloeckler, G.}, \bibinfo{author}{Hamilton, D.},
  \bibinfo{year}{1987}.
\newblock \bibinfo{title}{The hot plasma and radiation environment of the
  uranian magnetosphere}.
\newblock \bibinfo{journal}{Journal of Geophysical Research: Space Physics}
  \bibinfo{volume}{92}, \bibinfo{pages}{15283--15308}.
\bibitem[{McDougall and Barker(2011)}]{mcdougall2011getting}
\bibinfo{author}{McDougall, T.J.}, \bibinfo{author}{Barker, P.M.},
  \bibinfo{year}{2011}.
\newblock \bibinfo{title}{Getting started with {TEOS-10} and the {G}ibbs
  seawater {(GSW)} oceanographic toolbox}.
\newblock \bibinfo{journal}{SCOR/IAPSO WG} \bibinfo{volume}{127},
  \bibinfo{pages}{1--28}.
\bibitem[{Michaut and Manga(2014)}]{michaut2014domes}
\bibinfo{author}{Michaut, C.}, \bibinfo{author}{Manga, M.},
  \bibinfo{year}{2014}.
\newblock \bibinfo{title}{Domes, pits, and small chaos on {E}uropa produced by
  water sills}.
\newblock \bibinfo{journal}{Journal of Geophysical Research: Planets}
  \bibinfo{volume}{119}, \bibinfo{pages}{550--573}.
\bibitem[{Moffatt(1978)}]{moffatt1978field}
\bibinfo{author}{Moffatt, H.K.}, \bibinfo{year}{1978}.
\newblock \bibinfo{title}{Magnetic field generation in electrically conducting
  fluids}. volume~\bibinfo{volume}{2}.
\newblock \bibinfo{publisher}{Cambridge University Press}.
\bibitem[{Moore and Schubert(2000)}]{moore2000tidal}
\bibinfo{author}{Moore, W.B.}, \bibinfo{author}{Schubert, G.},
  \bibinfo{year}{2000}.
\newblock \bibinfo{title}{The tidal response of {E}uropa}.
\newblock \bibinfo{journal}{Icarus} \bibinfo{volume}{147},
  \bibinfo{pages}{317--319}.
\bibitem[{Parkinson(1983)}]{parkinson1983introduction}
\bibinfo{author}{Parkinson, W.D.}, \bibinfo{year}{1983}.
\newblock \bibinfo{title}{Introduction to Geomagnetism}.
\newblock \bibinfo{publisher}{Elsevier}.
\bibitem[{Postberg et~al.(2009)Postberg, Kempf, Schmidt, Brilliantov, Beinsen,
  Abel, Buck and Srama}]{postberg2009sodium}
\bibinfo{author}{Postberg, F.}, \bibinfo{author}{Kempf, S.},
  \bibinfo{author}{Schmidt, J.}, \bibinfo{author}{Brilliantov, N.},
  \bibinfo{author}{Beinsen, A.}, \bibinfo{author}{Abel, B.},
  \bibinfo{author}{Buck, U.}, \bibinfo{author}{Srama, R.},
  \bibinfo{year}{2009}.
\newblock \bibinfo{title}{Sodium salts in {E}-ring ice grains from an ocean
  below the surface of {E}nceladus}.
\newblock \bibinfo{journal}{Nature} \bibinfo{volume}{459},
  \bibinfo{pages}{1098--1101}.
\bibitem[{Rambaux and Castillo-Rogez(2013)}]{rambaux2013tides}
\bibinfo{author}{Rambaux, N.}, \bibinfo{author}{Castillo-Rogez, J.},
  \bibinfo{year}{2013}.
\newblock \bibinfo{title}{Tides on satellites of giant planets}, in:
  \bibinfo{editor}{Souchay, J.}, \bibinfo{editor}{Mathis, S.},
  \bibinfo{editor}{Tokieda, T.} (Eds.), \bibinfo{booktitle}{Tides in Astronomy
  and Astrophysics}. \bibinfo{publisher}{Springer-Verlag Berlin Heidelberg},
  pp. \bibinfo{pages}{167--200}.
\newblock \DOIprefix\doi{10.1007/978-3-642-32961-6_5}.
\bibitem[{Saur et~al.(2010)Saur, Neubauer and Glassmeier}]{saur2010induced}
\bibinfo{author}{Saur, J.}, \bibinfo{author}{Neubauer, F.M.},
  \bibinfo{author}{Glassmeier, K.H.}, \bibinfo{year}{2010}.
\newblock \bibinfo{title}{Induced magnetic fields in solar system bodies}.
\newblock \bibinfo{journal}{Space science reviews} \bibinfo{volume}{152},
  \bibinfo{pages}{391--421}.
\newblock \DOIprefix\doi{10.1007/s11214-009-9581-y}.
\bibitem[{Schilling et~al.(2007)Schilling, Neubauer and
  Saur}]{schilling2007time}
\bibinfo{author}{Schilling, N.}, \bibinfo{author}{Neubauer, F.M.},
  \bibinfo{author}{Saur, J.}, \bibinfo{year}{2007}.
\newblock \bibinfo{title}{Time-varying interaction of {E}uropa with the jovian
  magnetosphere: {C}onstraints on the conductivity of {E}uropa's subsurface
  ocean}.
\newblock \bibinfo{journal}{Icarus} \bibinfo{volume}{192},
  \bibinfo{pages}{41--55}.
\newblock \DOIprefix\doi{10.1016/j.icarus.2007.06.024}.
\bibitem[{Schmidt et~al.(2011)Schmidt, Blankenship, Patterson and
  Schenk}]{schmidt2011active}
\bibinfo{author}{Schmidt, B.}, \bibinfo{author}{Blankenship, D.},
  \bibinfo{author}{Patterson, G.}, \bibinfo{author}{Schenk, P.},
  \bibinfo{year}{2011}.
\newblock \bibinfo{title}{Active formation of `chaos terrain' over shallow
  subsurface water on {E}uropa}.
\newblock \bibinfo{journal}{Nature} \bibinfo{volume}{479},
  \bibinfo{pages}{502--505}.
\bibitem[{Schunk and Nagy(2009)}]{schunk2009ionospheres}
\bibinfo{author}{Schunk, R.}, \bibinfo{author}{Nagy, A.}, \bibinfo{year}{2009}.
\newblock \bibinfo{title}{Ionospheres: Physics, Plasma Physics, and Chemistry}.
\newblock \bibinfo{publisher}{Cambridge University Press}.
\bibitem[{Seufert et~al.(2011)Seufert, Saur and Neubauer}]{seufert2011multi}
\bibinfo{author}{Seufert, M.}, \bibinfo{author}{Saur, J.},
  \bibinfo{author}{Neubauer, F.M.}, \bibinfo{year}{2011}.
\newblock \bibinfo{title}{Multi-frequency electromagnetic sounding of the
  {G}alilean moons}.
\newblock \bibinfo{journal}{Icarus} \bibinfo{volume}{214},
  \bibinfo{pages}{477--494}.
\newblock \DOIprefix\doi{10.1016/j.icarus.2011.03.017}.
\bibitem[{Srivastava(1966)}]{srivastava1966theory}
\bibinfo{author}{Srivastava, S.P.}, \bibinfo{year}{1966}.
\newblock \bibinfo{title}{Theory of the magnetotelluric method for a spherical
  conductor}.
\newblock \bibinfo{journal}{Geophysical Journal International}
  \bibinfo{volume}{11}, \bibinfo{pages}{373--387}.
\bibitem[{Styczinski and Harnett(2021)}]{styczinski2021induced}
\bibinfo{author}{Styczinski, M.J.}, \bibinfo{author}{Harnett, E.M.},
  \bibinfo{year}{2021}.
\newblock \bibinfo{title}{Induced magnetic moments from a nearly spherical
  ocean}.
\newblock \bibinfo{journal}{Icarus} \bibinfo{volume}{354},
  \bibinfo{pages}{114020}.
\newblock \DOIprefix\doi{10.1016/j.icarus.2020.114020}.
\bibitem[{Tobie et~al.(2003)Tobie, Choblet and Sotin}]{tobie2003tidally}
\bibinfo{author}{Tobie, G.}, \bibinfo{author}{Choblet, G.},
  \bibinfo{author}{Sotin, C.}, \bibinfo{year}{2003}.
\newblock \bibinfo{title}{Tidally heated convection: {C}onstraints on
  {E}uropa's ice shell thickness}.
\newblock \bibinfo{journal}{Journal of Geophysical Research: Planets}
  \bibinfo{volume}{108}.
\newblock \DOIprefix\doi{10.1029/2003JE002099}.
\bibitem[{Tyler et~al.(1989)Tyler, Sweetnam, Anderson, Borutzki, Campbell,
  Eshleman, Gresh, Gurrola, Hinson, Kawashima, Kursinski, Levy, Lindal, Lyons,
  Marouf, Rosen, Simpson and Wood}]{tyler1989voyager}
\bibinfo{author}{Tyler, G.L.}, \bibinfo{author}{Sweetnam, D.N.},
  \bibinfo{author}{Anderson, J.D.}, \bibinfo{author}{Borutzki, S.E.},
  \bibinfo{author}{Campbell, J.K.}, \bibinfo{author}{Eshleman, V.R.},
  \bibinfo{author}{Gresh, D.L.}, \bibinfo{author}{Gurrola, E.M.},
  \bibinfo{author}{Hinson, D.P.}, \bibinfo{author}{Kawashima, N.},
  \bibinfo{author}{Kursinski, E.R.}, \bibinfo{author}{Levy, G.S.},
  \bibinfo{author}{Lindal, G.F.}, \bibinfo{author}{Lyons, J.R.},
  \bibinfo{author}{Marouf, E.A.}, \bibinfo{author}{Rosen, P.A.},
  \bibinfo{author}{Simpson, R.A.}, \bibinfo{author}{Wood, G.E.},
  \bibinfo{year}{1989}.
\newblock \bibinfo{title}{Voyager radio science observations of {N}eptune and
  {T}riton}.
\newblock \bibinfo{journal}{Science} \bibinfo{volume}{246},
  \bibinfo{pages}{1466--1473}.
\bibitem[{Vance et~al.(2018)Vance, Panning, St{\"a}hler, Cammarano, Bills,
  Tobie, Kamata, Kedar, Sotin, Pike, Lorenz, Huang, Jackson and
  Banerdt}]{vance2018geophysical}
\bibinfo{author}{Vance, S.D.}, \bibinfo{author}{Panning, M.P.},
  \bibinfo{author}{St{\"a}hler, S.}, \bibinfo{author}{Cammarano, F.},
  \bibinfo{author}{Bills, B.G.}, \bibinfo{author}{Tobie, G.},
  \bibinfo{author}{Kamata, S.}, \bibinfo{author}{Kedar, S.},
  \bibinfo{author}{Sotin, C.}, \bibinfo{author}{Pike, W.T.},
  \bibinfo{author}{Lorenz, R.}, \bibinfo{author}{Huang, H.H.},
  \bibinfo{author}{Jackson, J.M.}, \bibinfo{author}{Banerdt, B.},
  \bibinfo{year}{2018}.
\newblock \bibinfo{title}{Geophysical investigations of habitability in
  ice-covered ocean worlds}.
\newblock \bibinfo{journal}{Journal of Geophysical Research: Planets}
  \bibinfo{volume}{123}, \bibinfo{pages}{180--205}.
\newblock \DOIprefix\doi{10.1002/2017JE005341}.
\bibitem[{Vance et~al.(2021)Vance, Styczinski, Bills, Cochrane, Soderlund,
  G\'{o}mez-P\'{e}rez and Paty}]{vance2021magnetic}
\bibinfo{author}{Vance, S.D.}, \bibinfo{author}{Styczinski, M.J.},
  \bibinfo{author}{Bills, B.G.}, \bibinfo{author}{Cochrane, C.J.},
  \bibinfo{author}{Soderlund, K.M.}, \bibinfo{author}{G\'{o}mez-P\'{e}rez, N.},
  \bibinfo{author}{Paty, C.}, \bibinfo{year}{2021}.
\newblock \bibinfo{title}{Magnetic induction responses of {J}upiter's ocean
  moons including effects from adiabatic convection}.
\newblock \bibinfo{journal}{Journal of Geophysical Research: Planets}
  \bibinfo{volume}{126}, \bibinfo{pages}{e2020JE006418}.
\newblock \DOIprefix\doi{10.1029/2020JE006418}.
\bibitem[{Weiss et~al.(2020)Weiss, Colicci and Biersteker}]{weiss2020agu}
\bibinfo{author}{Weiss, B.P.}, \bibinfo{author}{Colicci, V.},
  \bibinfo{author}{Biersteker, J.B.}, \bibinfo{year}{2020}.
\newblock \bibinfo{title}{Searching for subsurface oceans on the moons of
  {U}ranus using magnetic induction}, in: \bibinfo{booktitle}{2020 AGU Fall
  Meeting}, pp. \bibinfo{pages}{P074--07}.
\bibitem[{Wigner(1931)}]{wigner1931gruppen}
\bibinfo{author}{Wigner, E.P.}, \bibinfo{year}{1931}.
\newblock \bibinfo{title}{Gruppentheorie und ihre Anwendung auf die
  Quantenmechanik der Atomspektren}.
\newblock \bibinfo{publisher}{Springer}.
\bibitem[{Zimmer et~al.(2000)Zimmer, Khurana and
  Kivelson}]{zimmer2000subsurface}
\bibinfo{author}{Zimmer, C.}, \bibinfo{author}{Khurana, K.K.},
  \bibinfo{author}{Kivelson, M.G.}, \bibinfo{year}{2000}.
\newblock \bibinfo{title}{Subsurface oceans on {E}uropa and {C}allisto:
  {C}onstraints from {G}alileo magnetometer observations}.
\newblock \bibinfo{journal}{Icarus} \bibinfo{volume}{147},
  \bibinfo{pages}{329--347}.
\newblock \DOIprefix\doi{10.1006/icar.2000.6456}.

\end{thebibliography}


\begin{thebibliography}{32}
\expandafter\ifx\csname natexlab\endcsname\relax\def\natexlab#1{#1}\fi
\providecommand{\url}[1]{\texttt{#1}}
\providecommand{\href}[2]{#2}
\providecommand{\path}[1]{#1}
\providecommand{\DOIprefix}{doi:}
\providecommand{\ArXivprefix}{arXiv:}
\providecommand{\URLprefix}{URL: }
\providecommand{\Pubmedprefix}{pmid:}
\providecommand{\doi}[1]{\href{http://dx.doi.org/#1}{\path{#1}}}
\providecommand{\Pubmed}[1]{\href{pmid:#1}{\path{#1}}}
\providecommand{\bibinfo}[2]{#2}
\ifx\xfnm\relax \def\xfnm[#1]{\unskip,\space#1}\fi
\bibitem[{Abramowitz and Stegun(1972)}]{abramowitzstegun}
\bibinfo{author}{Abramowitz, M.}, \bibinfo{author}{Stegun, I.A.},
  \bibinfo{year}{1972}.
\newblock \bibinfo{title}{Handbook of mathematical functions with formulas,
  graphs, and mathematical tables}. volume~\bibinfo{volume}{55}.
\newblock \bibinfo{edition}{Tenth} ed., \bibinfo{publisher}{US Government
  printing office}.
\bibitem[{Anderson et~al.(2001)Anderson, Jacobson, McElrath, Moore, Schubert
  and Thomas}]{anderson2001shape}
\bibinfo{author}{Anderson, J.D.}, \bibinfo{author}{Jacobson, R.A.},
  \bibinfo{author}{McElrath, T.P.}, \bibinfo{author}{Moore, W.B.},
  \bibinfo{author}{Schubert, G.}, \bibinfo{author}{Thomas, P.C.},
  \bibinfo{year}{2001}.
\newblock \bibinfo{title}{Shape, mean radius, gravity field, and interior
  structure of {C}allisto}.
\newblock \bibinfo{journal}{Icarus} \bibinfo{volume}{153},
  \bibinfo{pages}{157--161}.
\bibitem[{Anderson et~al.(1998)Anderson, Schubert, Jacobson, Lau, Moore and
  Sjogren}]{anderson1998europa}
\bibinfo{author}{Anderson, J.D.}, \bibinfo{author}{Schubert, G.},
  \bibinfo{author}{Jacobson, R.A.}, \bibinfo{author}{Lau, E.L.},
  \bibinfo{author}{Moore, W.B.}, \bibinfo{author}{Sjogren, W.L.},
  \bibinfo{year}{1998}.
\newblock \bibinfo{title}{Europa's differentiated internal structure:
  {I}nferences from four {G}alileo encounters}.
\newblock \bibinfo{journal}{Science} \bibinfo{volume}{281},
  \bibinfo{pages}{2019--2022}.
\newblock \DOIprefix\doi{10.1126/science.281.5385.2019}.
\bibitem[{Arfken et~al.(2012)Arfken, Weber and Harris}]{arfken2012mathematical}
\bibinfo{author}{Arfken, G.}, \bibinfo{author}{Weber, H.},
  \bibinfo{author}{Harris, F.}, \bibinfo{year}{2012}.
\newblock \bibinfo{title}{Mathematical Methods for Physicists}.
\newblock \bibinfo{edition}{Seventh} ed., \bibinfo{publisher}{Academic Press}.
\bibitem[{Arridge and Eggington(2021)}]{arridge2021electromagnetic}
\bibinfo{author}{Arridge, C.S.}, \bibinfo{author}{Eggington, J.W.},
  \bibinfo{year}{2021}.
\newblock \bibinfo{title}{Electromagnetic induction in the icy satellites of
  {U}ranus}.
\newblock \bibinfo{journal}{Icarus} ,
  \bibinfo{pages}{114562}\DOIprefix\doi{10.1016/j.icarus.2021.114562}.
\bibitem[{Brink and Satchler(1968)}]{brink1968angular}
\bibinfo{author}{Brink, D.M.}, \bibinfo{author}{Satchler, G.R.},
  \bibinfo{year}{1968}.
\newblock \bibinfo{title}{Angular momentum}.
\newblock \bibinfo{edition}{Second} ed., \bibinfo{publisher}{Clarendon Press}.
\bibitem[{Cochrane et~al.(2021)Cochrane, Vance, Nordheim, Styczinski, Masters
  and Regoli}]{cochrane2021search}
\bibinfo{author}{Cochrane, C.J.}, \bibinfo{author}{Vance, S.D.},
  \bibinfo{author}{Nordheim, T.A.}, \bibinfo{author}{Styczinski, M.},
  \bibinfo{author}{Masters, A.}, \bibinfo{author}{Regoli, L.H.},
  \bibinfo{year}{2021}.
\newblock \bibinfo{title}{In search of subsurface oceans within the uranian
  moons}.
\newblock \href{http://arxiv.org/abs/2105.06087}{\tt arXiv:2105.06087}.
  \bibinfo{note}{{S}ubmitted to Journal of Geophysical Research: Planets}.
\bibitem[{Condon and Shortley(1951)}]{condonshortley}
\bibinfo{author}{Condon, E.U.}, \bibinfo{author}{Shortley, G.H.},
  \bibinfo{year}{1951}.
\newblock \bibinfo{title}{The theory of atomic spectra}.
\newblock \bibinfo{publisher}{Cambridge University Press}.
\bibitem[{Dennery and Krzywicki(2012)}]{dennery2012mathematics}
\bibinfo{author}{Dennery, P.}, \bibinfo{author}{Krzywicki, A.},
  \bibinfo{year}{2012}.
\newblock \bibinfo{title}{Mathematics for physicists}.
\newblock \bibinfo{publisher}{Dover}.
\bibitem[{Edmonds(1996)}]{edmonds1996angular}
\bibinfo{author}{Edmonds, A.R.}, \bibinfo{year}{1996}.
\newblock \bibinfo{title}{Angular momentum in quantum mechanics}.
\newblock \bibinfo{publisher}{Princeton University Press}.
\bibitem[{Glover(2015)}]{glover2015geophysical}
\bibinfo{author}{Glover, P.W.J.}, \bibinfo{year}{2015}.
\newblock \bibinfo{title}{Geophysical properties of the near surface {E}arth:
  Electrical properties}, in: \bibinfo{editor}{Schubert, G.} (Ed.),
  \bibinfo{booktitle}{Treatise on Geophysics}. \bibinfo{edition}{second} ed..
  \bibinfo{publisher}{Elsevier}. volume~\bibinfo{volume}{11}, pp.
  \bibinfo{pages}{89--137}.
\newblock \DOIprefix\doi{10.1016/B978-0-444-53802-4.00189-5}.
\bibitem[{Hand and Chyba(2007)}]{hand2007empirical}
\bibinfo{author}{Hand, K.P.}, \bibinfo{author}{Chyba, C.F.},
  \bibinfo{year}{2007}.
\newblock \bibinfo{title}{Empirical constraints on the salinity of the europan
  ocean and implications for a thin ice shell}.
\newblock \bibinfo{journal}{Icarus} \bibinfo{volume}{189},
  \bibinfo{pages}{424--438}.
\newblock \DOIprefix\doi{10.1016/j.icarus.2007.02.002}.
\bibitem[{Hartkorn and Saur(2017)}]{hartkorn2017induction}
\bibinfo{author}{Hartkorn, O.}, \bibinfo{author}{Saur, J.},
  \bibinfo{year}{2017}.
\newblock \bibinfo{title}{Induction signals from {C}allisto's ionosphere and
  their implications on a possible subsurface ocean}.
\newblock \bibinfo{journal}{Journal of Geophysical Research: Space Physics}
  \bibinfo{volume}{122}, \bibinfo{pages}{11--677}.
\newblock \DOIprefix\doi{10.1002/2017JA024269}.
\bibitem[{Hartkorn et~al.(2017)Hartkorn, Saur and
  Strobel}]{hartkorn2017structure}
\bibinfo{author}{Hartkorn, O.}, \bibinfo{author}{Saur, J.},
  \bibinfo{author}{Strobel, D.F.}, \bibinfo{year}{2017}.
\newblock \bibinfo{title}{Structure and density of {C}allisto's atmosphere from
  a fluid-kinetic model of its ionosphere: Comparison with {H}ubble {S}pace
  {T}elescope and {G}alileo observations}.
\newblock \bibinfo{journal}{Icarus} \bibinfo{volume}{282},
  \bibinfo{pages}{237--259}.
\newblock \DOIprefix\doi{10.1016/j.icarus.2016.09.020}.
\bibitem[{Hemingway and Mittal(2019)}]{hemingway2019enceladus}
\bibinfo{author}{Hemingway, D.J.}, \bibinfo{author}{Mittal, T.},
  \bibinfo{year}{2019}.
\newblock \bibinfo{title}{Enceladus's ice shell structure as a window on
  internal heat production}.
\newblock \bibinfo{journal}{Icarus} \bibinfo{volume}{332},
  \bibinfo{pages}{111--131}.
\bibitem[{Hussmann et~al.(2006)Hussmann, Sohl and
  Spohn}]{hussmann2006subsurface}
\bibinfo{author}{Hussmann, H.}, \bibinfo{author}{Sohl, F.},
  \bibinfo{author}{Spohn, T.}, \bibinfo{year}{2006}.
\newblock \bibinfo{title}{Subsurface oceans and deep interiors of medium-sized
  outer planet satellites and large trans-neptunian objects}.
\newblock \bibinfo{journal}{Icarus} \bibinfo{volume}{185},
  \bibinfo{pages}{258--273}.
\bibitem[{Iess et~al.(2014)Iess, Stevenson, Parisi, Hemingway, Jacobson,
  Lunine, Nimmo, Armstrong, Asmar, Ducci and Tortora}]{iess2014gravity}
\bibinfo{author}{Iess, L.}, \bibinfo{author}{Stevenson, D.},
  \bibinfo{author}{Parisi, M.}, \bibinfo{author}{Hemingway, D.},
  \bibinfo{author}{Jacobson, R.}, \bibinfo{author}{Lunine, J.},
  \bibinfo{author}{Nimmo, F.}, \bibinfo{author}{Armstrong, J.},
  \bibinfo{author}{Asmar, S.}, \bibinfo{author}{Ducci, M.},
  \bibinfo{author}{Tortora, P.}, \bibinfo{year}{2014}.
\newblock \bibinfo{title}{The gravity field and interior structure of
  {E}nceladus}.
\newblock \bibinfo{journal}{Science} \bibinfo{volume}{344},
  \bibinfo{pages}{78--80}.
\bibitem[{Jackson(1999)}]{jackson1999classical}
\bibinfo{author}{Jackson, J.D.}, \bibinfo{year}{1999}.
\newblock \bibinfo{title}{Classical electrodynamics}.
\newblock \bibinfo{publisher}{John Wiley \& Sons}.
\newblock \DOIprefix\doi{10.1119/1.19136}.
\bibitem[{Khurana et~al.(2002)Khurana, Kivelson and
  Russell}]{khurana2002searching}
\bibinfo{author}{Khurana, K.K.}, \bibinfo{author}{Kivelson, M.G.},
  \bibinfo{author}{Russell, C.T.}, \bibinfo{year}{2002}.
\newblock \bibinfo{title}{Searching for liquid water in {E}uropa by using
  surface observatories}.
\newblock \bibinfo{journal}{Astrobiology} \bibinfo{volume}{2},
  \bibinfo{pages}{93--103}.
\newblock \DOIprefix\doi{10.1089/153110702753621376}.
\bibitem[{Marion and Heald(1980)}]{marion1980classical}
\bibinfo{author}{Marion, J.B.}, \bibinfo{author}{Heald, M.A.},
  \bibinfo{year}{1980}.
\newblock \bibinfo{title}{Classical Electromagnetic Radiation}.
\newblock \bibinfo{edition}{Second} ed., \bibinfo{publisher}{Academic Press},
  \bibinfo{address}{New York}.
\bibitem[{Moffatt(1978)}]{moffatt1978field}
\bibinfo{author}{Moffatt, H.K.}, \bibinfo{year}{1978}.
\newblock \bibinfo{title}{Magnetic field generation in electrically conducting
  fluids}. volume~\bibinfo{volume}{2}.
\newblock \bibinfo{publisher}{Cambridge University Press}.
\bibitem[{Parkinson(1983)}]{parkinson1983introduction}
\bibinfo{author}{Parkinson, W.D.}, \bibinfo{year}{1983}.
\newblock \bibinfo{title}{Introduction to Geomagnetism}.
\newblock \bibinfo{publisher}{Elsevier}.
\bibitem[{Rambaux and Castillo-Rogez(2013)}]{rambaux2013tides}
\bibinfo{author}{Rambaux, N.}, \bibinfo{author}{Castillo-Rogez, J.},
  \bibinfo{year}{2013}.
\newblock \bibinfo{title}{Tides on satellites of giant planets}, in:
  \bibinfo{editor}{Souchay, J.}, \bibinfo{editor}{Mathis, S.},
  \bibinfo{editor}{Tokieda, T.} (Eds.), \bibinfo{booktitle}{Tides in Astronomy
  and Astrophysics}. \bibinfo{publisher}{Springer-Verlag Berlin Heidelberg},
  pp. \bibinfo{pages}{167--200}.
\newblock \DOIprefix\doi{10.1007/978-3-642-32961-6_5}.
\bibitem[{Saur et~al.(2010)Saur, Neubauer and Glassmeier}]{saur2010induced}
\bibinfo{author}{Saur, J.}, \bibinfo{author}{Neubauer, F.M.},
  \bibinfo{author}{Glassmeier, K.H.}, \bibinfo{year}{2010}.
\newblock \bibinfo{title}{Induced magnetic fields in solar system bodies}.
\newblock \bibinfo{journal}{Space science reviews} \bibinfo{volume}{152},
  \bibinfo{pages}{391--421}.
\newblock \DOIprefix\doi{10.1007/s11214-009-9581-y}.
\bibitem[{Seufert et~al.(2011)Seufert, Saur and Neubauer}]{seufert2011multi}
\bibinfo{author}{Seufert, M.}, \bibinfo{author}{Saur, J.},
  \bibinfo{author}{Neubauer, F.M.}, \bibinfo{year}{2011}.
\newblock \bibinfo{title}{Multi-frequency electromagnetic sounding of the
  {G}alilean moons}.
\newblock \bibinfo{journal}{Icarus} \bibinfo{volume}{214},
  \bibinfo{pages}{477--494}.
\newblock \DOIprefix\doi{10.1016/j.icarus.2011.03.017}.
\bibitem[{Srivastava(1966)}]{srivastava1966theory}
\bibinfo{author}{Srivastava, S.P.}, \bibinfo{year}{1966}.
\newblock \bibinfo{title}{Theory of the magnetotelluric method for a spherical
  conductor}.
\newblock \bibinfo{journal}{Geophysical Journal International}
  \bibinfo{volume}{11}, \bibinfo{pages}{373--387}.
\bibitem[{Styczinski and Harnett(2021)}]{styczinski2021induced}
\bibinfo{author}{Styczinski, M.J.}, \bibinfo{author}{Harnett, E.M.},
  \bibinfo{year}{2021}.
\newblock \bibinfo{title}{Induced magnetic moments from a nearly spherical
  ocean}.
\newblock \bibinfo{journal}{Icarus} \bibinfo{volume}{354},
  \bibinfo{pages}{114020}.
\newblock \DOIprefix\doi{10.1016/j.icarus.2020.114020}.
\bibitem[{Tobie et~al.(2003)Tobie, Choblet and Sotin}]{tobie2003tidally}
\bibinfo{author}{Tobie, G.}, \bibinfo{author}{Choblet, G.},
  \bibinfo{author}{Sotin, C.}, \bibinfo{year}{2003}.
\newblock \bibinfo{title}{Tidally heated convection: {C}onstraints on
  {E}uropa's ice shell thickness}.
\newblock \bibinfo{journal}{Journal of Geophysical Research: Planets}
  \bibinfo{volume}{108}.
\newblock \DOIprefix\doi{10.1029/2003JE002099}.
\bibitem[{Tyler et~al.(1989)Tyler, Sweetnam, Anderson, Borutzki, Campbell,
  Eshleman, Gresh, Gurrola, Hinson, Kawashima, Kursinski, Levy, Lindal, Lyons,
  Marouf, Rosen, Simpson and Wood}]{tyler1989voyager}
\bibinfo{author}{Tyler, G.L.}, \bibinfo{author}{Sweetnam, D.N.},
  \bibinfo{author}{Anderson, J.D.}, \bibinfo{author}{Borutzki, S.E.},
  \bibinfo{author}{Campbell, J.K.}, \bibinfo{author}{Eshleman, V.R.},
  \bibinfo{author}{Gresh, D.L.}, \bibinfo{author}{Gurrola, E.M.},
  \bibinfo{author}{Hinson, D.P.}, \bibinfo{author}{Kawashima, N.},
  \bibinfo{author}{Kursinski, E.R.}, \bibinfo{author}{Levy, G.S.},
  \bibinfo{author}{Lindal, G.F.}, \bibinfo{author}{Lyons, J.R.},
  \bibinfo{author}{Marouf, E.A.}, \bibinfo{author}{Rosen, P.A.},
  \bibinfo{author}{Simpson, R.A.}, \bibinfo{author}{Wood, G.E.},
  \bibinfo{year}{1989}.
\newblock \bibinfo{title}{Voyager radio science observations of {N}eptune and
  {T}riton}.
\newblock \bibinfo{journal}{Science} \bibinfo{volume}{246},
  \bibinfo{pages}{1466--1473}.
\bibitem[{Vance et~al.(2021)Vance, Styczinski, Bills, Cochrane, Soderlund,
  G\'{o}mez-P\'{e}rez and Paty}]{vance2021magnetic}
\bibinfo{author}{Vance, S.D.}, \bibinfo{author}{Styczinski, M.J.},
  \bibinfo{author}{Bills, B.G.}, \bibinfo{author}{Cochrane, C.J.},
  \bibinfo{author}{Soderlund, K.M.}, \bibinfo{author}{G\'{o}mez-P\'{e}rez, N.},
  \bibinfo{author}{Paty, C.}, \bibinfo{year}{2021}.
\newblock \bibinfo{title}{Magnetic induction responses of {J}upiter's ocean
  moons including effects from adiabatic convection}.
\newblock \bibinfo{journal}{Journal of Geophysical Research: Planets}
  \bibinfo{volume}{126}, \bibinfo{pages}{e2020JE006418}.
\newblock \DOIprefix\doi{10.1029/2020JE006418}.
\bibitem[{Wigner(1931)}]{wigner1931gruppen}
\bibinfo{author}{Wigner, E.P.}, \bibinfo{year}{1931}.
\newblock \bibinfo{title}{Gruppentheorie und ihre Anwendung auf die
  Quantenmechanik der Atomspektren}.
\newblock \bibinfo{publisher}{Springer}.
\bibitem[{Zimmer et~al.(2000)Zimmer, Khurana and
  Kivelson}]{zimmer2000subsurface}
\bibinfo{author}{Zimmer, C.}, \bibinfo{author}{Khurana, K.K.},
  \bibinfo{author}{Kivelson, M.G.}, \bibinfo{year}{2000}.
\newblock \bibinfo{title}{Subsurface oceans on {E}uropa and {C}allisto:
  {C}onstraints from {G}alileo magnetometer observations}.
\newblock \bibinfo{journal}{Icarus} \bibinfo{volume}{147},
  \bibinfo{pages}{329--347}.
\newblock \DOIprefix\doi{10.1006/icar.2000.6456}.

\end{thebibliography}
	
\end{document}


\title[mode=title]{Supplemental Text for ``An analytic solution for evaluating the magnetic field induced from an arbitrary, asymmetric ocean world''}
	\shorttitle{Supplement: Magnetic induction from asymmetric ocean worlds}
	\shortauthors{M.J.~Styczinski et al.}
	
	\author[1,2]{Marshall J. Styczinski}[orcid=0000-0003-4048-5914]
	\cormark[1]
	\ead{mjstyczi@uw.edu}
	\credit{Conceptualization, calculations, software, writing}
	\cortext[cor1]{Corresponding author}
	\address[1]{Department of Physics, University of Washington, Box 351560, 3910 15th Ave NE, Seattle, WA 98195-1560, USA.}
	\address[2]{UW Astrobiology Program, University of Washington, Box 351580, 3910 15th Ave NE, Seattle, WA 98195-1580, USA.}
	
	\author[3]{Steven D. Vance}[orcid=0000-0002-4242-3293]
	\ead{svance@jpl.caltech.edu}
	\credit{Conceptualization, software, writing}
	\address[3]{Jet Propulsion Laboratory, California Institute of Technology, 4800 Oak Grove Dr, Pasadena, CA 91109-8001, USA.}
	
	\author[2,4]{Erika M. Harnett}[orcid=0000-0002-0025-7274]
	\ead{eharnett@uw.edu}
	\credit{Mentorship, editing}
	\address[4]{Department of Earth and Space Sciences, University of Washington, Box 351310, 4000 15th Ave NE, Seattle, WA 98195-1310, USA.}
	
	\author[3]{Corey J. Cochrane}[orcid=0000-0002-4935-1472]
	\ead{corey.j.cochrane@jpl.nasa.gov}
	\credit{External field models}
	
	\maketitle
	
	The final result from our derivation is complicated.
	To clarify the source of each piece of the result, we present each step of the derivation from first principles in \S\ref{supp:derivation}.
	Python code for evaluating our model is provided as Supplemental Material\footnote{Also available as a Zenodo archive: \url{https://zenodo.org/record/5002956}.}.
	Next, we present a table of values given as inputs to the \textit{PlanetProfile} framework in \S\ref{supp:ppTable} (\T\ref{table:suppPlanetProps}).
	In \S\ref{supp:gravity}, we give a detailed derivation of our method for translating $J_2$ and $C_{22}$ gravity coefficients into geometric perturbations of boundary shapes.
	The full results from our example models applying our results for magnetic induction of asymmetric conducting layers in icy moons are presented in \S\ref{supp:fullResults}.
	Therein, we include figures showing the asymmetric layer topography, induced field magnitude at the considered altitude for the asymmetric model, and differences in the induced field components and magnitude that result from the asymmetry in the conductivity structure.
	\S\ref{supp:wignerSY} compiles explicit formulas for the Wigner $3j$-symbols and the mixing coefficients $\SYrev$ and $\dSYrev$.
	Finally, in \S\ref{supp:transitions}, we describe strategies for mitigating numerical difficulties with applying the layer method that result from the need to evaluate ratios of differences of very large, complex numbers.
	All figures appear at the end of the document.
	
	\section{Derivation from first principles}\label{supp:derivation}
	
	Our goal is to apply Maxwell's laws at boundaries between regions and find the induced magnetic moments that satisfy the resulting equations.
	The same is true of the spherically symmetric, recursive solution developed by \citet{srivastava1966theory} and prominently applied by \citet{zimmer2000subsurface}, \citet{seufert2011multi}, and others.
	However, in our application, we describe the boundaries between regions of varying electrical conductivity by an arbitrary shape, rather than by a constant radius.
	The shape of the boundary has important consequences, and determines which magnetic moments are induced by a given excitation field applied by the parent planet.
	A derivation of the \citeauthor{srivastava1966theory} recursive method is presented in more modern notation by the oft-cited \citet{parkinson1983introduction}.
	The \citeauthor{parkinson1983introduction} derivation contains several errors and inconsistencies, so we include here a full detailing of our solution from first principles so that we may best identify the consequences of those errors.
	
	The excitation field applied to the conducting body has the form
	\begin{equation}
		\Bexc(\r,t) = \B_o(\r) + \sum_\mathrm{j}\B_{e,\mathrm{j}}(\r)e^{-i\omega_\mathrm{j} t} \label{eq:bexc},
	\end{equation}
	with static ($\B_o$) and dynamic ($\B_{e,\mathrm{j}}$) components that are complex in general.
	The measurable magnetic field is found by taking the real part of any complex expressions for the vector components.
	Superposition permits an independent handling of each excitation frequency, so we will focus on a single excitation frequency in our derivation.
	Our method may then be applied repeatedly for a combination of frequencies, and the results summed together.
	
	In regions free of electric currents, the magnetic field satisfies Laplace's equation and therefore may be described by the gradient of a scalar potential $\psi$:
	\begin{eqnarray}
		\nabla^2\B = 0 \label{eq:laplace}\\
		\B = -\nabla\psi \label{eq:psi}.
	\end{eqnarray}
	These equations are valid outside the conducting body (and outside any ionosphere) if we neglect currents in the magnetized plasma.
	Although currents in the plasma environment around the body are not generally negligible, the principle of superposition permits us to consider each contribution to the net electromagnetic response independently---the net magnetic field is then the sum from each individual contribution.
	In this work, we consider only the induced magnetic moments generated by the interaction of the primary excitation field with the conducting body, as this is the dominant interaction that induces magnetic fields from within solar system moons.
	
	In spherical polar coordinates $\r~=~(r,\theta,\phi)$, general solutions to \Eqs\ref{eq:laplace} and \ref{eq:psi} have the form \citep{jackson1999classical}
	\begin{equation}
		\psi(\r) = \sum_{n,m}\roB\left(\Benm\rova^n + \Binm\aovr^{n+1}\right)\Ynm(\theta,\phi)\label{eq:multipole},
	\end{equation}
	where $\Benm$ and $\Binm$ are complex coefficients for the excitation and induced magnetic fields, $\Ynm$ are spherical harmonics of degree $n$ and order $m$, and $\roB$ is a unit of radial distance, typically the outer radius of the conducting body.
	The $B^e$ potentials, proportional to positive powers of $r$, can only be generated from outside of the conducting body under examination.
	The $B^i$ terms in \Eq\ref{eq:multipole} are those of the multipole expansion, so each $\Binm$ is proportional to, and thus represents, an induced multipole moment.
	We assume the magnetic potential for the excitation field oscillates sinusoidally, so time dependence is added to \Eq\ref{eq:multipole} by multiplication of $\eiwt$ as in \Eq\ref{eq:bexc}.
	
	Within the conducting body, the dynamic excitation field induces electric fields that drive currents, so $\B$ cannot be represented by \Eq\ref{eq:psi} in this region.
	Instead, we must use a diffusion equation for $\B$, derived from combining Maxwell's laws with Ohm's law:
	\begin{equation}
		\lapl\B = \mu\sigma\dbdt\label{eq:diffusion}.
	\end{equation}
	For simplicity in deriving our model, we neglect movement of conducting material within the body (as in the case of ocean currents), which can itself induce secondary fields \citep{saur2010induced,vance2021magnetic}.
	As we are considering only the oscillatory magnetic field, taking the time derivative of $\B$ is equivalent to multiplication by $-i\omega$.
	We can thus rewrite \Eq\ref{eq:diffusion} in terms of a diffusion constant $k$, and arrive at a vector Helmholtz equation:
	\begin{eqnarray}
		\lapl\Bosc = -k^2\Bosc \label{eq:kdiff}\\
		k = \sqrt{i\omega\mu\sigma} \label{eq:k}.
	\end{eqnarray}
	The definition of $k$ includes an arbitrary phase---we may also have chosen to include the ($-$) underneath the square root, as in \citet{parkinson1983introduction}.
	This choice of phase will ultimately determine which differential equation we eventually find for the radial dependence of the internal magnetic field (\Eq\ref{eq:bessel}).
	
	In general, $\mu$ and $\sigma$ are functions of position and will vary throughout the body.
	On planetary scales, $\mu$ is well approximated by $\mu_o$, even for bodies containing large amounts of ferromagnetic materials \citep{saur2010induced}; we assume $\mu = \mu_o$ everywhere in this work.
	We further assume that $\sigma$ is uniform within each conducting region, and that each region is global in extent, {\ie}\ any path outwards from the center must pass through each region sequentially.
	
	General solutions to \Eq\ref{eq:kdiff} for the configuration at hand must be consistent with a poloidal field; since they are induced by an external field, there will be no toroidal field component \citep{moffatt1978field}.
	Poloidal fields take the following form:
	\begin{align}
		\B_\Psi &= \curl\curl(\Psi\r) \label{eq:polpot}\\
		B_{r,\Psi} &= \frac{1}{r}\left[\Lsq \right]\Psi \label{eq:brwithP}\\
		B_{\theta,\Psi} &= \frac{1}{r}\ddt\ddr(\Psi r) \label{eq:btwithP}\\
		B_{\phi,\Psi} &= \frac{1}{r\sn}\ddp\ddr(\Psi r) \label{eq:bpwithP},
	\end{align}
	where the poloidal potential $\Psi$ is a scalar function of position.
	Note that the expressions for the poloidal potential vector components given by \citet[][\Eq158]{parkinson1983introduction} contain a sign error in the $\hat{r}$ component.
	
	The quantity in square brackets in \Eq\ref{eq:brwithP} is the angular momentum operator $\Lop$, of which the spherical harmonics $\Ynm$ are eigenfunctions \citep{edmonds1996angular,jackson1999classical}:
	\begin{equation}
		\Lop\Ynm = n(n+1)\Ynm \label{eq:Leigen}.
	\end{equation}
	Note that \citet[][\Eq2.26]{moffatt1978field} presents \Eq\ref{eq:Leigen} with a sign error based on their definition of the angular momentum operator.
	If we suppose $\Psi$ is separable, we can expand it in spherical harmonics:
	\begin{equation}
		\Psi(r,\theta,\phi) = \sumnm\Anm\Rnm(r)\,\Ynm(\theta,\phi) \label{eq:polgeneral},
	\end{equation}
	where $\Anm$ are constant coefficients determined by the boundary conditions and $\Rnm$ are functions we must determine from other relations.
	As we later satisfy the boundary conditions with this functional form of $\Psi$, the uniqueness theorem confirms that this is \textit{the} physically correct representation \citep{dennery2012mathematics}, validating the supposition that $\Psi$ is separable.
	
	Inserting \Eq\ref{eq:polgeneral} into \Eqs\ref{eq:brwithP}--\ref{eq:bpwithP} and utilizing \Eq\ref{eq:Leigen} yields expressions for the components of the magnetic field within the conducting body in terms of $\R$:
	\begin{align}
		B_{r,\mathrm{int}} &= \sumnm\frac{\Anm}{r}\Rnm\, n(n+1)\Ynm \label{eq:brpol}\\
		B_{\theta,\mathrm{int}} &= \sumnm\frac{\Anm}{r} \ddr(r\Rnm)\dydth{nm} \label{eq:btpol}\\
		B_{\phi,\mathrm{int}} &= \sumnm\frac{\Anm}{r\sn} \ddr(r\Rnm)\dydph{nm} \label{eq:bppol}.
	\end{align}
	We can now make use of these expressions along with \Eq\ref{eq:kdiff} to find a differential equation for $\Rnm$.
	Linearity of the $\nabla^2$ operator allows us to consider only a single $n,m$ term, as the same equations will apply to all terms.
	The $\hat{r}$ component of \Eq\ref{eq:kdiff} reads as \citep{arfken2012mathematical}
	\begin{equation}
		\lapl B_r -\frac{2B_r}{r^2} -\frac{2}{r^2\sn}\ddt(\sn B_\theta) -\frac{2}{r^2\sn}\frac{\partial B_\phi}{\partial\phi} = -k^2 B_r \label{eq:helmholtz}.
	\end{equation}
	Inserting \Eqs\ref{eq:brpol}--\ref{eq:bppol} and again exploiting the angular momentum operator, we arrive at a Bessel equation for $\Rnm$:
	\begin{equation}
		\frac{1}{\Rnm}\ddr\left(r^2\frac{\partial\Rnm}{\partial r}\right) + k^2 r^2 - n(n+1) = 0\label{eq:bessel}.
	\end{equation}
	Solutions to this equation are linear combinations of the particular solutions, which are the spherical Bessel functions of the first and second kind, $j_n$ and $y_n$:
	\begin{align}
		\Rnm(r) &= \anm\jnkr + \bnm\ynkr \quad \text{or} \quad \Rnm(r) = \anm\left(\jnkr + \bPnm\ynkr\right) ~\text{with}~ \bPnm \equiv \bnm/\anm, \label{eq:sphbes}\\
		\jnkr &= (-kr)^n \left(\frac{1}{kr}\frac{d}{d(kr)}\right)^n \frac{\sin kr}{kr}, \label{eq:besselgen}\\ 
		\ynkr &= -(-kr)^n \left(\frac{1}{kr}\frac{d}{d(kr)}\right)^n \frac{\cos kr}{kr} \label{eq:neumanngen}.
	\end{align}
	$\anm$ and $\bnm$ (or $\anm$ and $\bPnm$) in \Eq\ref{eq:sphbes} are constants determined from the boundary conditions; they will play a critical role in determining the induced magnetic field.
	The second format for $\Rnm$ presented in \Eq\ref{eq:sphbes} is useful for expressing recursion relations for spherically symmetric bodies, but the first format must be used in solving the full boundary conditions for the asymmetric case.
	We will use both formats in our derivation.
	
	\subsection{Consequences of the choice of phase for $k$}\label{supp:errors}
	If, in defining the diffusion constant $k$ in \Eq\ref{eq:k}, we had chosen the alternate phase $k=\sqrt{-i\omega\mu\sigma}$, the first term in \Eq\ref{eq:bessel} would be negated.
	In that case, the resulting differential equation would be the \textit{modified} spherical Bessel equation, with solutions $\inkr$ and $\knkr$, the \textit{modified} spherical Bessel functions of the first and second kinds, respectively.
	These are similar to the spherical Bessel functions $\jn$ and $\yn$ and have similar properties; they are functions of $\sinh kr$ and $\cosh kr$ rather than $\sin kr$ and $\cos kr$.
	In the commonly cited derivation by \citet[][Ch.~5]{parkinson1983introduction}, this choice of phase for $k$ should yield the modified spherical Bessel functions for the radial dependence of the internal magnetic field, but these authors incorrectly arrive at \Eq\ref{eq:bessel}.
	Several authors \citep{zimmer2000subsurface,khurana2002searching,hand2007empirical,arridge2021electromagnetic} have repeated this error and applied a phase for $k$ inconsistent with use of the standard Bessel functions.
	This error leaves the real part of the argument $kr$ unchanged, while negating the imaginary part, equivalent to taking the complex conjugate.
	
	The real part of $kr$ determines the (real) exponential dependence of the Bessel functions, so the scale of the result is the same.
	The complex part determines the oscillation phase of the Bessel functions, and it becomes negated.
	In effect, the error described above causes the resulting solution for the induced magnetic field to \textit{lead} the excitation field by the amount of the phase delay $\upphi$, rather than lagging (as it must lag behind the excitation).
	Overall, the conclusions of the authors repeating this error are unaffected, but the expressions they apply all include a sign change in connecting the phase delay to the equations describing the induced magnetic field.
	For consistent comparisons to these important prior studies, we note by \Eq\ref{eqm:aeip} that the phase delay $\upphi$ is the \textit{negative} of the phase of the complex amplitude $\Amp^e_1$ we later derive.
	
	\subsection{General expressions for the magnetic field in each region}\label{supp:general}
	We must now use the expressions we have obtained to relate the magnetic field components at boundaries, as needed to solve Maxwell's equations.
	First, we note some considerations for the general radial dependence $\Rnm$ inside the conducting body: The solutions $y_n$ diverge at the origin, so $\bnm$ (or $\bPnm$) must always be zero for the innermost region.
	As our solution for $\Rnm$ now contains arbitrary coefficients, we absorb the coefficients $\Anm$ into $\anm$ and $\bnm$.
	
	For later convenience, we also require expressions for $\ddr(r\,\jnkr)$ and $\ddr(r\,\ynkr)$.
	\Eqs\ref{eq:besselgen} and \ref{eq:neumanngen} can be manipulated to obtain
	\begin{equation}
		\jd \equiv \ddr(r\jn) = (n+1)\jn - kr\,\jnP, \quad
		\yd \equiv \ddr(r\yn) = (n+1)\yn - kr\,\ynP, \label{eq:jderiv}
	\end{equation}
	which we now also define as $\jd$ and $\yd$, respectively.
	
	The details and formulation of the spherical harmonics are of central importance to this work.
	We use the fully normalized, complex spherical harmonics:
	\begin{align}
		\Ynm &= \sqrt{\frac{2n+1}{4\pi}\frac{(n-m)!}{(n+m)!}}\,\Pnmu(\cn)\,\eimp, \\
		\YnmM &= (-1)^m\Ynm^*, \label{eq:YnmConj}
	\end{align}
	where $\Pnmu$ are the associated Legendre functions with the Condon--Shortley phase.
	Although in the geomagnetics literature, the spherical harmonics are often expressed using real harmonics, in the Schmidt normalization, and without the Condon--Shortley phase, we have chosen a convention ubiquitous throughout the literature on angular momentum in quantum mechanics.
	From this literature we draw many helpful results, especially regarding the Clebsch--Gordan coefficients in \S\ref{supp:products}.
	
	We will also later need expressions for $\theta$ derivatives of $\Ynm$:
	\begin{align}
		\dYnm \equiv \dydth{nm} &= \frac{1}{\sn}\left( -\mixm Y_{n-1,m} + \mixp Y_{n+1,m} \right) \quad \mathrm{for}~ m \ge 0, \label{eq:dYdef}\\
		\mixm &= (n+1)\sqrt{\frac{n^2 - m^2}{(2n-1)(2n+1)}}, \quad \mixp = n\sqrt{\frac{(n+1)^2 - m^2}{(2n+1)(2n+3)}}, \label{eq:wdef}\\
		\dYnmM &= (-1)^m\dYnmConj \label{eq:dYconj}.
	\end{align}
	We thereby define $\dYnm$ similar to $\jd$ and $\yd$, as they all pertain to the tangential components; we will not need the $\phi$ derivatives in our derivation.
	The format of \Eq\ref{eq:dYdef}, obtained using recurrence relations for the associated Legendre functions \citep[\eg\ ][]{abramowitzstegun}, has been selected for optimal use of orthogonality relations to solve the boundary conditions later.
	
	We can now write general expressions for the magnetic field in all regions.
	From \Eqs\ref{eq:psi} and \ref{eq:multipole}, the external magnetic field follows
	\begin{align}
		B_{r,\mathrm{ext}} &= \sumnm\left[ -n\rova^{n-1}\Benm + (n+1)\aovr^{n+2}\Binm \right]\Ynm \label{eq:brext}\\
		B_{\theta,\mathrm{ext}} &= \sumnm\left[ -\rova^{n-1}\Benm -\aovr^{n+2}\Binm \right]\dYnm \label{eq:btext}\\
		B_{\phi,\mathrm{ext}} &= \sumnm\left[ -\rova^{n-1}\Benm - \aovr^{n+2}\Binm \right]\frac{1}{\sn}\dydph{nm} \label{eq:bpext}.
	\end{align}
	From \Eqs\ref{eq:brpol}--\ref{eq:bppol}, the internal magnetic field follows
	\begin{align}
		B_{r,\mathrm{int}} &= \sumnm\frac{\anm\jnkr + \bnm\ynkr}{r}\, n(n+1)\Ynm \label{eq:brint}\\
		B_{\theta,\mathrm{int}} &= \sumnm\frac{\anm\jdkr + \bnm\ydkr}{r} \dYnm \label{eq:btint}\\
		B_{\phi,\mathrm{int}} &= \sumnm\frac{\anm\jdkr + \bnm\ydkr}{r} \frac{1}{\sn}\dydph{nm} \label{eq:bpint}.
	\end{align}
	The tangential components $B_\theta$ and $B_\phi$ offer redundant information in matching the solutions across the boundaries, so we will restrict our focus to the $B_\theta$ component because it contains terms for all $n$ and $m$.
	
	Solving for the $\anm$, $\bnm$, and $\Binm$ coefficients in these equations is accomplished by applying Maxwell's laws at the common boundaries between each region.
	On each boundary surface, Maxwell's laws dictate that the normal component of $\B$ must be continuous, and the tangential components of $\B/\mu$ must be continuous whenever there are no surface currents confined to the boundary itself \citep{jackson1999classical}.
	As we assume $\mu=\mu_o$ within the body, $\B$ is continuous everywhere, and the components of the magnetic field for adjacent regions are equal on the boundary surface.
	
	\subsection{Internal boundary conditions}\label{supp:internal}
	At each boundary interior to the outer surface, the vector components of the magnetic field must match according to \Eqs\ref{eq:brint}--\ref{eq:bpint}.
	At a distance from the body center $\rb(\theta,\phi)$ describing the outer boundary surface of a lower layer $\low$ with wavenumber $\kl$ and an upper boundary $\upp$ with wavenumber $\ku$, the internal boundary conditions read
	\begin{align}
		B_r:\quad \sumnm n(n+1)\frac{\jnl + \bPnml \ynl}{\rb} \anml\Ynm &= \sumnm\frac{\jnu + \bPnmu \ynu}{\rb} \anmu\Ynm n(n+1) \label{eq:bcintr}\\
		B_\theta:\qquad\qquad \sumnm\frac{\jdl + \bPnml \ydl}{\rb} \anml\dYnm &= \sumnm\frac{\jdu + \bPnmu \ydu}{\rb} \anmu\dYnm \label{eq:bcintt}.
	\end{align}
	In general, $\rb$ may be a function of $\theta$ and $\phi$; this is the major focus of the present work.
	Formatting the linear combinations of $\jn$ and $\yn$ as we have done will, in the case of spherical symmetry, allow us to solve \Eqs\ref{eq:bcintr} and \ref{eq:bcintt} for $\bPnmu$ in terms of $\bPnml$, resulting in a recursion relation.
	
	Mutual orthogonality of the spherical harmonics may be exploited to extract terms in \Eq\ref{eq:bcintr} proportional to a desired harmonic $\Ynmp$.
	Multiplying both sides by the complex conjugate of the desired harmonic $\YnmpConj$ and integrating over a unit sphere is equivalent to replacing $\Ynmp$ with $\updelta_{n,n'}\updelta_{m,m'}$ (Kronecker delta functions), effectively discarding all other terms.
	A similar operation, multiplying both sides of \Eq\ref{eq:bcintt} by $\dYnmpConj\snsq$ and integrating over a unit sphere, yields somewhat different results but may be used to reach an analogous equation, as we now describe.
	
	Linear combinations of orthogonal functions are not, in general, mutually orthogonal.
	From \Eq\ref{eq:dYdef}, we can determine that $\dYnm\sn$ may not be orthogonal to $\dYnmp\sn$ when $m = m'$ and $n' = n+2$ or $n' = n-2$.
	One such example is $Y^\star_{31}$ and $Y^\star_{11}$: integrating $Y^{\star*}_{31}Y^\star_{11}\snsq$ over a unit sphere gives a non-zero result.
	This results from the overlap of the $Y_{n+1,m}$ term in \Eq\ref{eq:dYdef} for $n=1$ and the $Y_{n-1,m}$ term for $n=3$.
	When this operation is applied in the sums over $n$ and $m$ in the boundary conditions, coupled linear equations result for these overlapping values of $n$.
	Ultimately, the equations are separable because of the following factors:
	\begin{itemize}
		\item For all terms, $n \ge 1$ because $\dydth{00}=0$.
		\item We can determine which $\anm$ are zero from the radial boundary condition equations, which has terms proportional to mutually orthogonal functions.
		\item The result of the overlap integral is the same in each conducting region, and so appears on both sides of each boundary condition equation.
	\end{itemize}
	The first two items above bound the number of equations. 
	The final item may be used to scale and sum the equations so as to eliminate the terms proportional to $n' + 2$ or $n' - 2$, where the terms proportional to $\dYnmp$ are desired.
	The coefficient of the remaining $n'$ term is altered by the overlap terms, but all terms are multiplied by the same coefficient (combinations of $\mixmp$ and $\mixpp$), so it divides away.
	Therefore, analogous to the radial equation, we finally obtain a result equivalent to replacing $\dYnm$ by $\updelta_{n,n'}\updelta_{m,m'}$.
	These operations will be critical in collecting the new terms that arise from expanding the boundary radii in spherical harmonics.
	
	\subsubsection{Spherically symmetric case}\label{supp:intsph}
	If we assume spherical symmetry in the boundary surface at $\rb$, $\bPnm$ in \Eqs\ref{eq:bcintr} and \ref{eq:bcintt} is independent of $m$ and can be reduced using the orthogonality relations discussed above.
	Each value of $m$ yields equations identical to other $m$, so we set $m=0$ and drop that subscript on $\bP$.
	Multiplying both sides of \Eq\ref{eq:bcintr} by $\YnmpConj$ and integrating over a unit sphere yields
	\begin{equation}
		\anml\left(\jnl + \bPnl \ynl\right) = \anmu\left(\jnu + \bPnu \ynu\right). \label{eq:intsphr}
	\end{equation}
	with $n=n'$.
	Multiplying both sides of \Eq\ref{eq:bcintt} by $\dYnmpConj\snsq$ and integrating over a unit sphere similarly yields
	\begin{equation}
		\anml\left(\jdl + \bPnl \ydl\right) = \anmu\left(\jdu + \bPnu \ydu\right). \label{eq:intspht}
	\end{equation}
	
	Dividing these equations (\ref{eq:intsphr} and \ref{eq:intspht}) by each other, we can now solve for $\bPnu$ in terms of $\bPnl$ to obtain the desired recursion relations.
	The solution is
	\begin{equation}
		\bPnu = \frac{(\jnul\jdll - \jnll\jdul) + \bPnl(\jnul\ydll - \ynll\jdul)}{(\jnll\ydul - \ynul\jdll) + \bPnl(\ynll\ydul - \ynul\ydll)}, \label{eq:Lambdfull}
	\end{equation}
	or
	\begin{equation}
		\bPnu = \frac{\delul + \bPnl\epsul}{\betul + \bPnl\gamul} \label{eq:Lambd}
	\end{equation}
	with
	\begin{align}
		\alpul & \equiv \jnul\ydul - \ynul\jdul = \frac{1}{\ku\rb} \label{eq:ula}\\
		\betul &\equiv \jnll\ydul - \ynul\jdll \label{eq:ulb}\\
		\gamul &\equiv \ynll\ydul - \ynul\ydll\\
		\delul &\equiv \jnul\jdll - \jnll\jdul\\
		\epsul &\equiv \jnul\ydll - \ynll\jdul \label{eq:ule},\\[1em]
		&\jnll \equiv \jnl, \quad \ydul \equiv \ydu, \quad \mathit{etc.} \label{eq:shorthand}
	\end{align}
	Although $\alpul$ does not appear in \Eq\ref{eq:Lambd}, it will later appear in the asymmetric solutions.
	
	For $\Nl$ layers within the body, there are $\Nl-1$ internal boundaries, so \Eq\ref{eq:Lambd} must be applied $\Nl-1$ times to obtain $\bPn^\Nl$.
	Recall that for the innermost layer, $\bPn^1 = 0$, so the next layer above has $\bPn^2 = \delta_n^{2,1}/\beta_n^{2,1}$.
	The notation in \Eqs\ref{eq:Lambd}--\ref{eq:ule} is selected to be directly comparable to the recursion relations presented by \citet{parkinson1983introduction} for the spherically symmetric case.
	However, \Eq\ref{eq:Lambd} appears inverted because we have chosen $\bPn$ to be a coefficient for $\yn$ instead of $\jn$.
	In \S\ref{supp:nonsphBCs}, we will expand these results to first order about the boundary radius $\rb$ to obtain our results for asymmetric boundaries.
	
	\subsection{External boundary conditions}\label{supp:external}
	
	Combining \Eqs\ref{eq:brext}--\ref{eq:bpint}, we obtain the boundary conditions that apply at the outermost $\left(\Nl^\mathrm{th}\right)$ conducting boundary of any shape:
	\begin{align}
		B_r:~ \sumnm  n(n+1) \anmo\frac{\jnkro + \bPnmo\ynkro}{\ro}\Ynm =& \sumnm \left[-n\roova^{n-1}\Benm +(n+1)\aovro^{n+2}\Binm \right]\Ynm \label{eq:bcr}\\
		B_\theta:~ \sumnm \anmo\frac{\jdkro + \bPnmo\ydkro}{\ro}\dYnm =& \sumnm\left[ -\roova^{n-1}\Benm -\aovro^{n+2}\Binm\right]\dYnm \label{eq:bct},
	\end{align}
	where $\ro(\theta,\phi)$ is the distance from the center of the body to the outermost conducting surface, with a nominal mean value of $\roB$.
	$\koB = \ko$ is that of the outermost conducting layer in the ocean.
	As with the internal boundary conditions, the only tangential component we consider is $B_\theta$, as $B_\phi$ offers redundant information.
	
	\subsubsection{Spherically symmetric case}\label{supp:extsph}
	In the case of spherical symmetry in the outer boundary surface, $\ro = \roB$.
	We can again exploit the orthogonality of the spherical harmonics to extract terms proportional to each individual $\Ynm$ or $\dYnm$.
	As described in \S\ref{supp:intsph}, spherical symmetry in the interior layers results in all $\bPnm$ independent of $m$, but the same does not apply to the external coefficients if we wish to describe a general excitation field.
	Multiplying \Eq\ref{eq:bcr} by $\YnmpConj$ and inserting $\ro = \roB$, then integrating over a unit sphere we obtain
	\begin{equation}
		n(n+1)\frac{\anmo}{\roB}\left(\jnkR + \bPno\ynkR\right) = -n\Benm + (n+1)\Binm
	\end{equation}
	with $n=n'$.
	Multiplying \Eq\ref{eq:bct} by $\dYnmpConj\snsq$ and inserting $\ro = \roB$, then integrating over a unit sphere we obtain
	\begin{equation}
		\frac{\anmo}{\roB}\left(\jdkR + \bPno\ydkR\right) = -\Benm - \Binm,
	\end{equation}
	again with $n=n'$.
	If the interior boundaries are all spherically symmetric as well, $\bPno$ may be derived from the recursion relation (\Eq\ref{eq:Lambd}) using the desired interior layer model.
	
	Eliminating $\anmo$ as in the internal boundary conditions, we can solve for the unknown $\Binm$ in terms of $\Benm$.
	In this spherically symmetric case, the solution is
	\begin{equation}
		\Binm = \frac{n}{n+1}\frac{\beto + \bPno\gamo}{\delo + \bPno\epso} \Benm \label{eq:ghisph}
	\end{equation}
	with
	\begin{align}
		\beto &\equiv \jdo - (n+1)\jno &= -\koB\roB\jnPkR \label{eq:ghb}\\
		\gamo &\equiv \ydo - (n+1)\yno &= -\koB\roB\ynPkR \\
		\delo &\equiv n\jno + \jdo &= \koB\roB\jnMkR\label{eq:ghd} \\
		\epso &\equiv n\yno + \ydo &= \koB\roB\ynMkR \label{eq:ghe},\\[1em]
		&\jno \equiv \jn(\koB\roB), \quad \mathit{etc.}
	\end{align}
	Our notation here differs slightly from that of \citet{parkinson1983introduction}, in that we factor out $n/(n+1)$ from the other terms in \Eq\ref{eq:ghisph}.
	This change allows us to readily make comparisons to the response of a perfectly conducting ocean.
	We collect the remaining parameters into another quantity, the complex response amplitude $\Ampne$:
	\begin{equation}
		\Ampne \equiv \frac{\beto + \bPno\gamo}{\delo + \bPno\epso} \qquad = -\frac{\jnPkR + \bPno\ynPkR}{\jnMkR + \bPno\ynMkR}, \label{eq:Ae}
	\end{equation}
	so that
	\begin{equation}
		\Binm = \frac{n}{n+1}\Ampne\Benm \label{eq:Aeinaction}
	\end{equation}
	describes the magnetic field induced by the body.
	\Eq\ref{eq:Aeinaction} gives the primary response to the excitation field.
	The results of this work all represent perturbations to the spherically symmetric case; as a consequence, the complex response amplitude $\Ampne$ appears in each result.
	At large $|kr|$, $\Ampne$ is asymptotic to $(1+0i)$ for all $n$.
	$\Ampne$ is independent of $m$, a result that will also hold in the asymmetric case, though the form of \Eq\ref{eq:Aeinaction} will change.
	
	In the commonly studied case of a uniform excitation field, with $n=1$, $\Ampne$ may be expressed in terms of the real amplitude $A$ and phase delay $\upphi$ \citep[{\eg}\ as defined by][]{zimmer2000subsurface}:
	\begin{equation}
		\Amp^e_1 = Ae^{-i\upphi}, \label{eq:ampPhase} 
	\end{equation}
	allowing for ready comparison with prior work.
	\Eq\ref{eq:Ae} may therefore be used with a proposed layered structure model to easily calculate the response amplitude and lag phase in the case of spherical symmetry.
	Python programs we created for this purpose are provided as Supplemental Material.
	
	Note that the negative exponent in \Eq\ref{eq:ampPhase} results from our definition of $k$ in \Eq\ref{eq:k}, as described further in \S\ref{supp:errors}.
	As a consequence of the error in the \citet{parkinson1983introduction} derivation, both our input wavenumber $k$ and our result for the complex response amplitude $\Amp$ are equal to the complex conjugate of the analogous quantities from prior work, hence the negative exponent in \Eq\ref{eq:ampPhase}.
	We define the relationship between $\Amp^e_1$, $A$, and $\upphi$ as in \Eq\ref{eq:ampPhase} to facilitate comparison with the rich set of prior work related to this topic.
	
	\subsection{Near-spherical boundary shapes}\label{supp:nonspherical}
	We must now define surfaces $\rbtp$ for near-spherical boundaries that we will insert into the internal and external boundary condition equations.
	Expanding each surface in spherical harmonics allows us to make use of relations well-known from problems involving addition of angular momenta from quantum mechanics, which we need because we will be multiplying harmonics together.
	We therefore choose boundary surfaces of the form
	\begin{equation}
		\rbtp = \rbavg + \bdb\sumpq\gpq\Ypq(\theta,\phi). \label{eq:rio}
	\end{equation}
	$\rbavg$ is the mean radius of boundary $\bdy$, $\bdb$ is the amplitude of deviation from spherical symmetry, $\gpq$ is a dimensionless constant that indicates the relative amount of each harmonic represented in the boundary surface, and $\Ypq$ are fully normalized spherical harmonics of degree $\shapen$ and order $\shapem$.
	We use the index $\low$ to indicate that this surface describes the outer boundary of the lower region.
	
	Surfaces described by \Eq\ref{eq:rio} are near-spherical in that we make the approximation $\bdb \ll \rbavg$ for all $\bdy$.
	Equivalently, $\bdb/\rbavg \ll 1$, and we retain terms up to first order in $\bdb/\rbavg$ only.
	This approximation enables us to use a Taylor expansion in the boundary conditions that truncates quickly, adding only one term containing a product of two spherical harmonics.
	A product of spherical harmonics may be expressed as a linear combination of different harmonics \citep{wigner1931gruppen,condonshortley}.
	The multiplication of harmonics therefore results in ``mixing'' of harmonics in the excitation field from $n=1$ into other $n$, so a uniform excitation field induces magnetic moments of quadrupole order or higher for this shape, in addition to altering the original dipole moments.
	
	\subsubsection{Taylor expansion of boundary shapes}\label{supp:taylor}
	Let us now insert our near-spherical $\rb$ into the expressions for the magnetic field in \Eqs\ref{eq:brext}--\ref{eq:bpint}.
	To first order, a Taylor expansion of a function $f(r)$ about $\rb$ has terms
	\begin{equation}
		f(\rb) \approx f(\rbavg) + (\rb-\rbavg)\frac{\partial f(r)}{\partial r}\Bigr|_{r=\rbavg} = f(\rbavg) + \bdb\Big[\sumpq\gpq\Ypq(\theta,\phi)\Big]\frac{\partial f(r)}{\partial r}\Bigr|_{r=\rbavg}.
	\end{equation}
	The $r^n$ power series that multiply $\Benm$ and $\Binm$ in \Eqs\ref{eq:brext}--\ref{eq:bpext} then have the form
	\begin{equation}
		\rb^n \approx \rbavg^n\left(1 + n\,\frac{\bdb}{\rbavg}\sumpq\gpq\Ypq\right). \label{eq:rpower}
	\end{equation}
	The interior field terms in \Eqs\ref{eq:brint}--\ref{eq:bpint} become
	\begin{align}
		\frac{\jnkrb + \bPnmb\ynkrb}{\rb} \approx \frac{1}{\rbavg}\bigg( \jnkrbz + \bPnml\ynkrbz & \label{eq:taylor1}\\
		+ \frac{\bdb}{\rbavg}\bigg[\sumpq\gpq\Ypq\bigg] \bigg[ \jdkrbz + \bPnml\ydkrbz &- 2\big(\jnkrbz + \bPnml\ynkrbz\big) \bigg] \bigg) \nonumber
	\end{align}
	and
	\begin{align}
		\frac{\jdkrb + \bPnmb\ydkrb}{\rb} \approx \frac{1}{\rbavg}\bigg( \jdkrbz + \bPnml\ydkrbz & \label{eq:taylor2}\\
		+ \frac{\bdb}{\rbavg}\bigg[\sumpq\gpq\Ypq\bigg] \bigg[ \big(\jnkrbz + \bPnml\ynkrbz\big) &\big(n(n+1)-k^2 \rbavg^2\big) - \big(\jdkrbz + \bPnml\ydkrbz\big) \bigg] \bigg),\nonumber
	\end{align}
	where $k$ may correspond to that above or below the boundary (consistent throughout the expression), except at the outer boundary where all $k$ take the value of the top conducting layer $k_\Nl$.
	With the above expressions, we can now evaluate both the internal and external boundary conditions at the perturbed, near-spherical boundaries between layers.
	
	\subsubsection{Products of spherical harmonics}\label{supp:products}
	The final piece required to express the internal and external boundary conditions in our method is to convert the products of spherical harmonics that result from the Taylor expansion into sums of other harmonics.
	We wish to express
	\begin{align}
		\Ynm\Ypq &= \sumnmp\SY\Ynmp, \label{eq:SYdef}
	\end{align}
	allowing us to replace these products where they appear so that we may retrieve the resulting magnetic moments through orthogonality of the spherical harmonics.
	$\SYchar$ are constant coefficients with subscripts indicating the input excitation harmonic $n,m$, the boundary shape harmonic $\shapen,\shapem$, and the output expansion harmonic $n',m'$.
	We also desire a similar expression for the analogous $\SYchar^\star$ that will allow us to replace $\dYnm\Ypq$, but it cannot be expressed as simply (see \S\ref{supp:nonortho}).
	Using fully normalized spherical harmonics $Y$ that incorporate the Condon--Shortley phase, $\SY$ are the Clebsch--Gordan coefficients.
	Ultimately, we will multiply by $\YnmpConj$ or $\dYnmpConj\snsq$ and integrate over a unit sphere; with this intent in mind, we will express $\SYchar$ in terms of the Wigner $3j$-symbols $\wigGen$, which are proportional to the Clebsch--Gordan coefficients.
	The $3j$-symbols satisfy the useful identity
	\begin{equation}
		\int_0^{2\pi}\int_0^{\pi}\Ynm\Ypq\Ynmp\sn d\theta d\phi = \sqrt{\frac{(2n+1)(2\shapen+1)(2n'+1)}{4\pi}}\wigZYnm\wigYnmM \label{eq:3Yproduct}
	\end{equation}
	which follows from a related identity from \citet{brink1968angular}:
	\begin{equation}
		\int_0^{2\pi}\int_0^{\pi} (-1)^{m\updelta_{m|m|}}\Pnm\eimp ~(-1)^{m\updelta_{\shapem|\shapem|}}\Ppq\eiqp ~(-1)^{m\updelta_{m'|m'|}}\Pnmp\eimpp \sn d\theta d\phi = 4\pi\wigZYnm\wigYnm, \label{eq:barPproduct}
	\end{equation}
	where $\Pnm$ are Schmidt semi-normalized associated Legendre functions without the Condon--Shortley phase.
	The general expression for the Wigner $3j$-symbols is given in \S\ref{supp:wignerSY}.
	The selection rules, which dictate the combinations of harmonics yielding non-zero terms, are summarized as follows \citep{brink1968angular}:
	\begin{itemize}
		\item $|n - \shapen| \le n' \le n + \shapen$, often called the ``triangular condition.''
		\item $m' = -(m + \shapem)$.
		The general expression for the $3j$-symbols is unchanged under $m' \rightarrow -m'$ (see \Eq\ref{eq:wigner}).
		From \Eq\ref{eq:YnmConj}, we see that \Eq\ref{eq:3Yproduct} is unchanged if we replace $\Ynmp$ with $(-1)^{m'} Y^*_{n',-m'}$.
		We therefore use the selection rule $m' = m + \shapem$ for clarity in our derivation and negate $m'$ in the final $3j$-symbol, as we have done in \Eq\ref{eq:3Yproduct}.
		\item When $m = \shapem = m' = 0$, the $3j$-symbols are non-zero only if $n+\shapen+n'$ is even.
		Therefore, only $n' = n+\shapen,~ n+\shapen-2,~ n+\shapen-4,~ \dots~ |n - \shapen|$ give nonzero terms.
	\end{itemize}
	The results of the integration in each of \Eqs\ref{eq:3Yproduct} and \ref{eq:barPproduct} are proportional to $\wigZYnm$, so all products are subject to all of these conditions.
	
	Finally, we now obtain a direct expression for the product coefficients $\SY$ to replace the products $\Ynm\Ypq$ in the radial boundary conditions.
	Multiplying both sides of \Eq\ref{eq:SYdef} by $\YnmpConj$ and integrating over a unit sphere, we obtain
	\begin{align}
		\SY &= (-1)^{m'}\wigNorm{n}{\shapen}{n'}\wig{n}{0}{\shapen}{0}{n'}{0}\wig{n}{m}{\shapen}{\shapem}{n'}{-m'} \label{eq:SYresult}\\
		&\quad \mathrm{for} \quad |n - \shapen| \le n' \le n + \shapen, \quad n + \shapen + n' ~\mathrm{is~even,} \quad\mathrm{and} \quad m' = m + \shapem, \nonumber\\
		\SY &= 0 \quad \mathrm{otherwise.}
	\end{align}
	Using the selection rules and the expressions in \Eqs\ref{eq:wigner} and \ref{eq:wignerzero}, we can derive an explicit formula for the nonzero terms in $\SY$ (see \Eq\ref{eq:SYdirect}):
	\begin{align}
		&\SY = (-1)^\evind\sqrt{\frac{(2n+1)(2\shapen+1)(2n'+1) ~ (n'+m')!(n'-m')!}{4\pi ~  (n+m)!(n-m)!(\shapen+\shapem)!(\shapen-\shapem)!}} ~
		\frac{\left(n+\shapen-\evind\right)!}{\evind!(n-\evind)!(\shapen-\evind)!} \quad\times \label{eq:SYresultexplicit}\\
		& \frac{(2n-2\evind)!(2\shapen-2\evind)!}{(2n'+1+2\evind)!} ~
		\sum_{\kv=\kv^-}^{\kv^+}\frac{(-1)^\kv(2\evind)!}{\kv!(2\evind-\kv)!}~\frac{(n+m)!(n-m)!(\shapen+\shapem)!(\shapen-\shapem)!}{\big(n+m-(2\evind-\kv)\big)!\big(n-m-\kv\big)!\big(\shapen+\shapem-\kv\big)!\big(\shapen-\shapem-(2\evind-\kv)\big)!} \nonumber
	\end{align}
	for $\evind = \frac{1}{2}(n + p - n')$, $\kv^- = \max(0,~2\evind-(n+m),~2\evind-(\shapen-\shapem))$, and $\kv^+ = \min(2\evind,~n-m,~\shapen+\shapem)$.
	These expressions allow us to use the replacement rule $\Ynm\Ypq\rightarrow \sumnmpq\SY$.
	
	We also require a similar expression that allows us to make the same replacement for the products $\dYnm\Ypq$ in the tangential boundary conditions.
	Although we are able to make this replacement owing to separability of the linear equations, lack of orthogonality in $\dYnm$ results in an extremely complicated expression for $\dSY$.
	A finite expression may nevertheless be obtained under the condition that we assume a maximum $n$ in the excitation field $\nmax$ and a maximum $\shapen$ in the boundary shape $\shapemax$.
	For $n>\nmax$, $\Benm=0$ and for $p>\shapemax$, $\gpq=0$, so the sums over these indices truncate.
	
	From \Eqs\ref{eq:dYdef}, \ref{eq:3Yproduct}, and \ref{eq:SYresult}, we obtain
	\begin{align}
		&\int_0^{2\pi}\int_0^{\pi} \dYnm\Ypq\dYnmpConj\sin^3\theta d\theta d\phi = \SYw{}, \label{eq:dSYdef}\\
		&\SYw{} \equiv \mixm\mixmp\SYarg{n-1\,m\shapen\shapem}{n'-1\,m'} +  \mixp\mixpp\SYarg{n+1\,m\shapen\shapem}{n'+1\,m'} -  \mixm\mixpp\SYarg{n-1\,m\shapen\shapem}{n'+1\,m'} -  \mixp\mixmp\SYarg{n+1\,m\shapen\shapem}{n'-1\,m'},
	\end{align}
	where $\mix^\pm_{nm}$ are defined in \Eq\ref{eq:wdef}.
	We have defined $\SYw{}$ for compactness of notation; $n'$ is the only index which varies for this quantity in subsequent expressions, although all other indices $n,m,\shapen,\shapem,m'$ are implied and required to evaluate it.
	This quantity roughly represents the amount of overlap between the product $\dYnm\Ypq$ and the desired harmonic $\dYnmp$.
	
	To next express $\dSY$ compactly, we must also define the following quantities, which are continued fractions:
	\begin{align}
		\Fcont^\pm_{n,\Fsub} \equiv \frac{\mix^\pm_{n\pm\Fsub}\mix^\mp_{n\pm(\Fsub+2)}}{\left(\mix^-_{n\pm(\Fsub+2)}\right)^2 + \left(\mix^+_{n\pm(\Fsub+2)}\right)^2 - \mix^\pm_{n\pm(\Fsub+2)}\mix^\mp_{n\pm(\Fsub+4)}\Fcont^\pm_{n,\Fsub+2}},
	\end{align}
	where $\Fsub=0,2,4,\dots$ and a subscript $m$ is implied on all $\mix$.
	The recursion in $\Fcontnk^-$ continues until $\Fsub$ exceeds $n-|m|-2$ and the recursion in $\Fcontnk^+$ continues until $\Fsub$ exceeds $\nmax+\shapemax-n$.
	When $\Fsub$ exceeds these respective values, $\Fcontnk^\pm=0$.
	$\Fcontnk^-$ is also zero for $n<3$ because there are no $\dYnmp$ below $n=2$ that can mix into $n=2$, and the same is true for $n=1$.
	These bounding values result from non-orthogonality of the $\dYnm$.
	For a term of degree $n''$ below the considered value $n'$, non-orthogonality is possible only for $n'' = n'-2$, $n'-4$, {\etc}, but $\dYnmpp$ is defined and non-zero only for $n'' > 0$ and $n'' \ge m''$.
	This constraint limits the non-zero values of $\Fcont^-_{n',0}$ to $n' \ge 3$ and $n' \ge m'+2$.
	The $n''$ above the considered $n'$ that yield non-orthogonal terms are similarly limited by the selection rules for $\SY$, {\ie}\ $n''$ cannot exceed $\nmax+\shapemax$ because no combination of $\dYnm\Ypq \rightarrow \dSY\dYnmp$ results in $n' > \nmax+\shapemax$ and we have assumed the series in $n'$ may be truncated at $n'_\mathrm{max}=\nmax+\shapemax$.
	
	We obtain solvable linear equations by multiplying both sides of the tangential boundary condition equations (\Eqs\ref{eq:bcintt} and \ref{eq:bct}) by $\dYnmpConj\snsq$, then integrating over a unit sphere.
	Solving the resulting equations for the terms initially proportional to $\dYnmp$, we finally obtain the replacement rule $\dYnm\Ypq \rightarrow \sumnmpq\dSY$, with
	\begin{align}
		&\dSY = \frac{ \SYw{} + \sum_{\evind=1}^{\dg^-_\mathrm{max}}\SYw{-2\evind}\prod_{\kv=0}^{\evind-1}\Fcont_{n',2\kv}^- + \sum_{\evind=1}^{\dg^+_\mathrm{max}}\SYw{+2\evind}\prod_{\kv=0}^{\evind-1}\Fcont_{n',2\kv}^+ }{ \left(\mix^-_{n'}\right)^2 +  \left(\mix^+_{n'}\right)^2 - \mix^-_{n'}\mix^+_{n'-2}\Fcont_{n',0}^-  - \mix^+_{n'}\mix^-_{n'+2}\Fcont_{n',0}^+ },\\
		&\dg^-_\mathrm{max} \equiv \begin{cases} \fl\left(\frac{1}{2}\left(n'-|m'|\right) \right) &\text{if }n' \ge 3\\
			0 &\text{otherwise}\end{cases}, \quad 
		\dg^+_\mathrm{max} \equiv \fl\left(\frac{1}{2}\left(\nmax+\shapemax-n'\right)\right),
	\end{align}
	where $\fl(a)$ is the nearest integer less than or equal to $a$.
	
	\subsubsection{Consequences of non-orthogonality of $\dYnm$}\label{supp:nonortho}
	In our previous work \citep{styczinski2021induced}, based on several considered cases we assumed that certain linear combinations of the $\dYnm$ could be made mutually orthogonal.
	Although this assumption seems to be true when the shape harmonics $\Ypq$ are limited to $\shapemax=2$, in this work we have endeavored to find a more general solution that may be applied to all $\dYnm$ and for any $\shapemax$.
	The primary consequence of non-orthogonality of the $\dYnm$ is coupling between the boundary condition equations.
	This coupling is why $\dSY$ is not equal to $\SYw{}$, and may be non-zero  for any allowed $n'$ and $m'$ that differ from $(n+\shapen)$ by an even integer.
	The end result is that some harmonics in the near-spherical boundaries can produce induced moments with high degree $n'$.
	For example, for a boundary surface containing harmonics up to degree $\shapemax=8$, as we consider in our Miranda model in \S\ref{sec:resultsMiranda}, the $Y_{21}$ boundary harmonic combines with the $Y_{11}$ excitation harmonic to produce induced moments of degree $n' = 1$, $3$, $5$, $7$, and $9$.
	In this work, though we calculate all of the induced magnetic moments, we only plot the magnetic fields for $n'$ up to $4$, as the induced moments are already negligible at that degree, and the fields from these moments also shrink faster with distance from the body.
	Although the induced fields from the high-degree moments are likely to be small at spacecraft distances, they are nevertheless predicted by our model, a direct result from the non-orthogonality of the $\dYnm$.
	
	\subsection{Boundary conditions with a near-spherical boundary shape}\label{supp:nonsphBCs}
	
	Solving all the boundary conditions with near-spherical boundaries relies on simultaneous Taylor expansions of each boundary.
	For clarity in the end solution, we will now exchange the arbitrary indices $n$ and $n'$, so that $n'$ refers to the ``input'' harmonics pertaining to the excitation field and $n$ refers to the ``output'' harmonics that index the induced magnetic moments.
	We must also now use $\bnm = \anm\bPnm$ for the asymmetric layer coefficients, as a solution cannot be obtained otherwise.
	The procedures described in \S\ref{supp:nonspherical} result in the following boundary conditions for asymmetric bodies:
	\begin{align}
		&B_{r,\mathrm{int}}:\label{eq:bcrintasym}\\[-0.5em]
		&n(n+1)\left(\anml\jnll + \bnml\ynll\right) + \sumnmpq \frac{\bdb\gpq}{\rbavg}\SYrev\, n'(n'+1)\anmlp\left((\jdllp + \bPnlp\ydllp) - 2(\jnllp + \bPnlp\ynllp) \right) =\nonumber\\
		&n(n+1)\left(\anmu\jnul + \bnmu\ynul\right) + \sumnmpq \frac{\bdb\gpq}{\rbavg}\SYrev\, n'(n'+1)\anmup\left((\jdulp + \bPnup\ydulp) - 2(\jnulp + \bPnup\ynulp) \right) \nonumber\\[0.5em]
		&B_{\theta,\mathrm{int}}:\label{eq:bctintasym}\\[-0.5em]
		&\anml\jdll + \bnml\ydll + \sumnmpq\frac{\bdb\gpq}{\rbavg}\dSYrev\, \anmlp\left( (n'(n'+1) - \kl^2\rbavg^2)(\jnllp + \bPnlp\ynllp) - (\jdllp + \bPnlp\ydllp) \right) =\nonumber\\
		&\anmu\jdul + \bnmu\ydul + \sumnmpq\frac{\bdb\gpq}{\rbavg}\dSYrev\, \anmup\left( (n'(n'+1) - \ku^2\rbavg^2)(\jnulp + \bPnup\ynulp) - (\jdulp + \bPnup\ydulp) \right) \nonumber\\[0.5em]
		&B_{r,\mathrm{ext}}:\label{eq:bcrextasym}\\[-0.5em]
		&\frac{n(n+1)}{\roB}\left(\anmo\jno + \bnmo\yno\right) + \sumnmpq \frac{\bdo\gpqo}{\roB}\SYrev\, \frac{n'(n'+1)}{\roB}\anmop\left( (\jdop + \bPnop\ydop) - 2(\jnop + \bPnop\ynop) \right) = \nonumber\\
		&-n\Benm + (n+1)\Binm + \sumnmpq\frac{\bdo\gpqo}{\roB}\SYrev\, \left( -n'(n'-1)\Benmp - (n'+1)(n'+2)\Binmp \right) \nonumber\\[0.5em]
		&B_{\theta,\mathrm{ext}}:\label{eq:bctextasym}\\[-0.5em]
		&\frac{1}{\roB}\left( \anmo\jdo + \bnmo\ydo \right) + \sumnmpq\frac{\bdo\gpqo}{\roB}\dSYrev\, \frac{\anmop}{\roB}\left( (n'(n'+1) - \koB^2\roB^2)(\jnop + \bPnop\ynop) - (\jdop + \bPnop\ydop) \right) = \nonumber\\
		&-\Benm - \Binm + \sumnmpq\frac{\bdo\gpqo}{\roB}\dSYrev\left( -(n'-1)\Benmp + (n'+2)\Binmp \right)\nonumber
	\end{align}
	
	The new terms are all contained within the series in each equation.
	Because these terms are all multiplied by a factor $\bdb/\rb$, we can insert the spherically symmetric solutions for every constant within the series, since we assume terms to second order in $\bdb/\rb$ are negligible.
	Inserting expressions from \Eqs\ref{eq:intsphr} and \ref{eq:ghisph}, we find that every single term within the series in the radial boundary conditions has a matching term on both sides, and so they all cancel.
	In the tangential equations, the only terms that survive are those that multiply an additional factor of $k$ because it is different above and below the boundary. Thus, the boundary condition equations reduce to:
	\begin{align}
		&\anml\jnll + \bnml\ynll = \anmu\jnul + \bnmu\ynul \label{eq:bcrintasymreduced}\\
		&\anml\jdll + \bnml\ydll = \anmu\jdul + \bnmu\ydul + \sumnmpq\frac{\bdb\gpq}{\rbavg}\dSYrev\, \abarl\left(\jnllp + \bPnlp\ynllp\right)\left(\kl^2\rb^2 - \ku^2\rb^2\right) \label{eq:bctintasymreduced}\\
		&\frac{n(n+1)}{\roB}\left(\anmo\jno + \bnmo\yno\right) = -n\Benm + (n+1)\Binm \label{eq:bcrextasymreduced}\\
		&\frac{1}{\roB}\left(\anmo\jdo + \bnmo\ydo\right) = -\Benm - \Binm + \sumnmpq\frac{\bdo\gpqo}{\roB}\dSYrev\, \frac{2n'+1}{n'+1}\Benmp\, \frac{\xiop + \bPnop\rhop}{\delop + \bPno\epsop} \label{eq:bctextasymreduced},
	\end{align}
	with
	\begin{align}
		\xiop &\equiv -(kR)^2\jnkR \label{eq:xidef}\\
		\rhop &\equiv -(kR)^2\ynkR \label{eq:rhodef},
	\end{align}
	and the bar over $\abarl$ to indicate that it is identically the solution from the spherically symmetric case.
	The symmetric boundary condition equations give us
	\begin{equation}
		\abari \equiv -\frac{2n+1}{n+1}\,\frac{R\Benm}{\delo + \bPno\epso} \prod_{\jl=\il+1}^{\Nl}\frac{\jn^{\jl,\jl-1} + \bPnj\yn^{\jl,\jl-1}}{\jn^{\jl-1,\jl-1} + \bPnjM\yn^{\jl-1,\jl-1}} \label{eq:anmsph}.
	\end{equation}
	
	With these expressions, and recalling that $\bconst^1_{nm}=0$ for all $n$ and $m$, \Eqs\ref{eq:bcrintasymreduced}--\ref{eq:bctextasymreduced} are at last a solvable linear system of equations.
	Ultimately, for $\Binm$ we obtain a solution
	\begin{equation}
		\Binm = \frac{n}{n+1}\Ampne\Benm + n\sum_{\il=1}^{\Nl}\Ampnti\Kni\Dnmi, \label{eq:Binm}
	\end{equation}
	where we have defined
	\begin{align}
		\Ampnti &\equiv \frac{\jnii + \bPni\ynii}{\delo + \bPno\epso}, \label{eq:ampt}\\
		\Kni &\equiv \prod_{\jl=\il+1}^\Nl \frac{\alpjj}{\betaj + \bPnjM\gammj}, \qquad \Kno = 1, \label{eq:Kni}\\
		\Dnmi &\equiv
		\begin{dcases}
			\sumnmpq\frac{\bdi\gpqi}{\rbi}\dSYrev\, \frac{\abari}{\roB}\left(\jniip + \bPnip\yniip\right)\left(\ki^2\rbi^2 - \kiP^2\rbi^2\right) & \mathrm{for}~\il<\Nl, \\
			\sumnmpq\frac{\bdo\gpqo}{\roB}\dSYrev\, \frac{2n'+1}{n'+1}\Ampnpd\Benmp & \mathrm{for}~\il=\Nl,
		\end{dcases} \label{eq:Dnmi}\\
		\Ampnpd &\equiv \frac{\xiop + \bPnop\rhop}{\delop + \bPnop\epsop} \label{eq:ampd}.
	\end{align}
	As with $\Ampne$, the product $\Ampnto\Ampnpd$ is asymptotic to $(1+0i)$ in the limit $|\koB\roB|\rightarrow\infty$ for all $n$ and $n'$.
	Each of these complex amplitude quantities is labeled with a superscript to indicate their relationship to other relevant quantities: $\Amp^e$ multiplies the excitation field, $\Amp^t$ multiplies the tangential first-order term in the non-spherical expansion, and $\Amp^\star$ multiplies the mixing coefficients $\SYchar^\star$.
	The radial first-order term that would multiply a quantity $\Amp^r$ analogous to $\Amp^t$ is identically zero, so it does not appear.
	\Eq\ref{eq:Binm} is our final result, and may be used to evaluate the induced magnetic field for any arbitrary layered conducting body, so long as the near-spherical approximation holds for each boundary.

	\section{Table of interior structure parameters}\label{supp:ppTable}
	
	In Table \ref{table:suppPlanetProps}, we list the chosen bulk properties for the four satellites investigated here.
	The base PlanetProfile models assume spherical symmetry.
	The entire set of assumed parameters can be found in the Matlab files for the individual models, which can also be used to reproduce the models themselves.
	
	\begin{table}[h]
		\centering
		\begin{tabular}{llllll}
			Body    & $T_\mathrm{s}~(\si{K})$  & $R~(\si{km})$  & $M~(\si{kg})$ & $C/MR^2$  & Reference \\
			\hline
			Europa  & $110$   & $1561.0$    & $4.7991\times10^{22}$ & $0.346\pm0.005$   & \citet{anderson1998europa}\\
			Callisto    & $110$   & $2410.3$    & $1.4819\times10^{23}$ & $0.3549\pm0.0042$   & \citet{anderson2001shape} \\
			Miranda & $60$    & $235.8$ & $6.4\times10^{19}$    & $0.346$ & \citet{hussmann2006subsurface} \\
			Triton  & $38$    & $1353.4$    & $2.14\times10^{22}$   & $0.315$ & \citet{hussmann2006subsurface} \\
			\hline
		\end{tabular}
		\caption{
			Model parameters used to determine interior conductivity profiles using \textit{PlanetProfile.}
			$T_s$: surface temperature; $R$: radius of body surface; $M$: total body mass; $C/MR^2$: axial coefficient of moment of inertia.
			Moments of inertia for Miranda and Triton are assumed for consistency with prior models, as no measurements are available.
			For Triton the value was increased by 0.005 for consistency with the model study of \citet{cochrane2021search}.
		}\label{table:suppPlanetProps}
	\end{table}

	\section{Gravitational deformation in satellites; application to Europa}\label{supp:gravity}
	Tidal forces applied by gravity from the parent planet and centrifugal acceleration from spin rotation will deform satellites, primarily in the $\shapen=2$ spherical harmonic shapes \citep{rambaux2013tides}.
	\citet{anderson1998europa} used Doppler shifts available from precise radio tracking of \textit{Galileo} by the Deep Space Network to infer the $\shapen=2$ gravity coefficients for Europa.
	These authors favored best-fit values of $C_{20} = \num{-435.5e-6}$, $C_{22} = \num{131.0e-6}$ from their analysis.
	Assuming the body is in hydrostatic equilibrium, these authors also found the axial moment of inertia for Europa to be approximately $C/MR^2 = 0.346$.
	
	Under the assumption of hydrostatic equilibrium, the Radau--Darwin approximation allows us to relate the axial moment of inertia to the secular (non-time-varying) Love number $k_f$ \citep{rambaux2013tides}:
	\begin{equation}
		\frac{C}{MR^2} = \frac{2}{3}\left[ 1 - \frac{2}{5}\sqrt{\frac{4 - k_f}{1 + k_f}} \right] \label{eq:radau}.
	\end{equation}
	The fluid Love number $h_f = k_f + 1$.
	This quantity relates the gravitational tides to the geometric deformation of the body by \citep{rambaux2013tides,hemingway2019enceladus}
	\begin{equation}
		r_\mathrm{surf}(\theta,\phi) = h_f \frac{V(\theta,\phi)}{g_\mathrm{surf}} \label{eq:gravityShape},
	\end{equation}
	where $r_\mathrm{surf}$ is the shape of the body surface, $V$ is the gravitational potential, and $g_\mathrm{surf} = GM/R^2$ is the mean gravitational acceleration at the surface.
	At the surface of the body, the gravitational moments $V_{\shapen\shapem}$ are proportional to the coefficients $C_{\shapen\shapem}$ with a proportionality factor of $GM/R$ \citep{rambaux2013tides}.
	Inserting gravitational terms into \Eq\ref{eq:gravityShape}, we arrive at the simple conversion
	\begin{equation}
		H_{\shapen\shapem} = h_f C_{\shapen\shapem} R \label{eq:gravityConvert},
	\end{equation}
	where $H$ are Schmidt semi-normalized spherical harmonic coefficients for the body surface shape---analogous to $\gpqo$ (\Eq\ref{eq:rio}), but in a different normalization that matches the one in which the gravity coefficients are given.
	\Eq\ref{eq:gravityConvert} will allow us to find the shape of the body surface as perturbed by tidal forces.
	
	We can now determine the values to apply to Europa.
	Solving \Eq\ref{eq:radau} for $k_f$ and replacing with $h_f$, we obtain
	\begin{equation}
		h_f = 1 + \frac{4 - u}{1 + u}, \quad u \equiv \left[ \frac{5}{2}\left( 1 - \frac{3}{2}\frac{C}{MR^2} \right) \right]^2 \label{eq:hf},
	\end{equation}
	with $h_f = 2.044$ for Europa from $C/MR^2 = 0.346$ as determined by \citet{anderson1998europa}.
	Combining these results with \Eq\ref{eq:gravityConvert} and the $C_{20}, C_{22}$ gravity coefficients from \citet{anderson1998europa}, we obtain Schmidt semi-normalized shape coefficients of
	\begin{align}
		H_{20} &= -1.390\siu{km} \label{eq:H20},\\
		H_{22} &=  0.418\siu{km} \label{eq:H22}.
	\end{align}
	These values describe the equilibrium shape of the icy surface of Europa as perturbed by tides.
	In order to preserve the ice shell thickness estimates we model after \citet{tobie2003tidally}, we apply the same gravitational shape perturbation to the surface and ocean layers, without scaling.
	
	Although Enceladus also has a triaxial ellipsoid shape perturbed by gravity \citep{iess2014gravity}, we choose not to apply this analysis to our Enceladus-like interior for Miranda because gravity data are not yet available for Miranda.
	Thus, we limit our Miranda models to the asymmetry in the ice--ocean boundary.

	\section{Supplemental figures for example applications}\label{supp:fullResults}
\begin{figure}
		\centering
		\begin{subfigure}[b]{0.495\textwidth}
			\caption{Europa Seawater ice shell thickness ($\mathrm{km}$), $\D=22\siu{km}$
			}
			\centering
			\includegraphics[width=\textwidth]{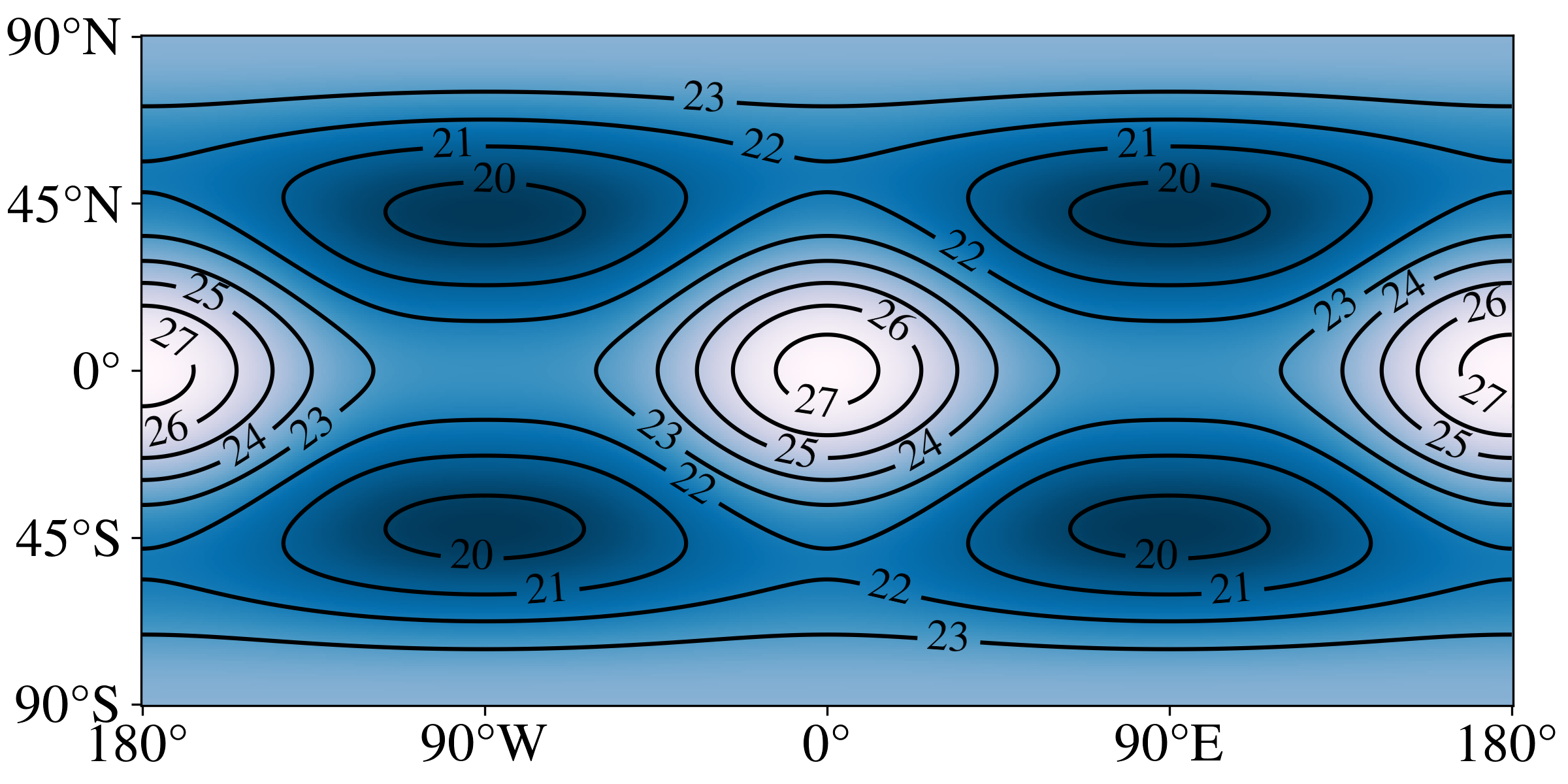}
			\label{fig:suppEuropaHighContour}
		\end{subfigure}
		\hfill
		\begin{subfigure}[b]{0.495\textwidth}
			\caption{Asymmetric model $|\Bind|$ ($\mathrm{nT}$) at $r=1.02R_E$
			}
			\centering
			\includegraphics[width=\textwidth]{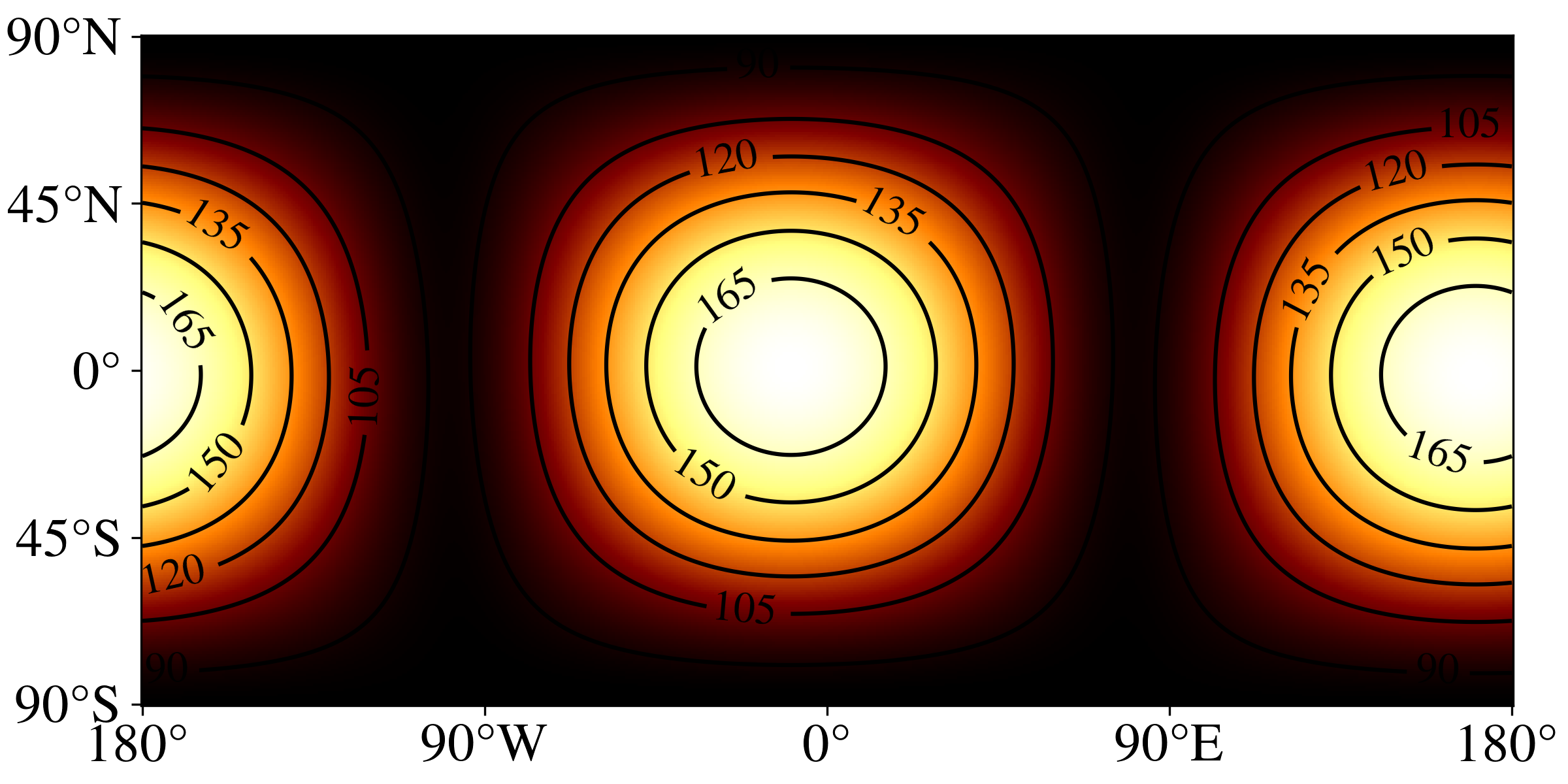}
			\label{fig:suppEuropaHighMag}
		\end{subfigure}\\
		\begin{subfigure}[b]{0.495\textwidth}
			\caption{$B_x$ difference ($\mathrm{nT}$) vs.~symmetric at $r=1.02R_E$
			}
			\centering
			\includegraphics[width=\textwidth]{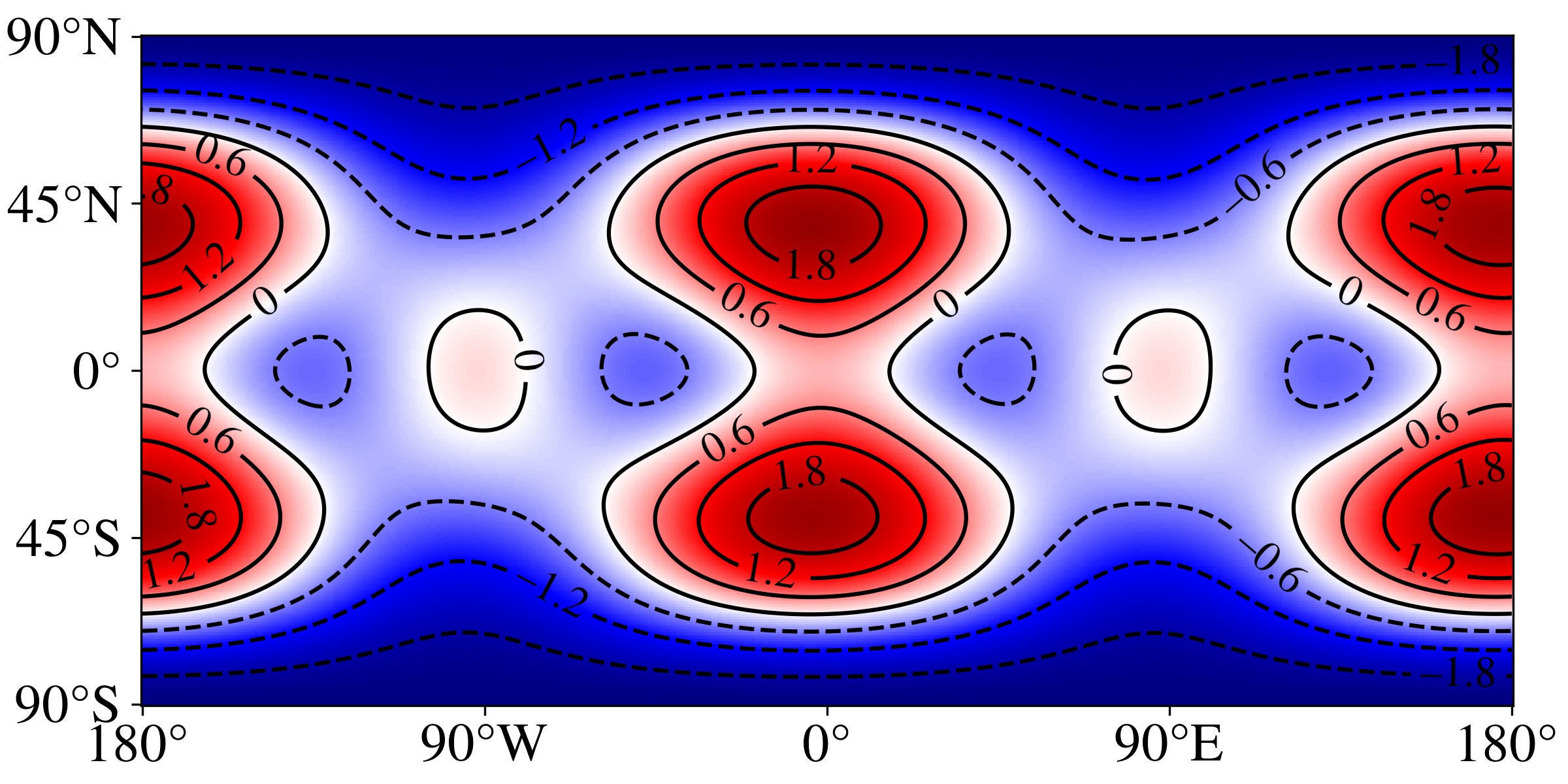}
			\label{fig:suppEuropaHighXdiff}
		\end{subfigure}
		\hfill
		\begin{subfigure}[b]{0.495\textwidth}
			\caption{$|\Bind|$ difference ($\mathrm{nT}$) vs.~symmetric at $r=1.02R_E$
			}
			\centering
			\includegraphics[width=\textwidth]{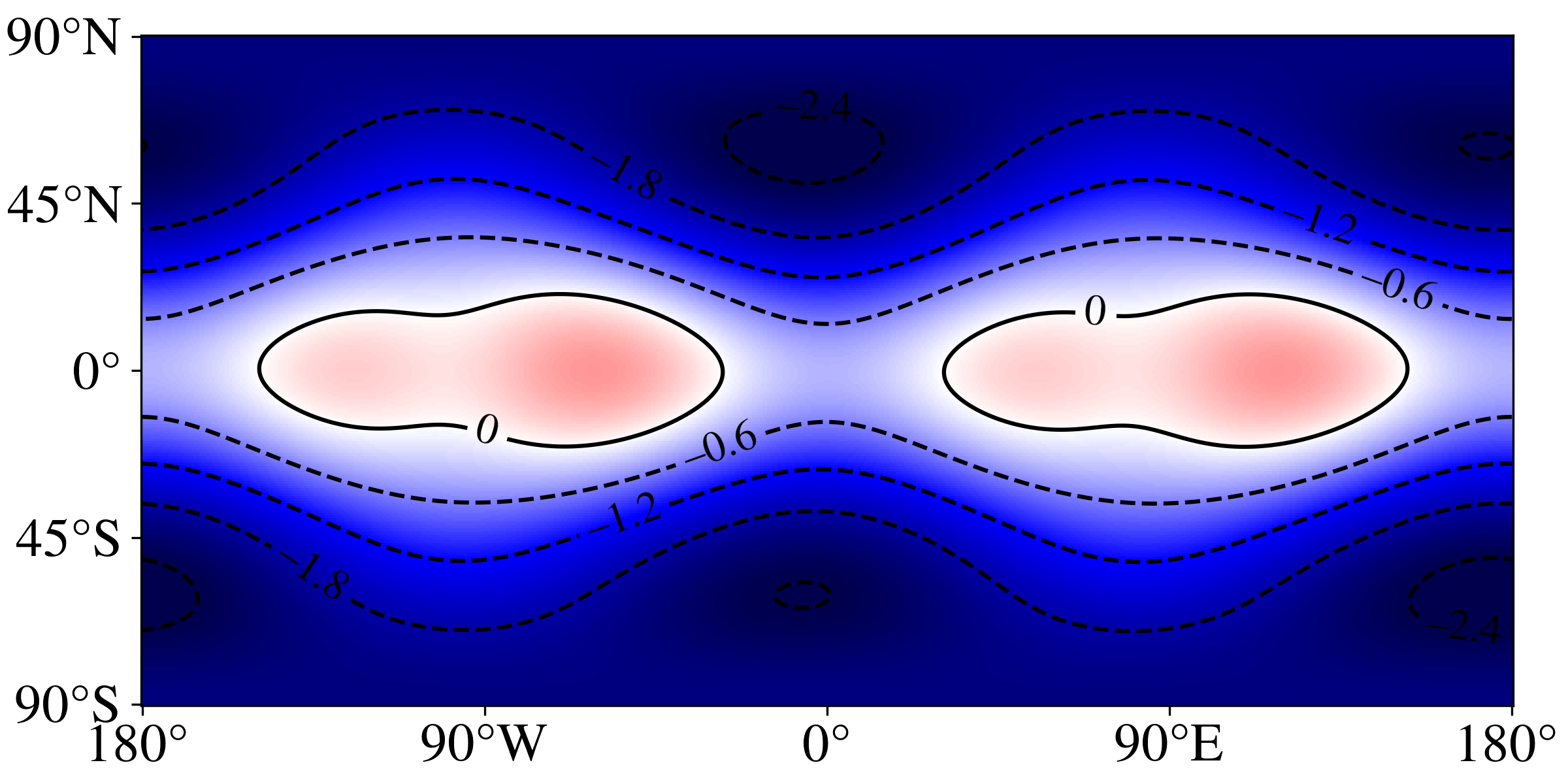}
			\label{fig:suppEuropaHighMagDiff}
		\end{subfigure}\\
		\begin{subfigure}[b]{0.495\textwidth}
			\caption{$B_y$ difference ($\mathrm{nT}$) vs.~symmetric at $r=1.02R_E$
			}
			\centering
			\includegraphics[width=\textwidth]{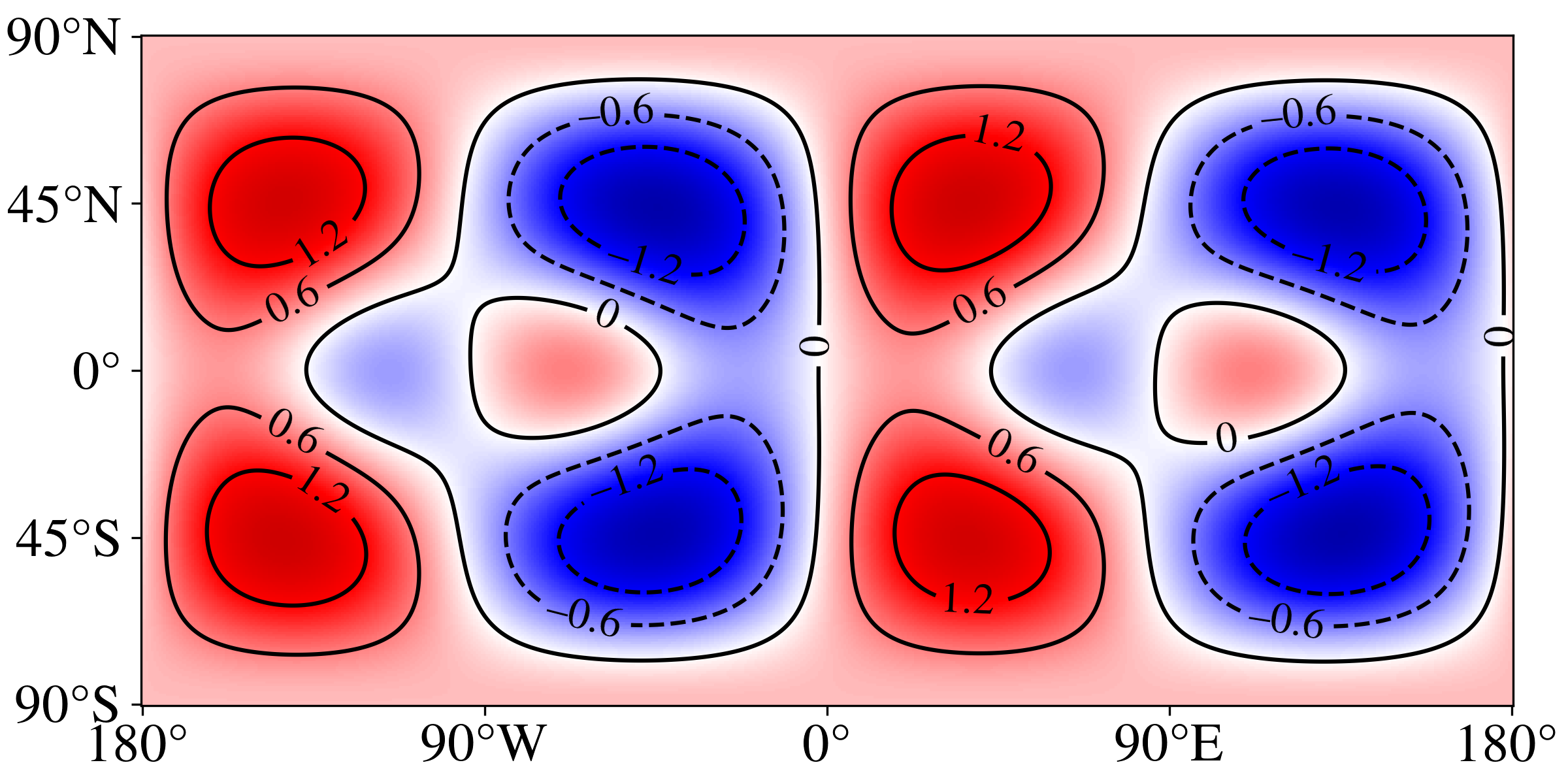}
			\label{fig:suppEuropaHighYdiff}
		\end{subfigure}
		\hfill
		\begin{subfigure}[b]{0.495\textwidth}
			\caption{$B_z$ difference ($\mathrm{nT}$) vs.~symmetric at $r=1.02R_E$
			}
			\centering
			\includegraphics[width=\textwidth]{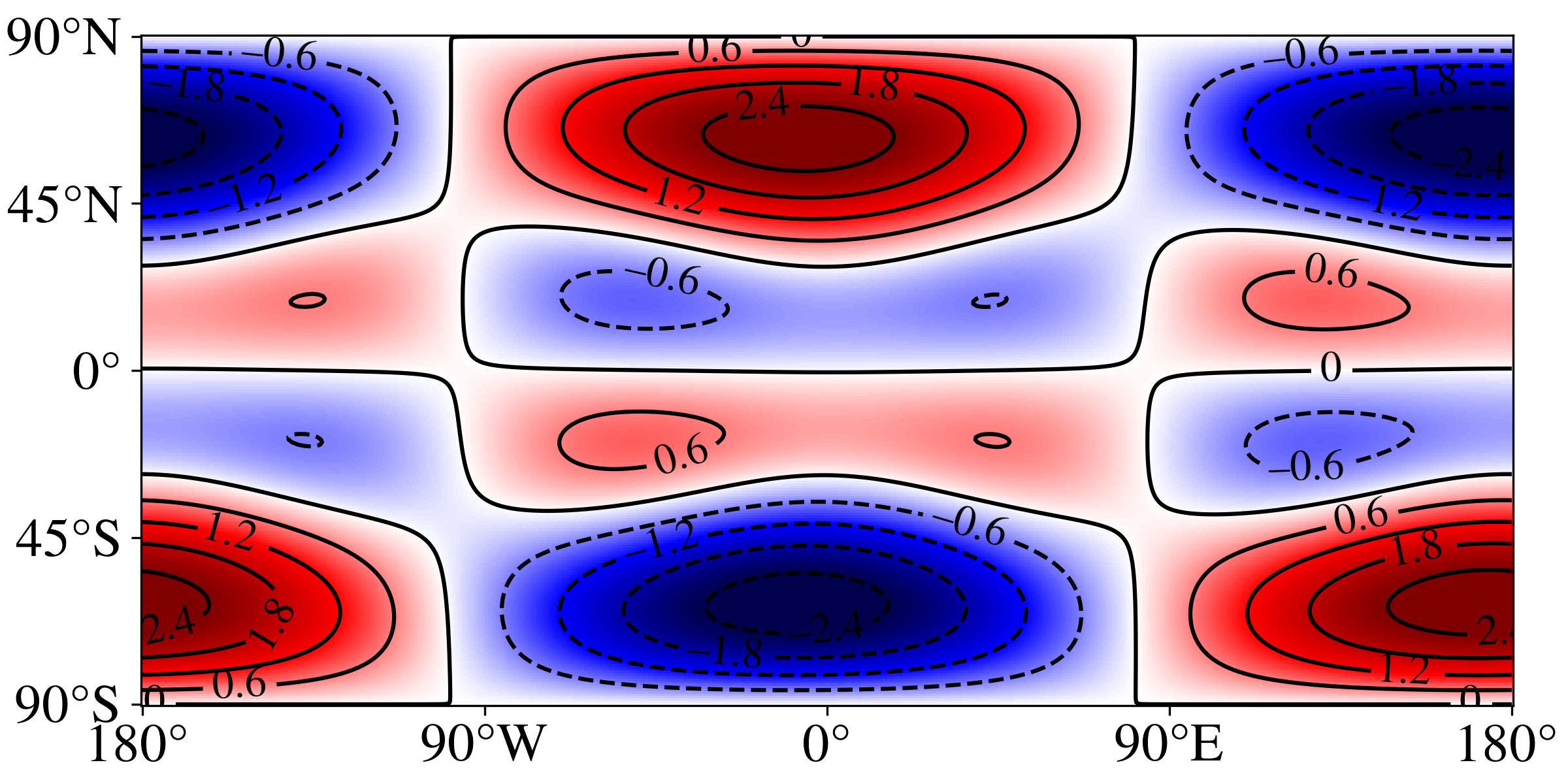}
			\label{fig:suppEuropaHighZdiff}
		\end{subfigure}
		\caption{
			Europa model with an asymmetric ice--ocean boundary approximating the results of \citet{tobie2003tidally}, whose analysis was based on modeling tidal heating and thermodynamics, and static gravity inferred by \citet{anderson1998europa}.
			Compare \F\ref{fig:suppEuropaHighContour} to \F12a of \citet{tobie2003tidally}.
			Average ice shell thickness is $22.5\siu{km}$.
			In this model, a Seawater composition is assumed for the ocean; conduction in the ionosphere is ignored.
			Electrical conductivities are calculated using the \textit{PlanetProfile} geophysical modeling framework \citep{vance2021magnetic}.
			Magnetic fields are evaluated at the J2000 epoch and at $25\siu{km}$ altitude, as the upcoming \textit{Europa Clipper} mission plans several flybys of $25\siu{km}$ or less at closest approach.
			Only the synodic period is modeled here for simplicity.
			The difference in the magnetic field resulting from asymmetry is over $2\siu{nT}$ in some locations, and is likely to have a measurable influence on \textit{Europa Clipper} investigations using data from these near flybys.
			The differences in induced field are not static but move and oscillate throughout the synodic period.
			Animations for the difference in $x$ component and magnitude are included as Supplemental Material.
		}
		\label{fig:suppEuropaHigh}
	\end{figure}

\begin{figure}
		\centering
		\begin{subfigure}[b]{0.495\textwidth}
			\caption{Europa 10\% Seawater ice shell thickness ($\mathrm{km}$), $\D=22\siu{km}$
			}
			\centering
			\includegraphics[width=\textwidth]{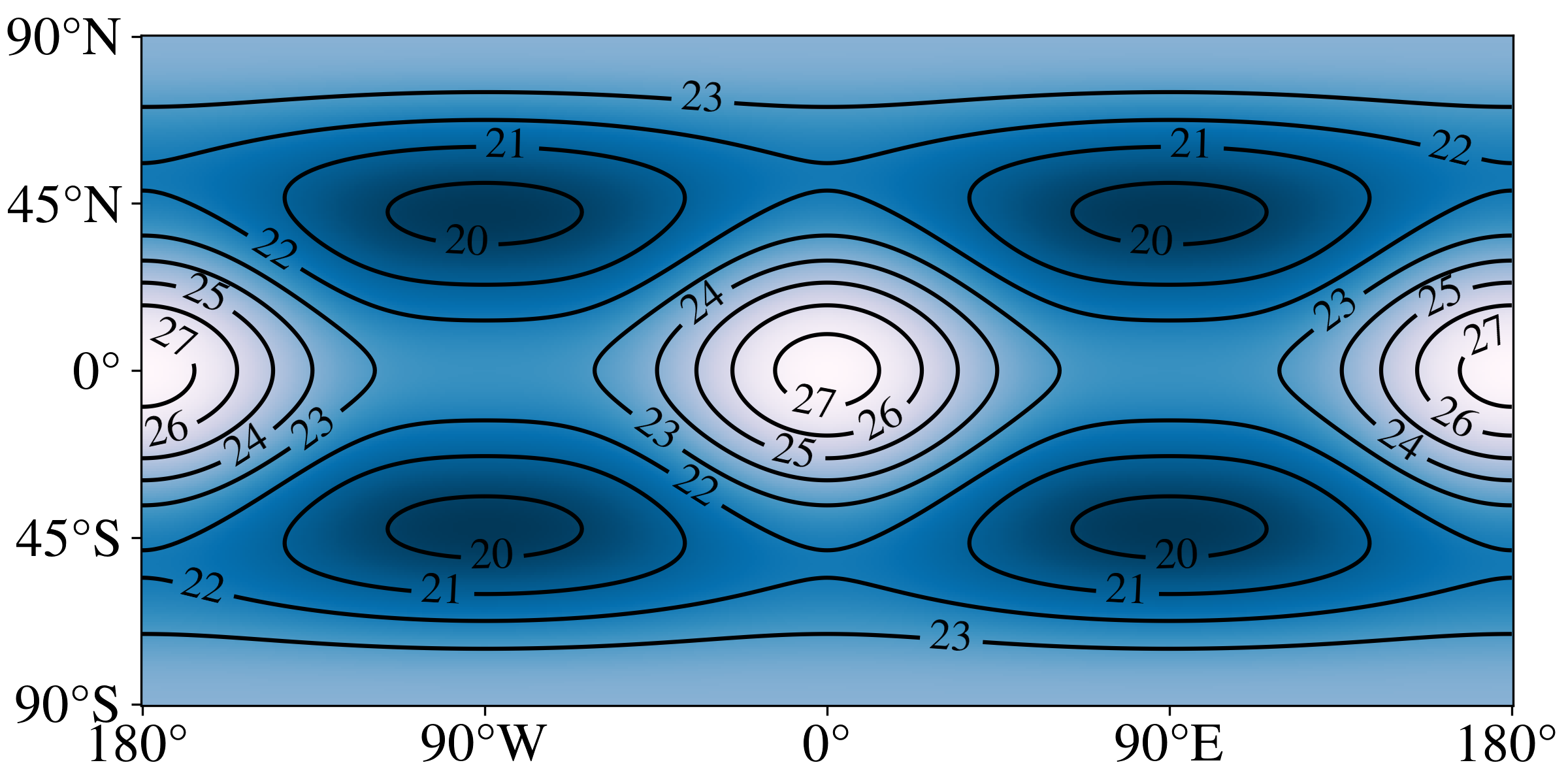}
			\label{fig:suppEuropaLowContour}
		\end{subfigure}
		\hfill
		\begin{subfigure}[b]{0.495\textwidth}
			\caption{Asymmetric model $|\Bind|$ ($\mathrm{nT}$) at $r=1.02R_E$
			}
			\centering
			\includegraphics[width=\textwidth]{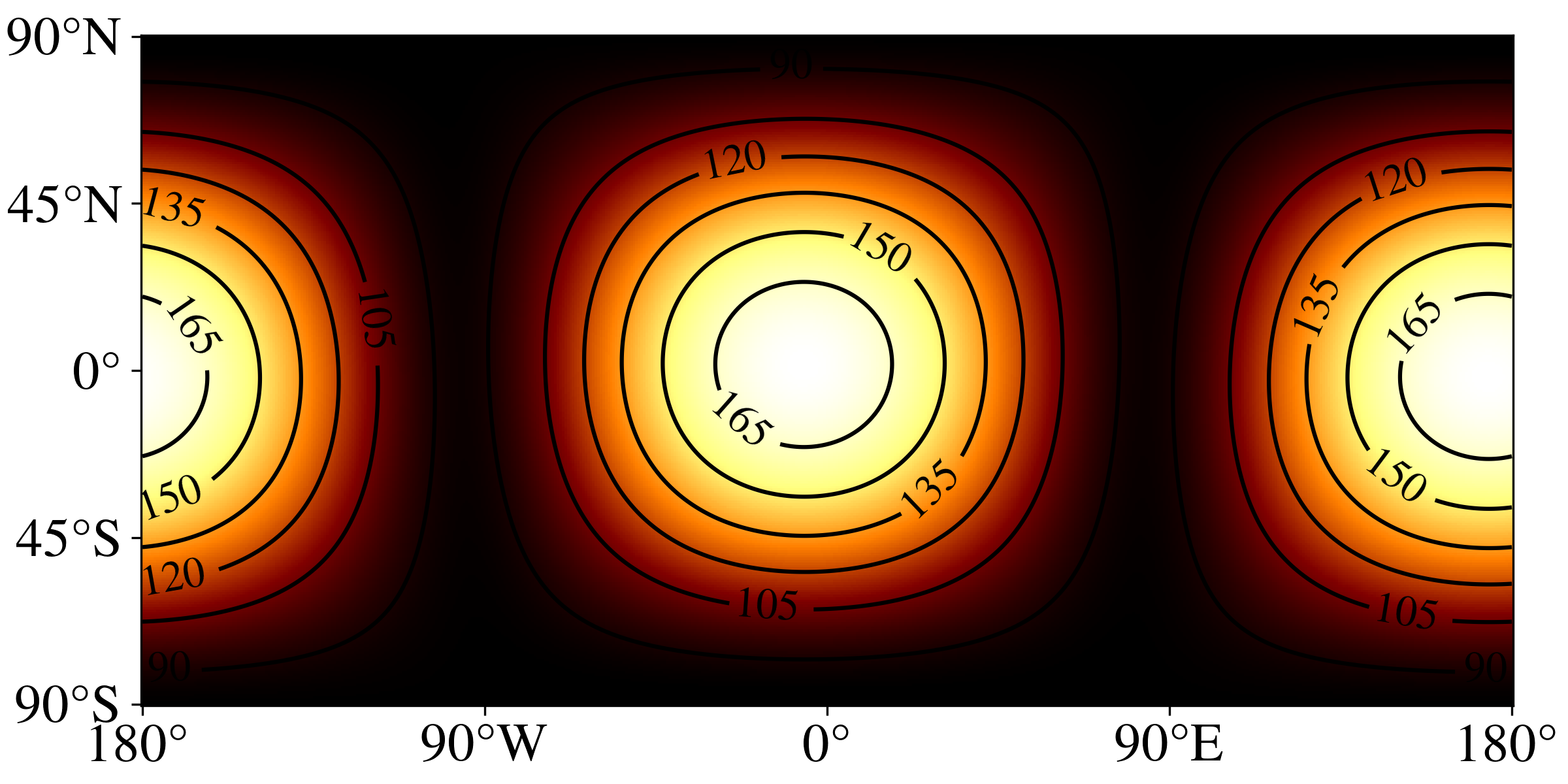}
			\label{fig:suppEuropaLowMag}
		\end{subfigure}\\
		\begin{subfigure}[b]{0.495\textwidth}
			\caption{$B_x$ difference ($\mathrm{nT}$) vs.~symmetric at $r=1.02R_E$
			}
			\centering
			\includegraphics[width=\textwidth]{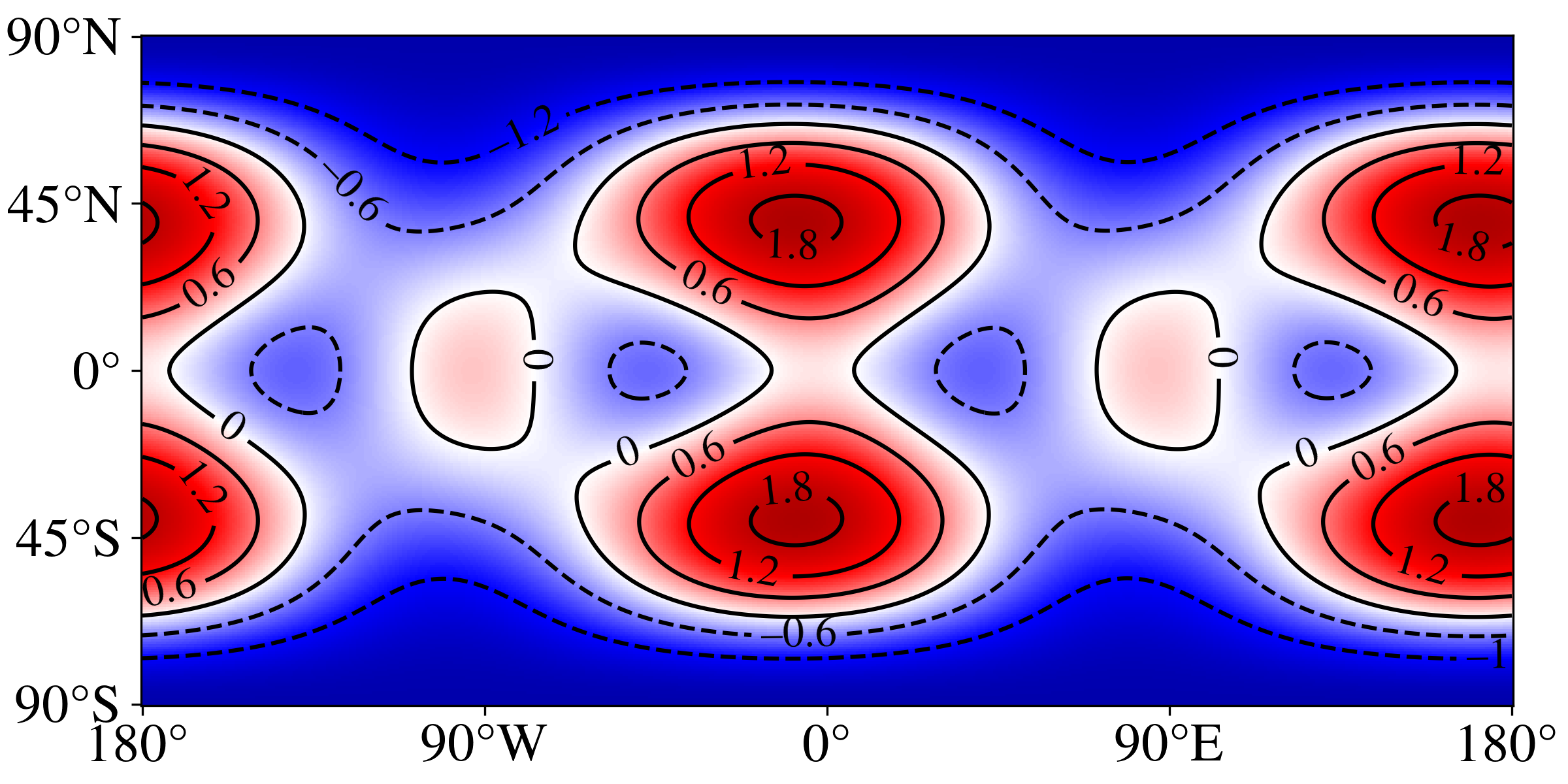}
			\label{fig:suppEuropaLowXdiff}
		\end{subfigure}
		\hfill
		\begin{subfigure}[b]{0.495\textwidth}
			\caption{$|\Bind|$ difference ($\mathrm{nT}$) vs.~symmetric at $r=1.02R_E$
			}
			\centering
			\includegraphics[width=\textwidth]{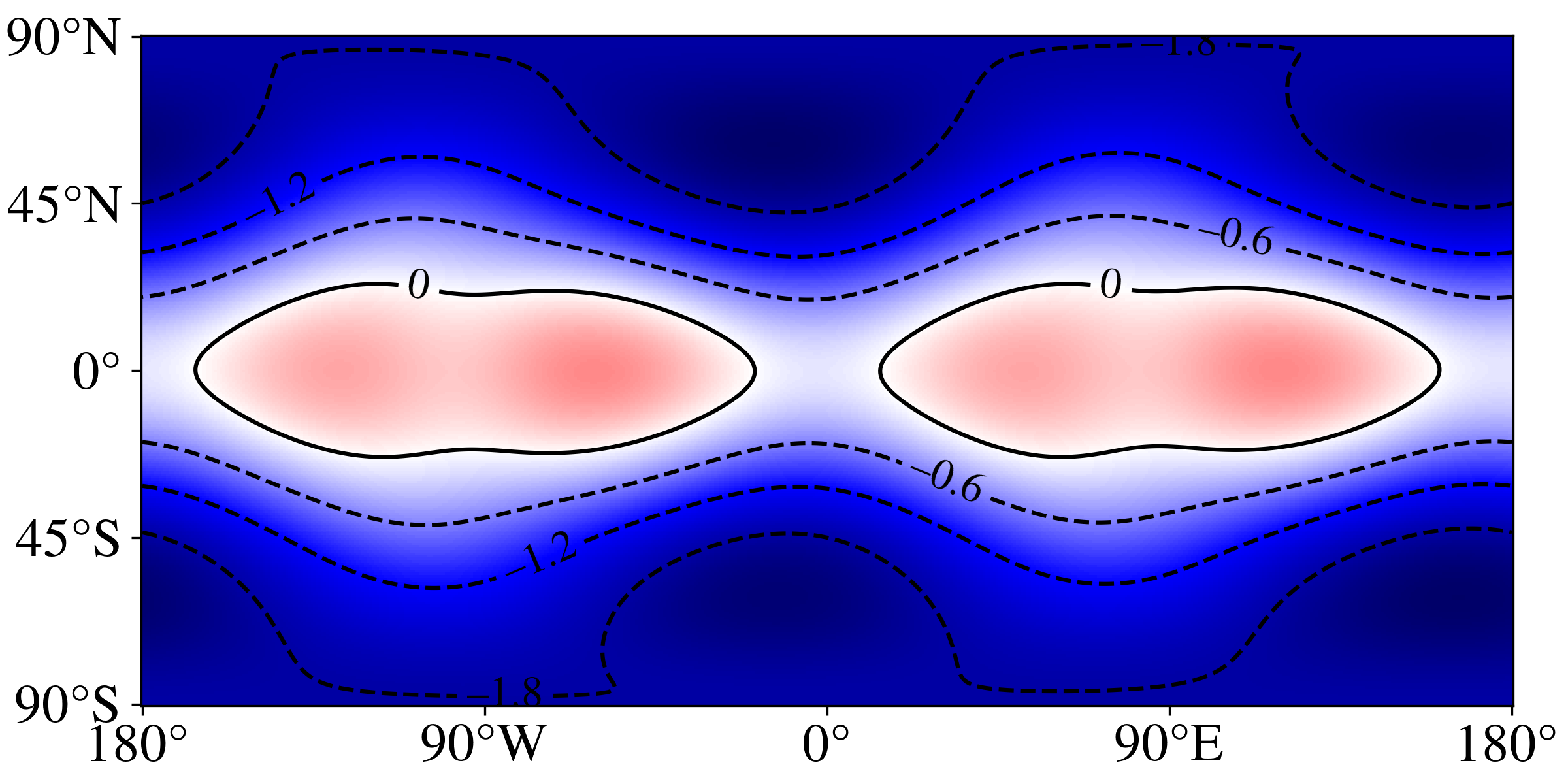}
			\label{fig:suppEuropaLowMagDiff}
		\end{subfigure}\\
		\begin{subfigure}[b]{0.495\textwidth}
			\caption{$B_y$ difference ($\mathrm{nT}$) vs.~symmetric at $r=1.02R_E$
			}
			\centering
			\includegraphics[width=\textwidth]{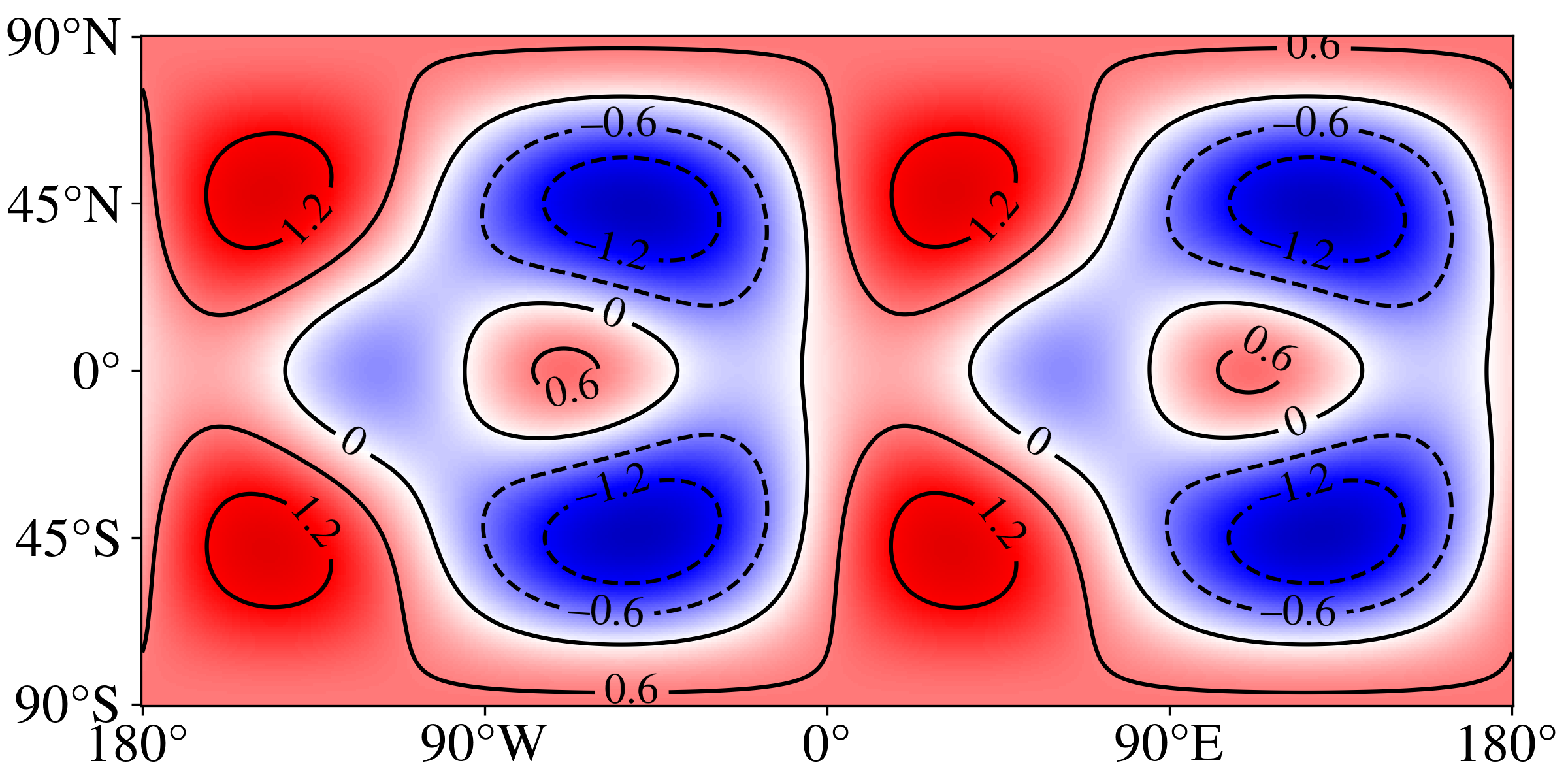}
			\label{fig:suppEuropaLowYdiff}
		\end{subfigure}
		\hfill
		\begin{subfigure}[b]{0.495\textwidth}
			\caption{$B_z$ difference ($\mathrm{nT}$) vs.~symmetric at $r=1.02R_E$
			}
			\centering
			\includegraphics[width=\textwidth]{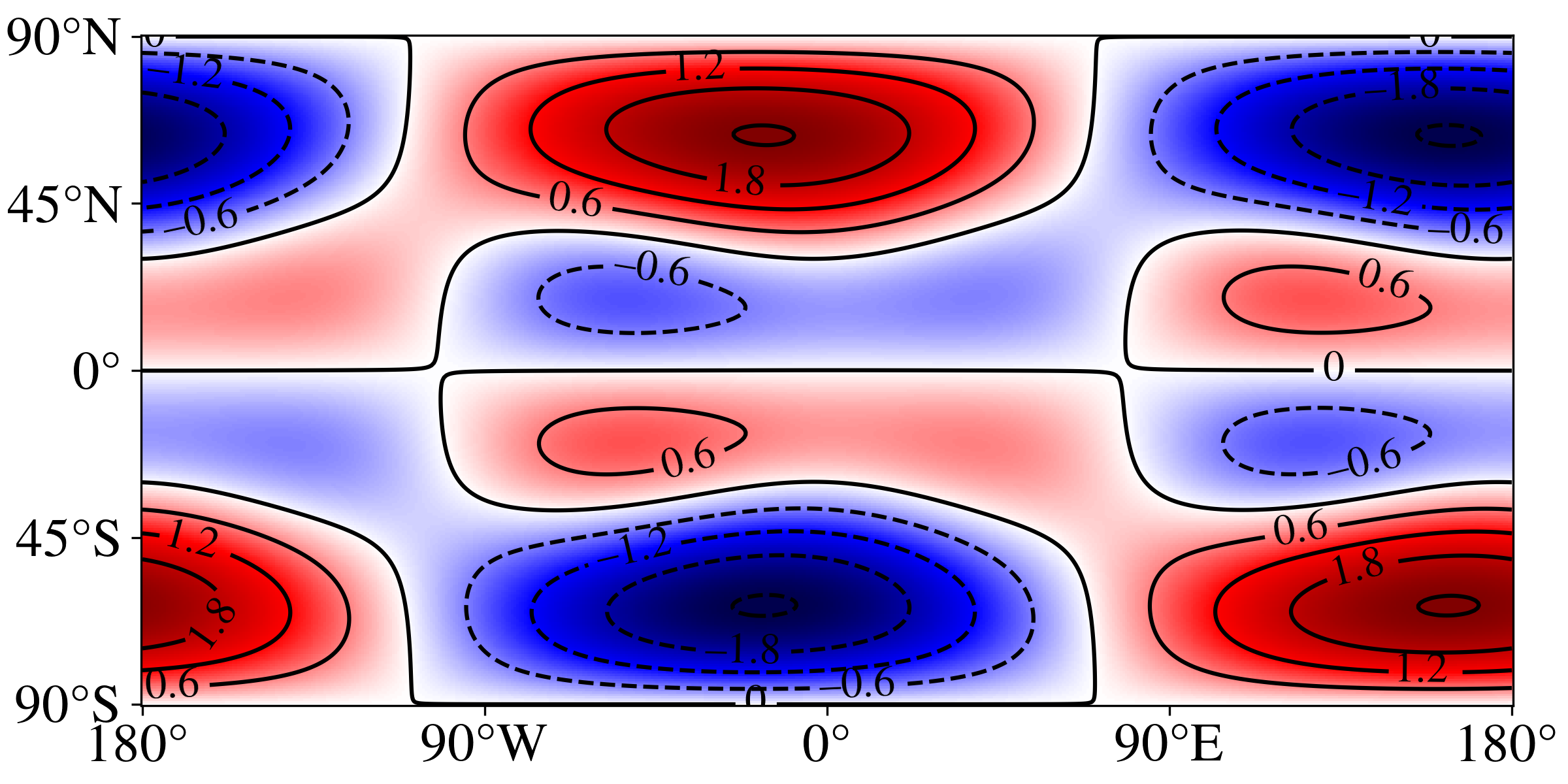}
			\label{fig:suppEuropaLowZdiff}
		\end{subfigure}
		\caption{
			Europa model very similar to that in \F\ref{fig:suppEuropaHigh} but with an ocean composition $1/10$ the salinity of Seawater.
			Magnetic fields are again evaluated at the J2000 epoch and at $25\siu{km}$ altitude.
			Comparison to \F\ref{fig:suppEuropaHigh} shows smaller differences and a slight phase difference.
			However, the differences in the magnetic field resulting from asymmetry are still well over $1\siu{nT}$ in some places even in this lower-salinity case, a consequence of Europa's relatively large size.
			The colormaps and contours for the difference plots are fixed to match those of \F\ref{fig:suppEuropaHigh}.
			Animations for the difference in $x$ component and magnitude as they vary throughout the synodic period are included as Supplemental Material.
		}
		\label{fig:suppEuropaLow}
	\end{figure}

\begin{figure}
		\centering
		\begin{subfigure}[b]{0.495\textwidth}
			\caption{Europa comparison ice shell thickness ($\mathrm{km}$), $\D=22\siu{km}$
			}
			\centering
			\includegraphics[width=\textwidth]{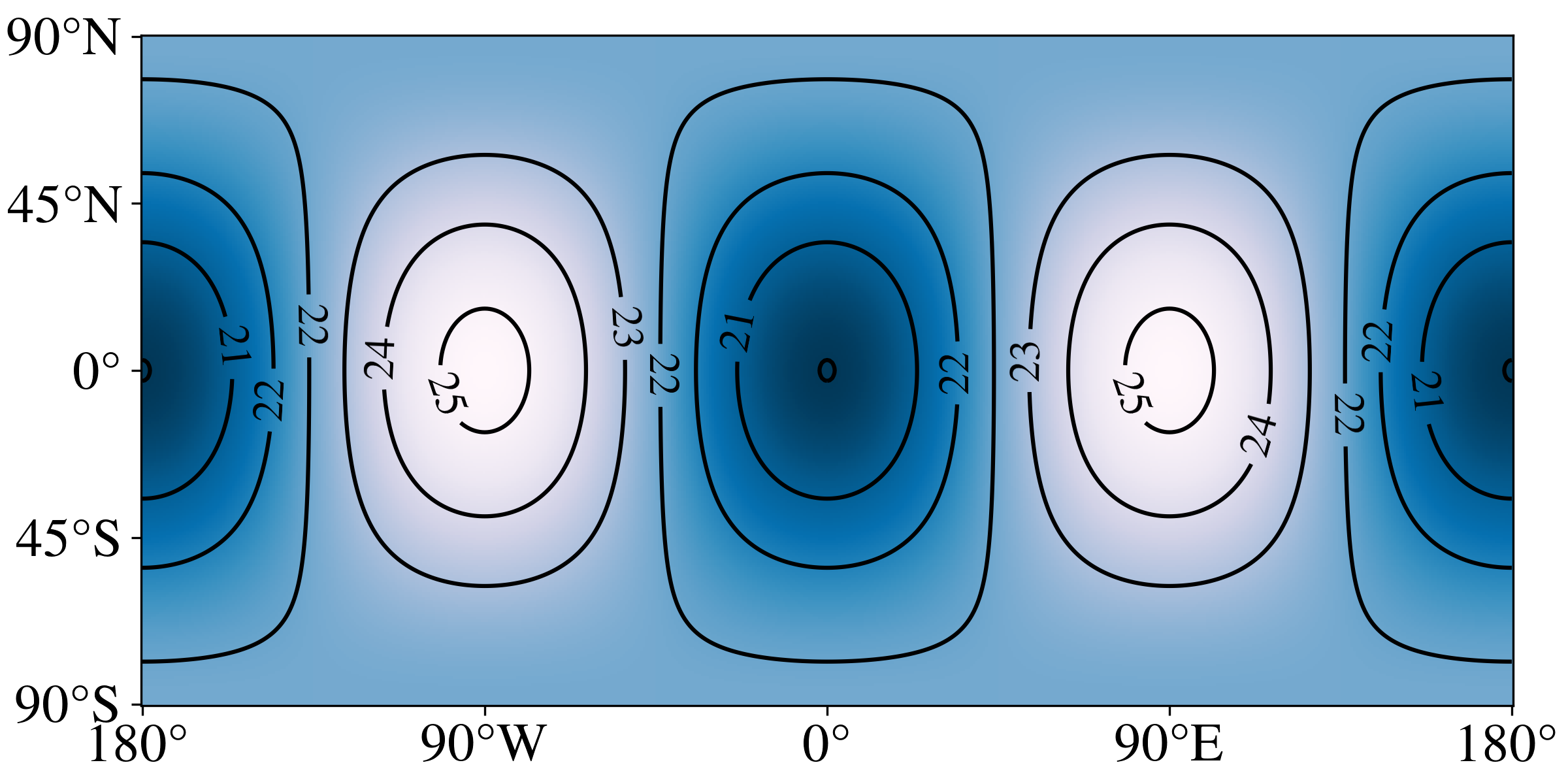}
			\label{fig:suppEuropaPrevContour}
		\end{subfigure}
		\hfill
		\begin{subfigure}[b]{0.495\textwidth}
			\caption{Asymmetric model $|\Bind|$ ($\mathrm{nT}$) at $r=R_E$
			}
			\centering
			\includegraphics[width=\textwidth]{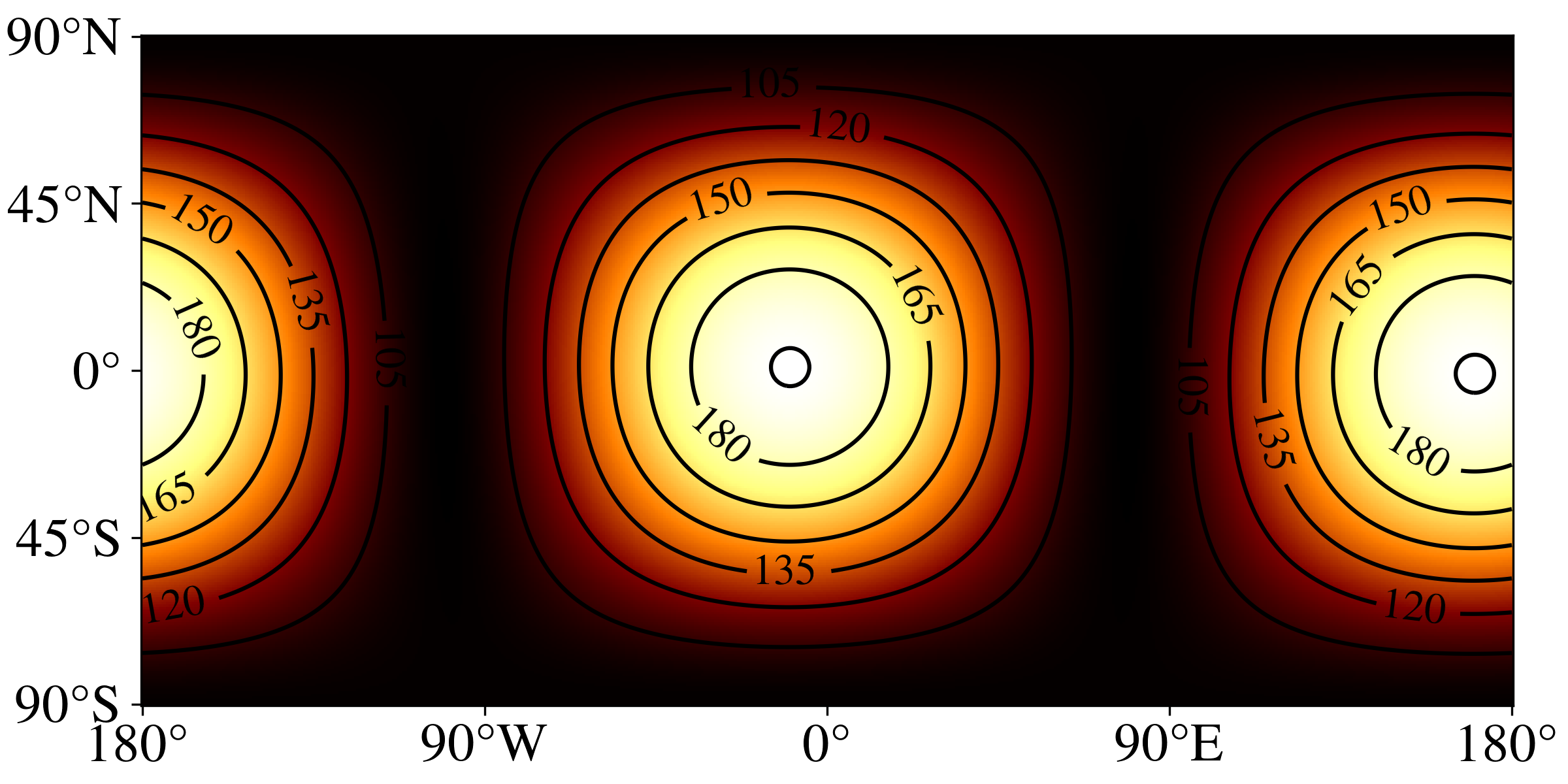}
			\label{fig:suppEuropaPrevMag}
		\end{subfigure}\\
		\begin{subfigure}[b]{0.495\textwidth}
			\caption{$B_x$ difference ($\mathrm{nT}$) vs.~symmetric at $r=R_E$
			}
			\centering
			\includegraphics[width=\textwidth]{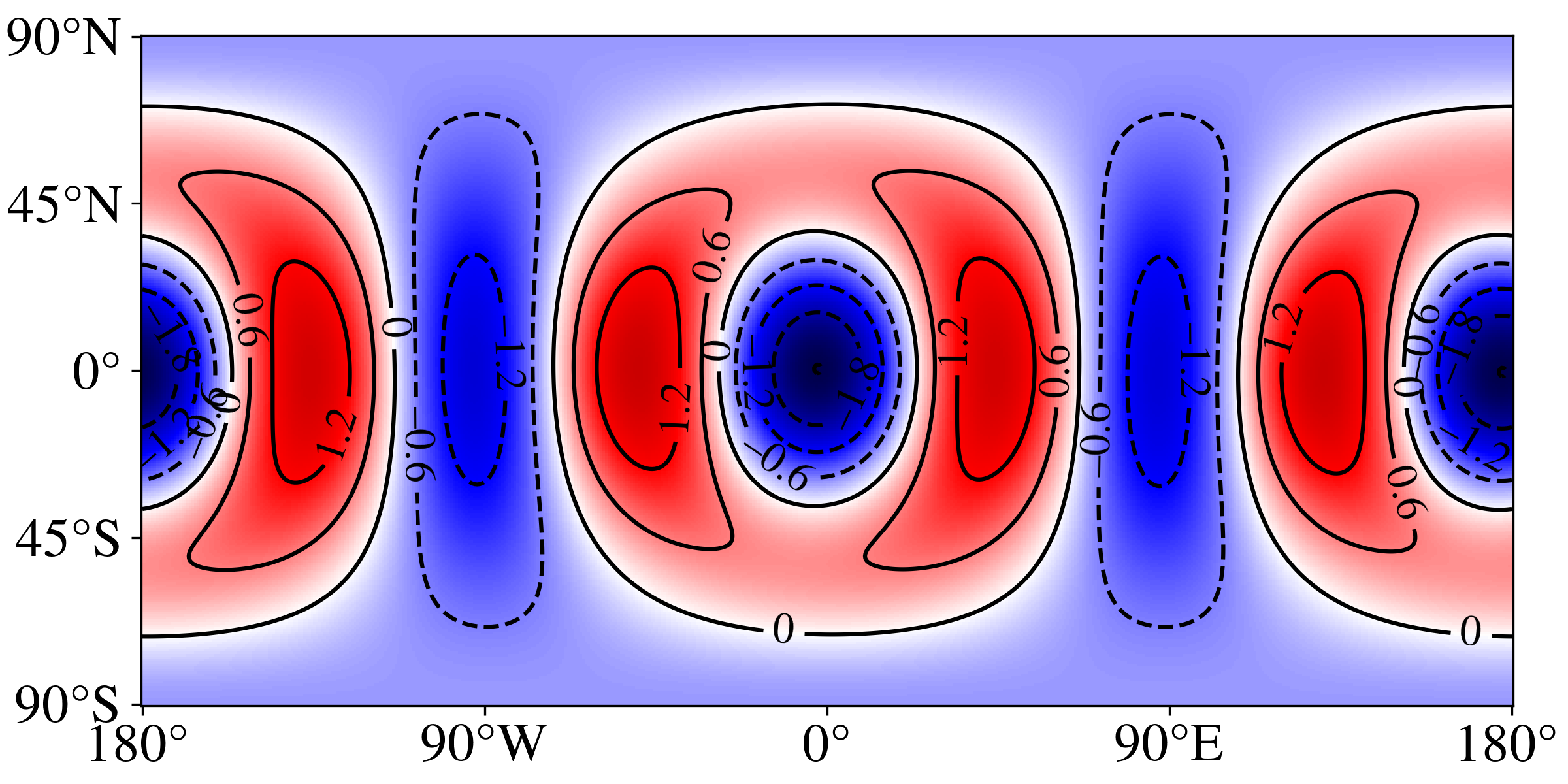}
			\label{fig:suppEuropaPrevXdiff}
		\end{subfigure}
		\hfill
		\begin{subfigure}[b]{0.495\textwidth}
			\caption{$|\Bind|$ difference ($\mathrm{nT}$) vs.~symmetric at $r=R_E$
			}
			\centering
			\includegraphics[width=\textwidth]{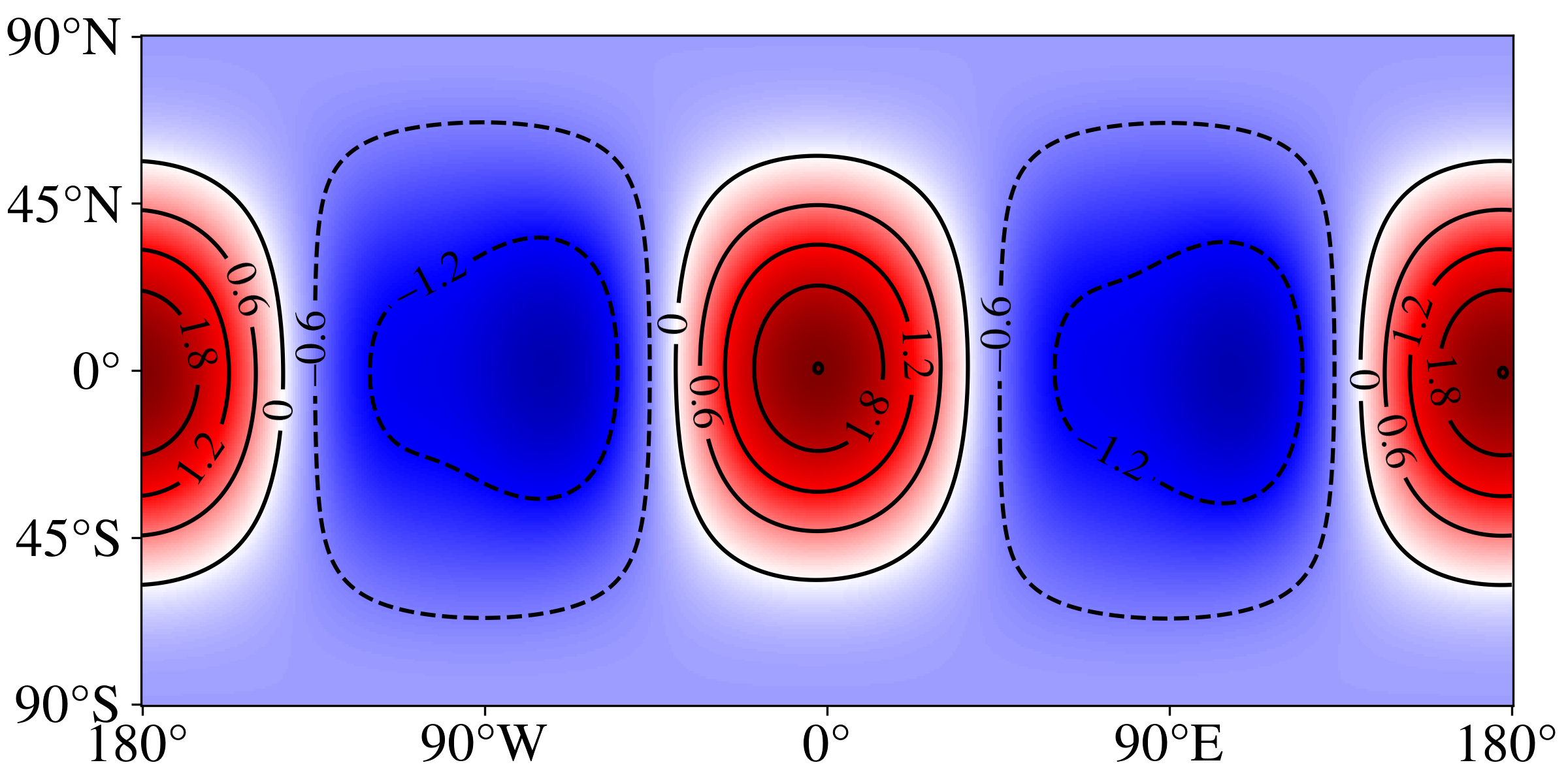}
			\label{fig:suppEuropaPrevMagDiff}
		\end{subfigure}\\
		\begin{subfigure}[b]{0.495\textwidth}
			\caption{$B_y$ difference ($\mathrm{nT}$) vs.~symmetric at $r=R_E$
			}
			\centering
			\includegraphics[width=\textwidth]{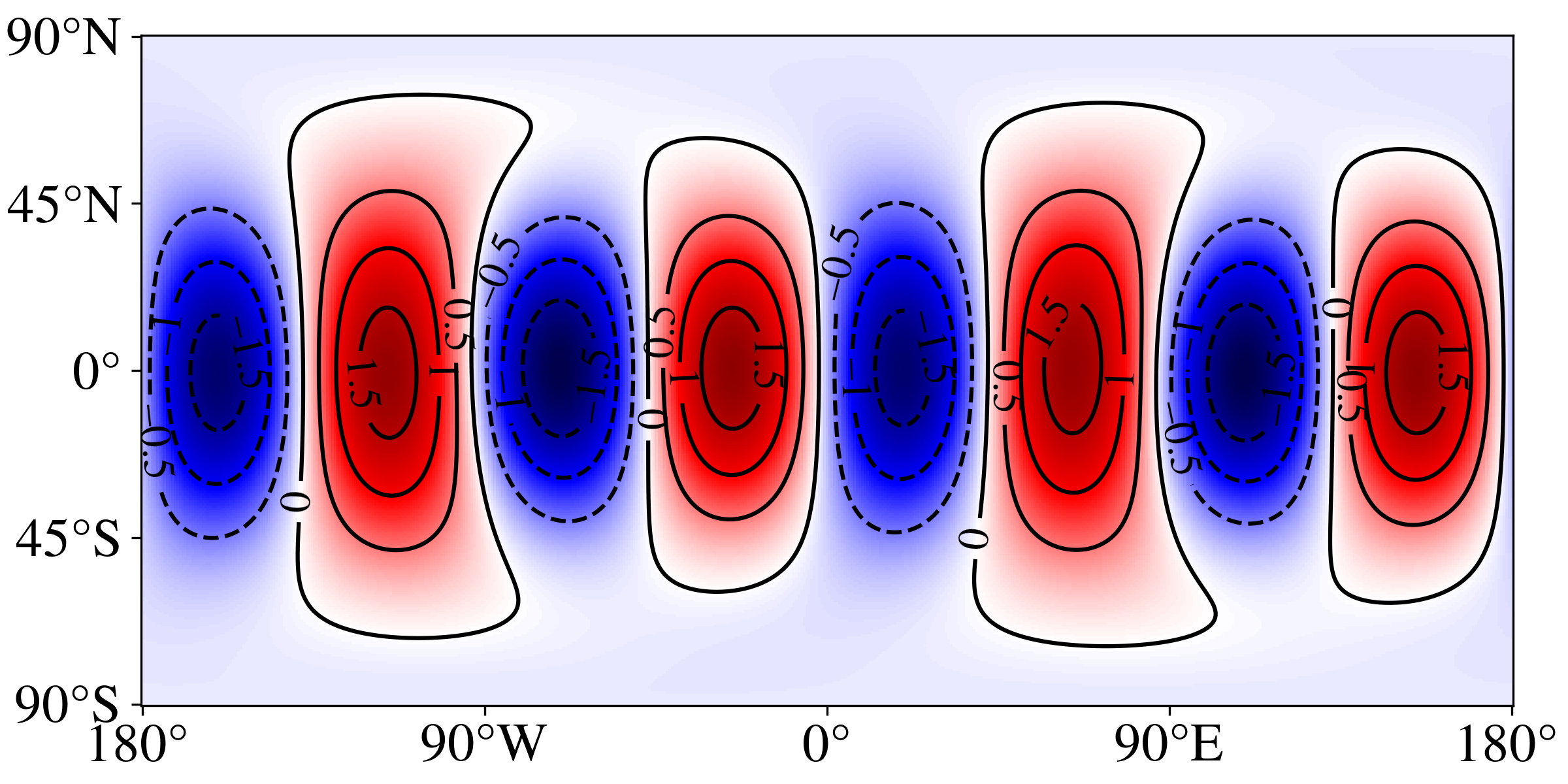}
			\label{fig:suppEuropaPrevYdiff}
		\end{subfigure}
		\hfill
		\begin{subfigure}[b]{0.495\textwidth}
			\caption{$B_z$ difference ($\mathrm{nT}$) vs.~symmetric at $r=R_E$
			}
			\centering
			\includegraphics[width=\textwidth]{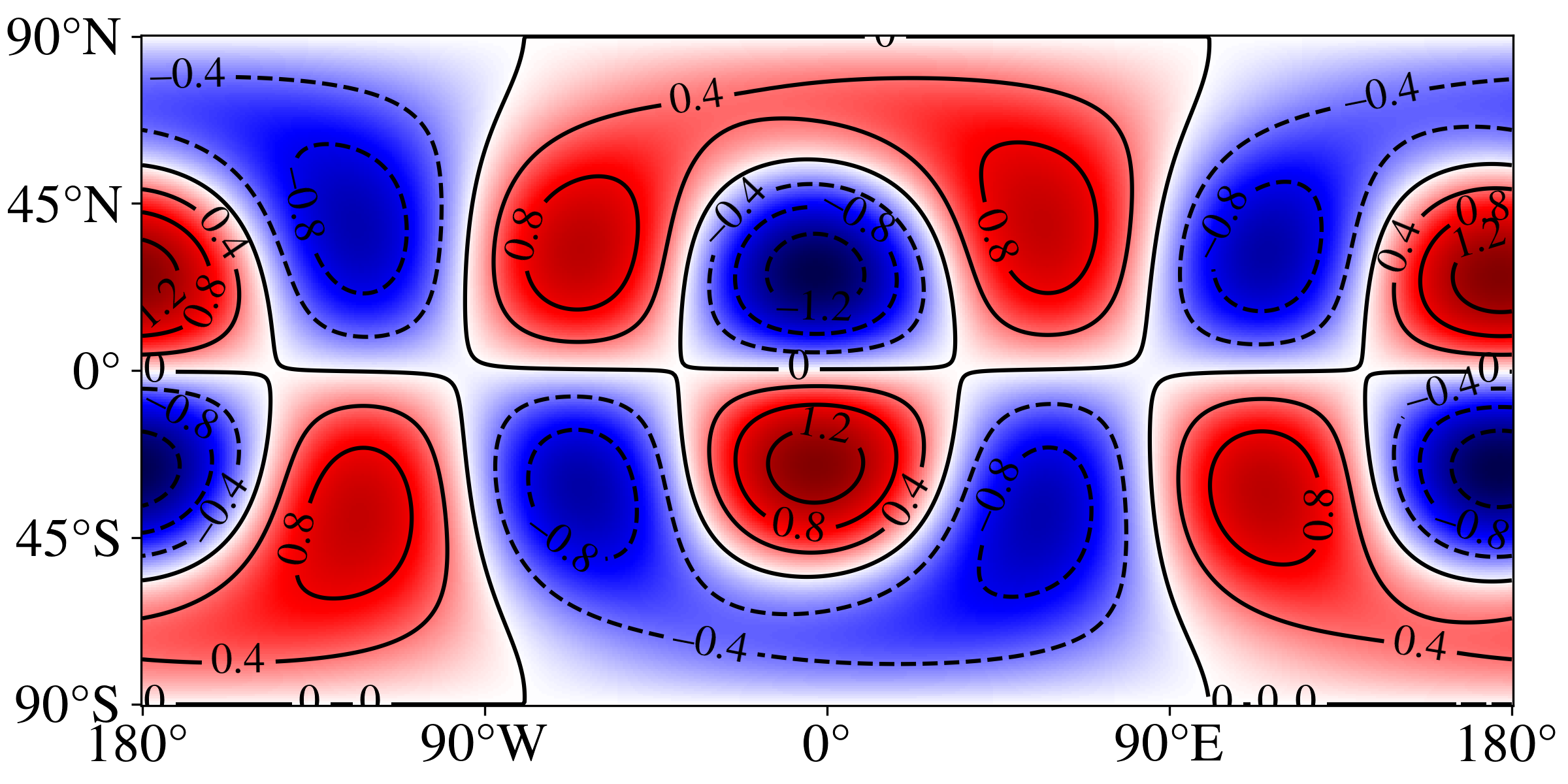}
			\label{fig:suppEuropaPrevZdiff}
		\end{subfigure}
		\caption{
			Simplified Europa model also based on the results of \citet{tobie2003tidally}, but with only a single real harmonic represented in the boundary shape and with very high ocean conductivity.
			The asymmetry model is identical to that considered in our previous work \citep{styczinski2021induced}; in this case, the conductivity is very high ($2,750\siu{S/m}$) so that the result can be compared to the high-conductivity limit previously studied.
			Conduction in the ionosphere is ignored.
			Magnetic fields are evaluated at the J2000 epoch on the surface of Europa.
			For simplicity, only the synodic period is modeled here.
			As can be seen in the time series for $B_x$ at the sub-jovian point plotted in \F\ref{fig:tSeriesLander} (main text), the more rigorous model derived in this work predicts a larger difference compared to the spherically symmetric case, several times larger than our previous estimates.
			Animations for the difference in $x$ component and magnitude as they vary throughout the synodic period are included as Supplemental Material.
		}
		\label{fig:suppEuropaPrev}
	\end{figure}

\begin{figure}
		\centering
		\begin{subfigure}[b]{0.495\textwidth}
			\caption{Miranda ice shell thickness ($\mathrm{km}$), $\D=50\siu{km}$
			}
			\centering
			\includegraphics[width=\textwidth]{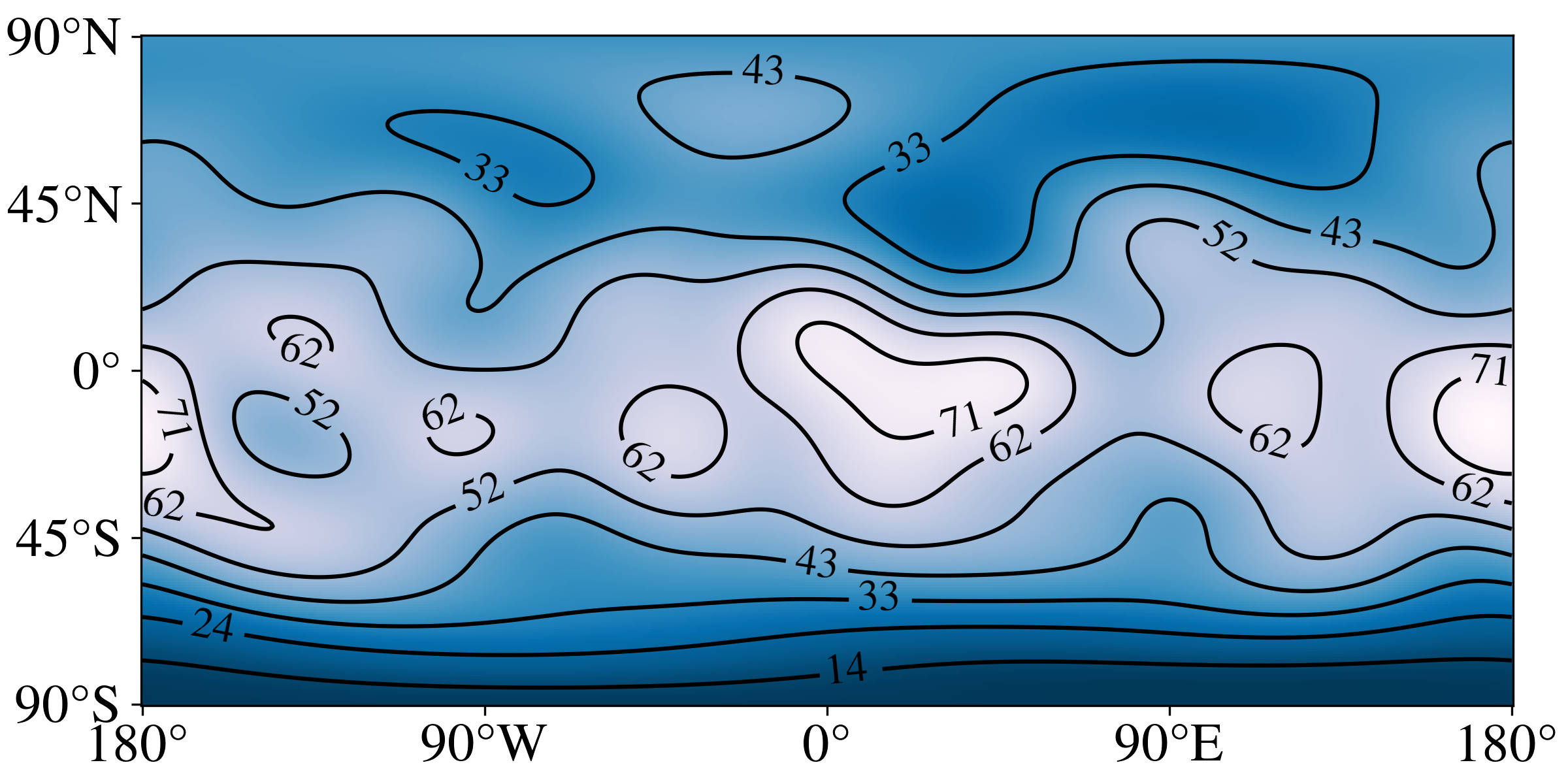}
			\label{fig:suppMirandaContour}
		\end{subfigure}
		\hfill
		\begin{subfigure}[b]{0.495\textwidth}
			\caption{Asymmetric model $|\Bind|$ ($\mathrm{nT}$) at $r=2R_M$
			}
			\centering
			\includegraphics[width=\textwidth]{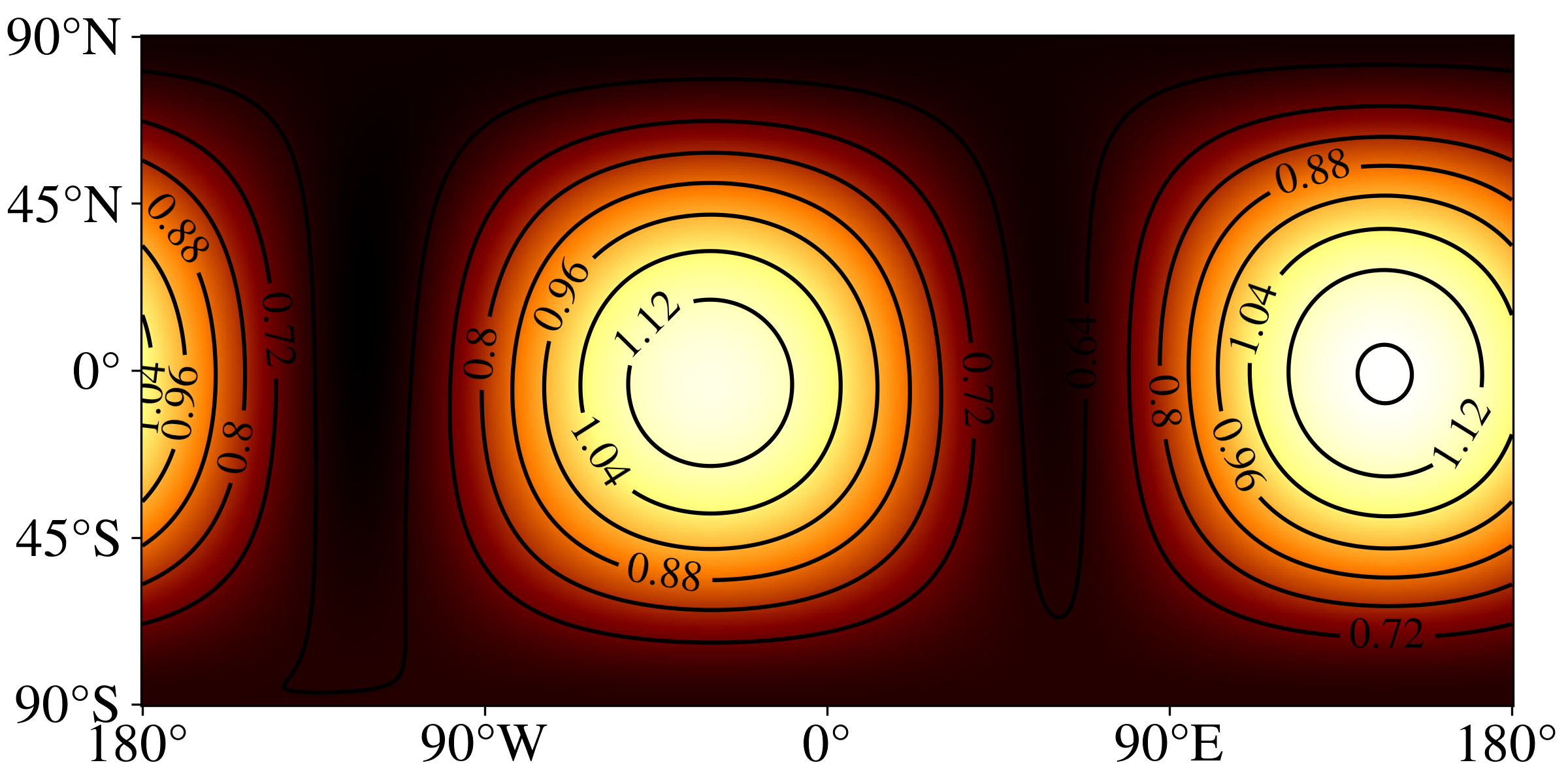}
			\label{fig:suppMirandaMag}
		\end{subfigure}\\
		\begin{subfigure}[b]{0.495\textwidth}
			\caption{$B_x$ difference ($\mathrm{nT}$) vs.~symmetric at $r=2R_M$
			}
			\centering
			\includegraphics[width=\textwidth]{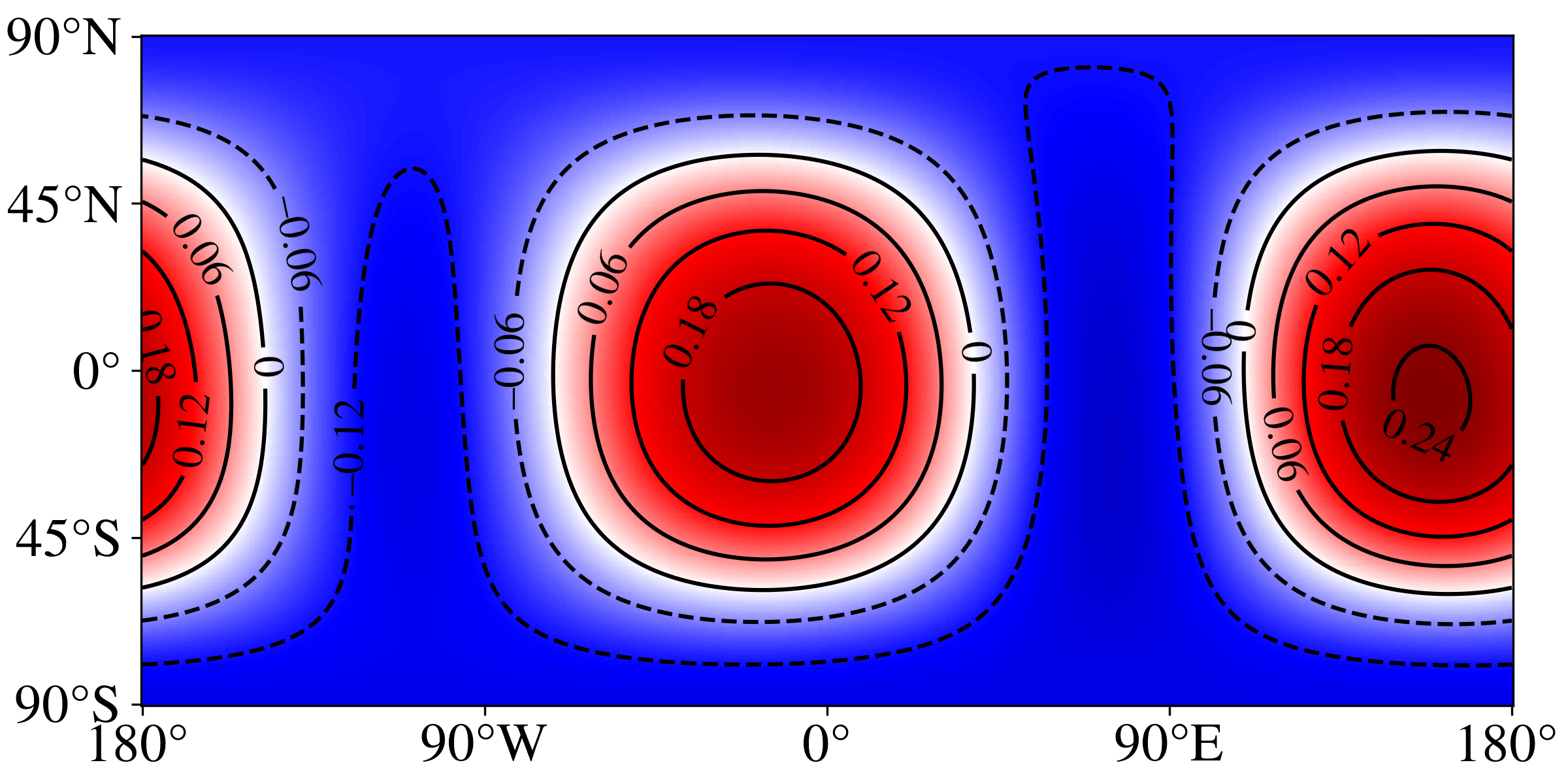}
			\label{fig:suppMirandaXdiff}
		\end{subfigure}
		\hfill
		\begin{subfigure}[b]{0.495\textwidth}
			\caption{$|\Bind|$ difference ($\mathrm{nT}$) vs.~symmetric at $r=2R_M$
			}
			\centering
			\includegraphics[width=\textwidth]{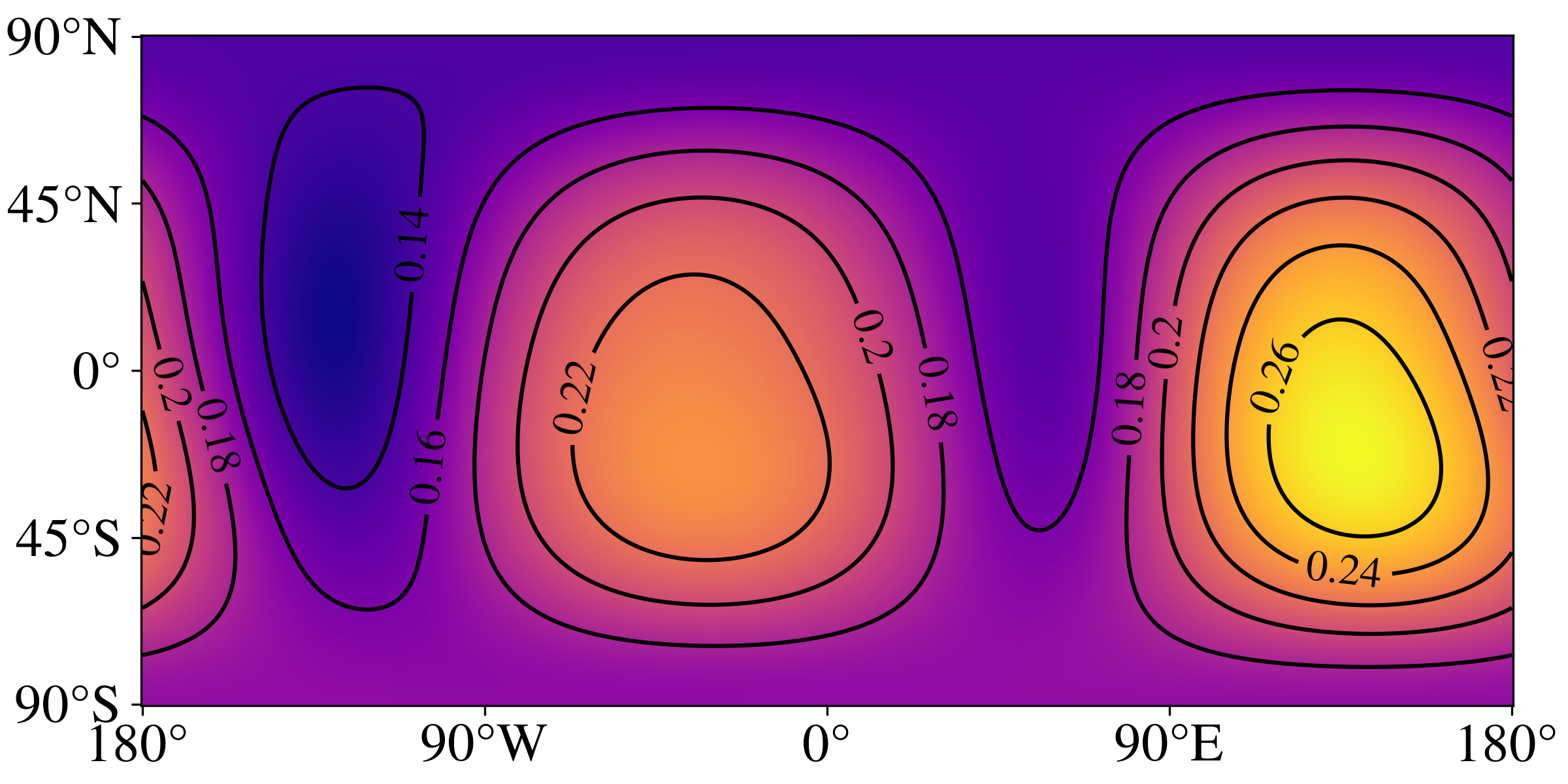}
			\label{fig:suppMirandaMagDiff}
		\end{subfigure}\\
		\begin{subfigure}[b]{0.495\textwidth}
			\caption{$B_y$ difference ($\mathrm{nT}$) vs.~symmetric at $r=2R_M$
			}
			\centering
			\includegraphics[width=\textwidth]{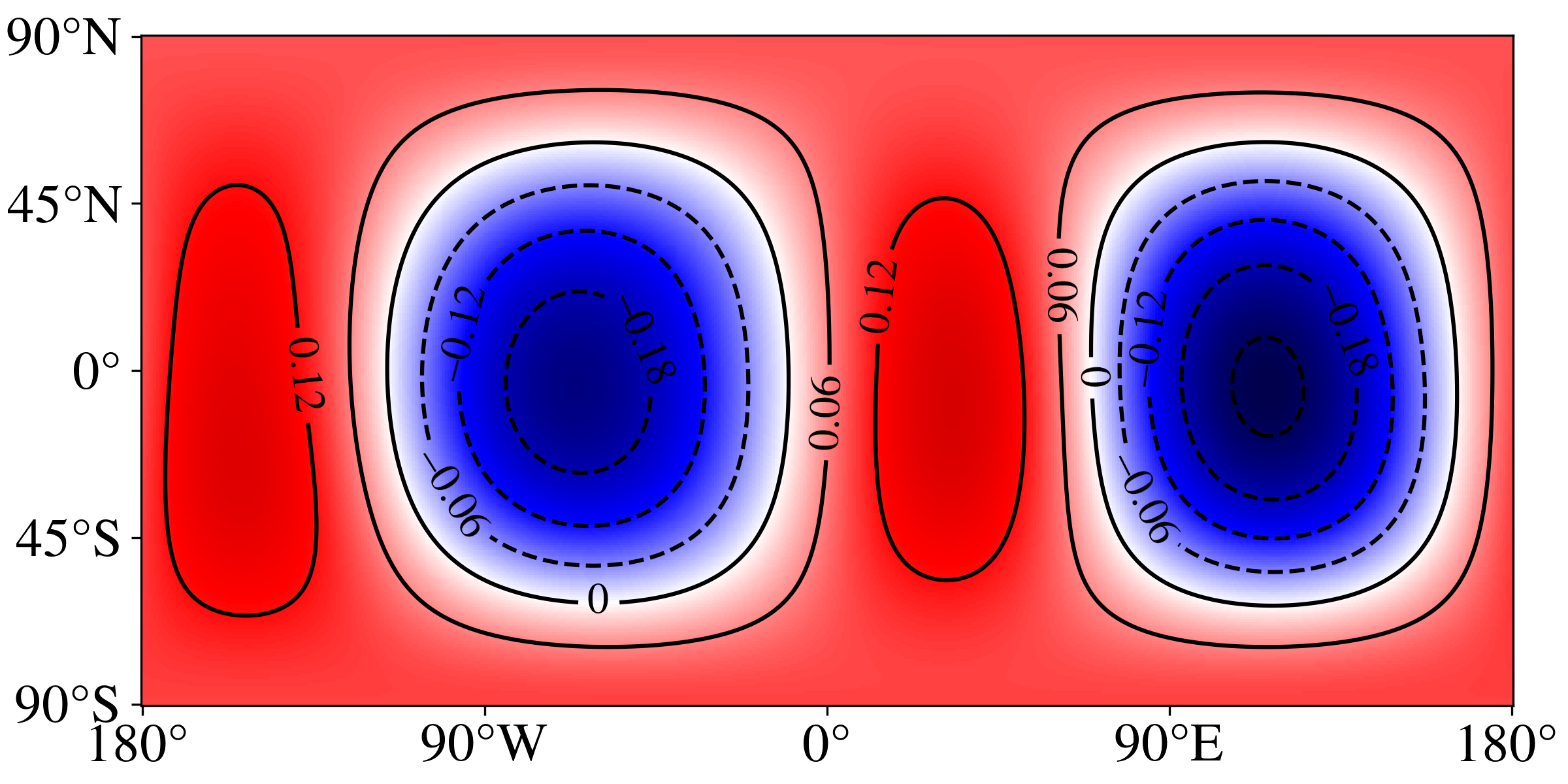}
			\label{fig:suppMirandaYdiff}
		\end{subfigure}
		\hfill
		\begin{subfigure}[b]{0.495\textwidth}
			\caption{$B_z$ difference ($\mathrm{nT}$) vs.~symmetric at $r=2R_M$
			}
			\centering
			\includegraphics[width=\textwidth]{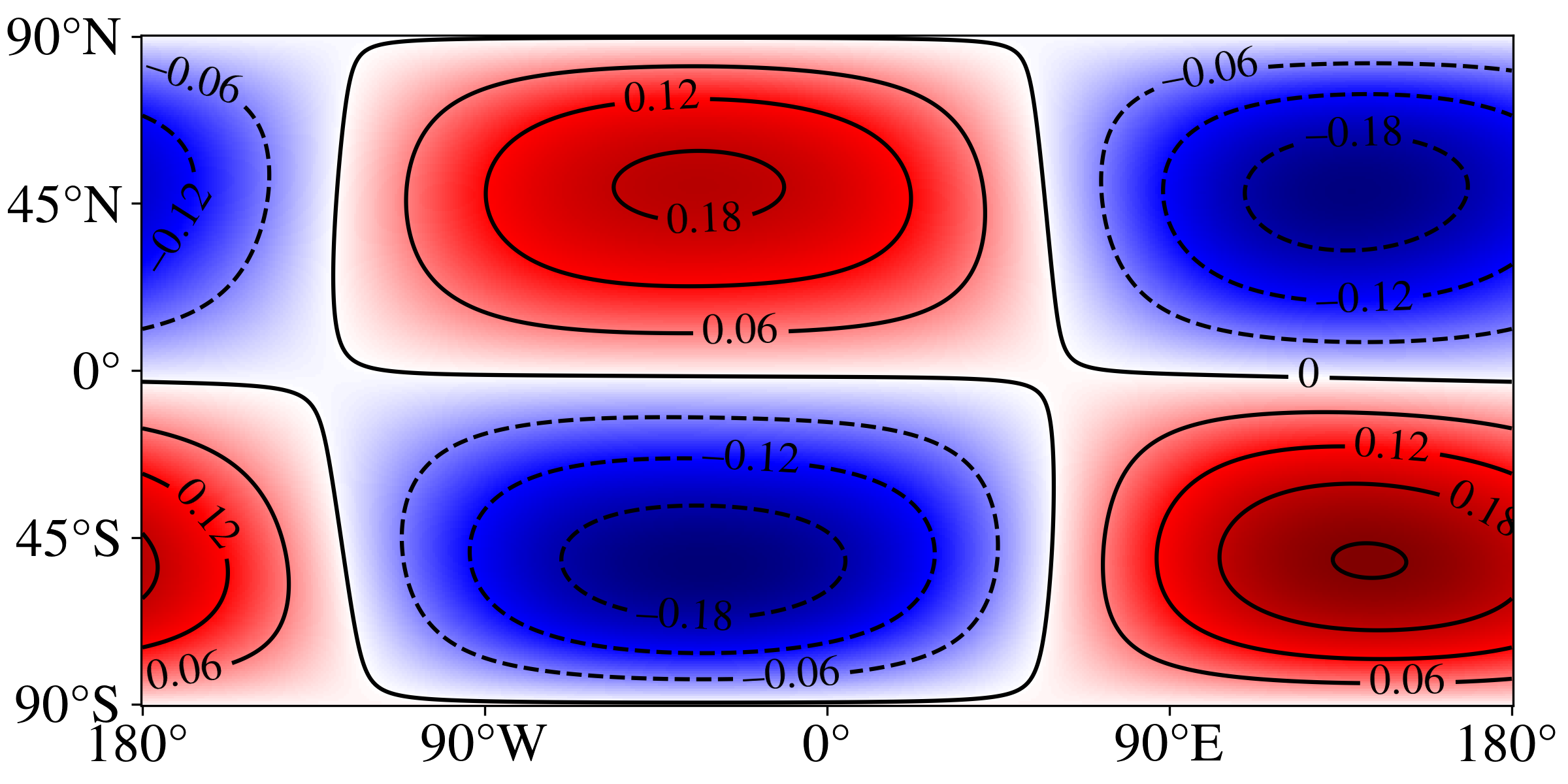}
			\label{fig:suppMirandaZdiff}
		\end{subfigure}
		\caption{
			Miranda model with an ice--ocean boundary shape based on the asymmetric Enceladus ice shell inferred by \citet{hemingway2019enceladus} from isostatic compensation and gravity measurements.
			The asymmetry in the ice shell has been scaled up from the $21\siu{km}$ average thickness model of Enceladus to the $50\siu{km}$ average thickness ice shell we suppose for Miranda, to serve as an upper limit of expected asymmetry in demonstrating application of our results.
			Enceladus model data are courtesy D.~Hemingway; compare \F\ref{fig:suppMirandaContour} to \F11d of \citet{hemingway2019enceladus}.
			The difference in induced field magnitude relative to spherically symmetric is always positive because the largest change to the induced magnetic moments is in the dipole moment.
			More conducting ocean material is closer to the surface in the asymmetric model, enhancing the largest moments.
			Magnetic fields are evaluated at the J2000 epoch, this time at $r=2R_M$, a plausible flyby altitude for a future mission to the Uranus system.
			A Seawater composition is assumed for the ocean, and a $100\siu{km}$ uniform ionosphere is assumed, with a total ionospheric conductance of $800\siu{S}$ based on comparison with the plasma environment of Callisto \citep{hartkorn2017induction}.
			Electrical conductivities in the ocean are calculated using the \textit{PlanetProfile} geophysical modeling framework \citep{vance2021magnetic}.
			For simplicity, only the synodic period is modeled here.
			Animations for the difference in $x$ component and magnitude as they vary throughout the synodic period are included as Supplemental Material.
		}
		\label{fig:suppMiranda}
	\end{figure}

\begin{figure}
		\centering
		\begin{subfigure}[b]{0.495\textwidth}
			\caption{Callisto ionosphere thickness ($\mathrm{km}$), $\D=100\siu{km}$
			}
			\centering
			\includegraphics[width=\textwidth]{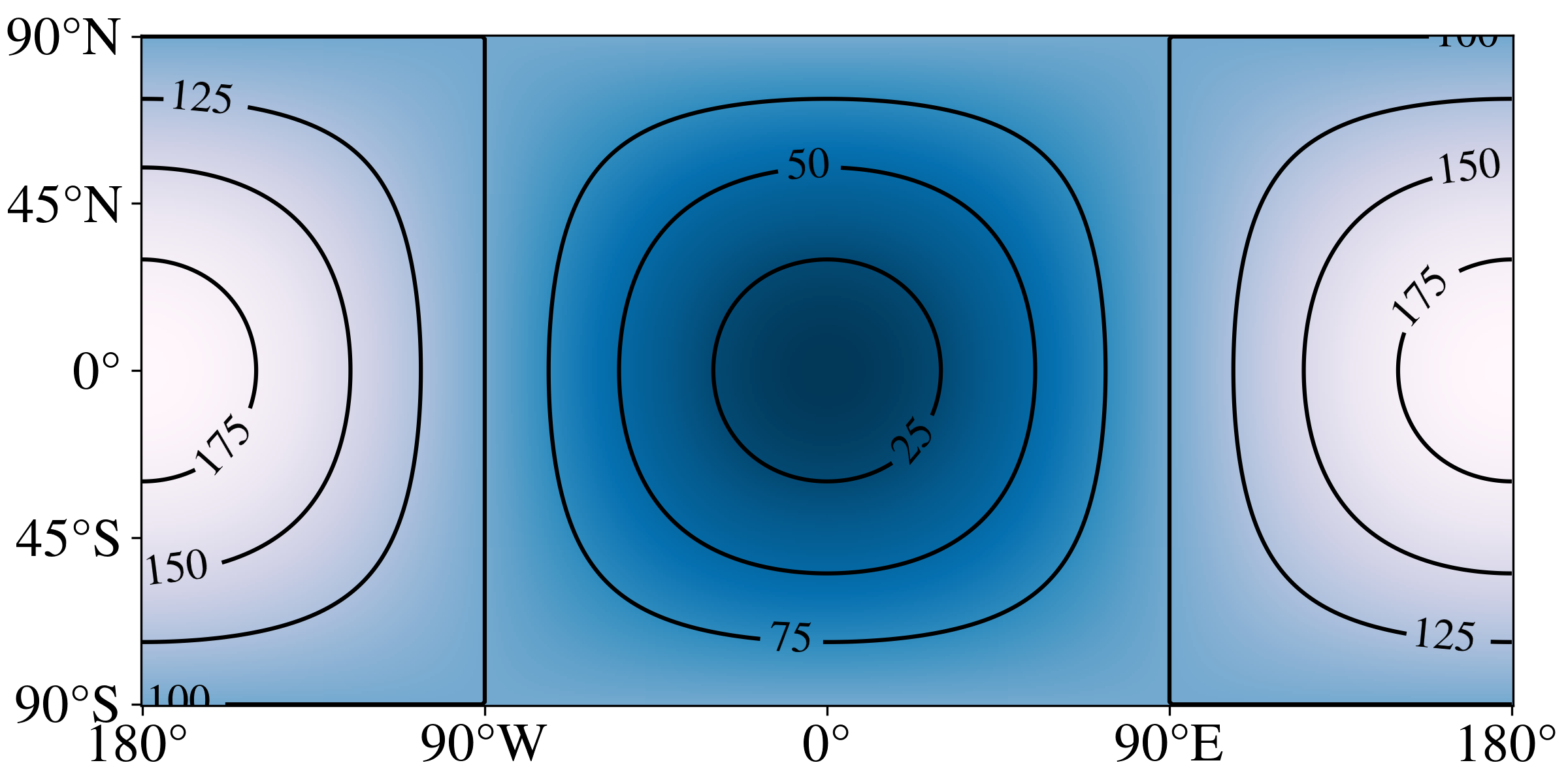}
			\label{fig:suppCallistoContour}
		\end{subfigure}
		\hfill
		\begin{subfigure}[b]{0.495\textwidth}
			\caption{Asymmetric model $|\Bind|$ ($\mathrm{nT}$) at $r=2R_C$
			}
			\centering
			\includegraphics[width=\textwidth]{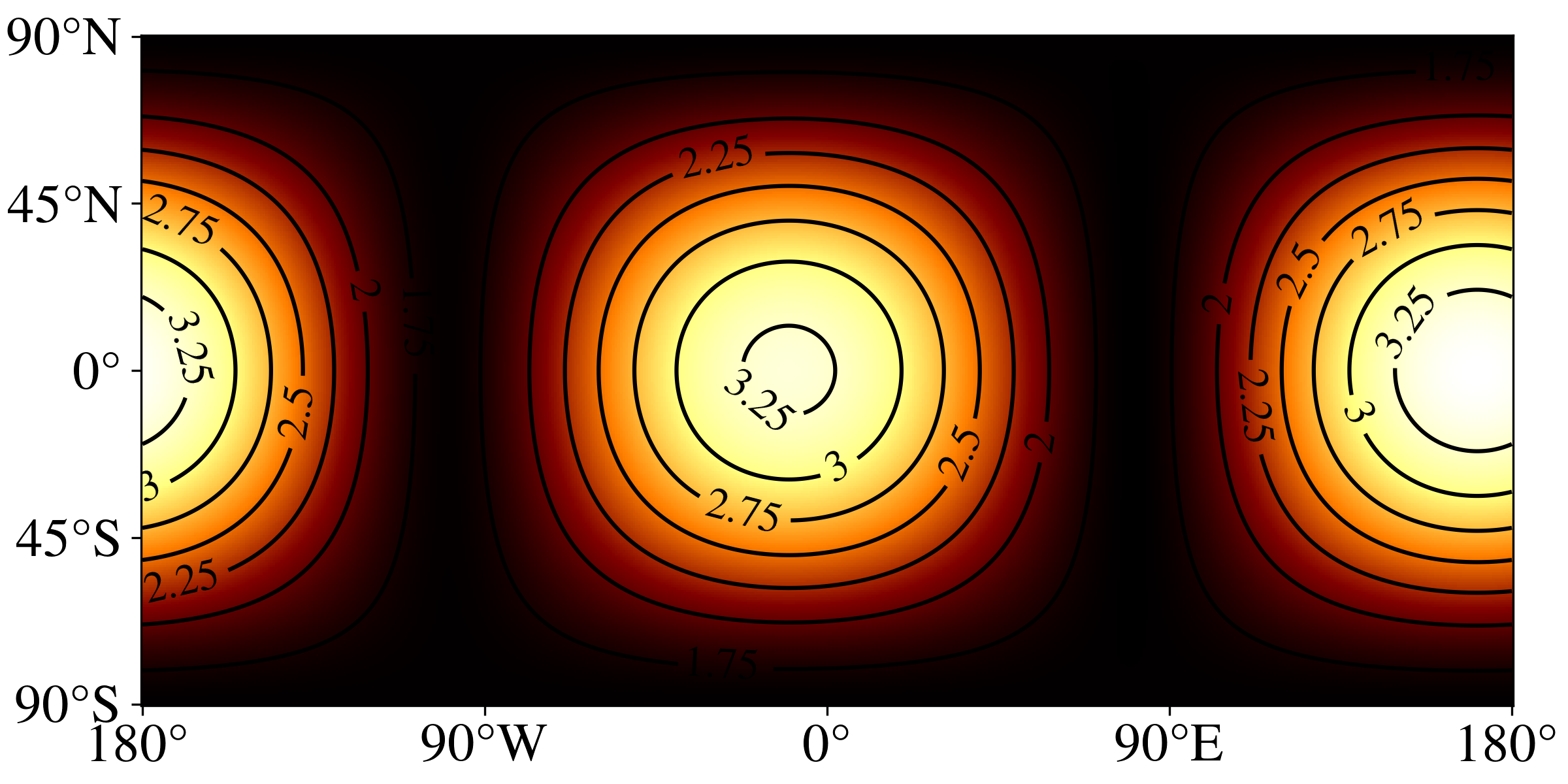}
			\label{fig:suppCallistoMag}
		\end{subfigure}\\
		\begin{subfigure}[b]{0.495\textwidth}
			\caption{$B_x$ difference ($\mathrm{nT}$) vs.~symmetric at $r=2R_C$
			}
			\centering
			\includegraphics[width=\textwidth]{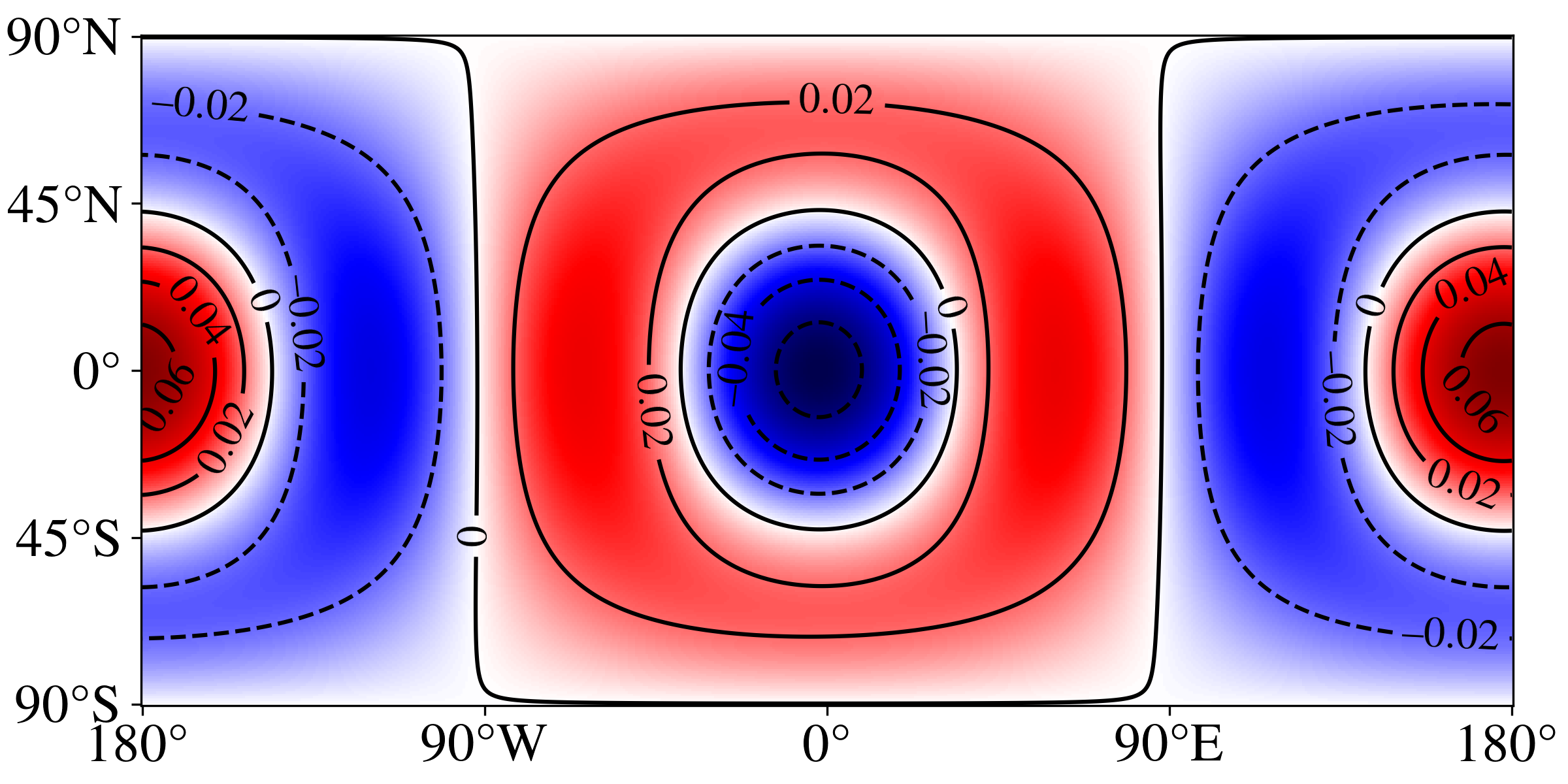}
			\label{fig:suppCallistoXdiff}
		\end{subfigure}
		\hfill
		\begin{subfigure}[b]{0.495\textwidth}
			\caption{$|\Bind|$ difference ($\mathrm{nT}$) vs.~symmetric at $r=2R_C$
			}
			\centering
			\includegraphics[width=\textwidth]{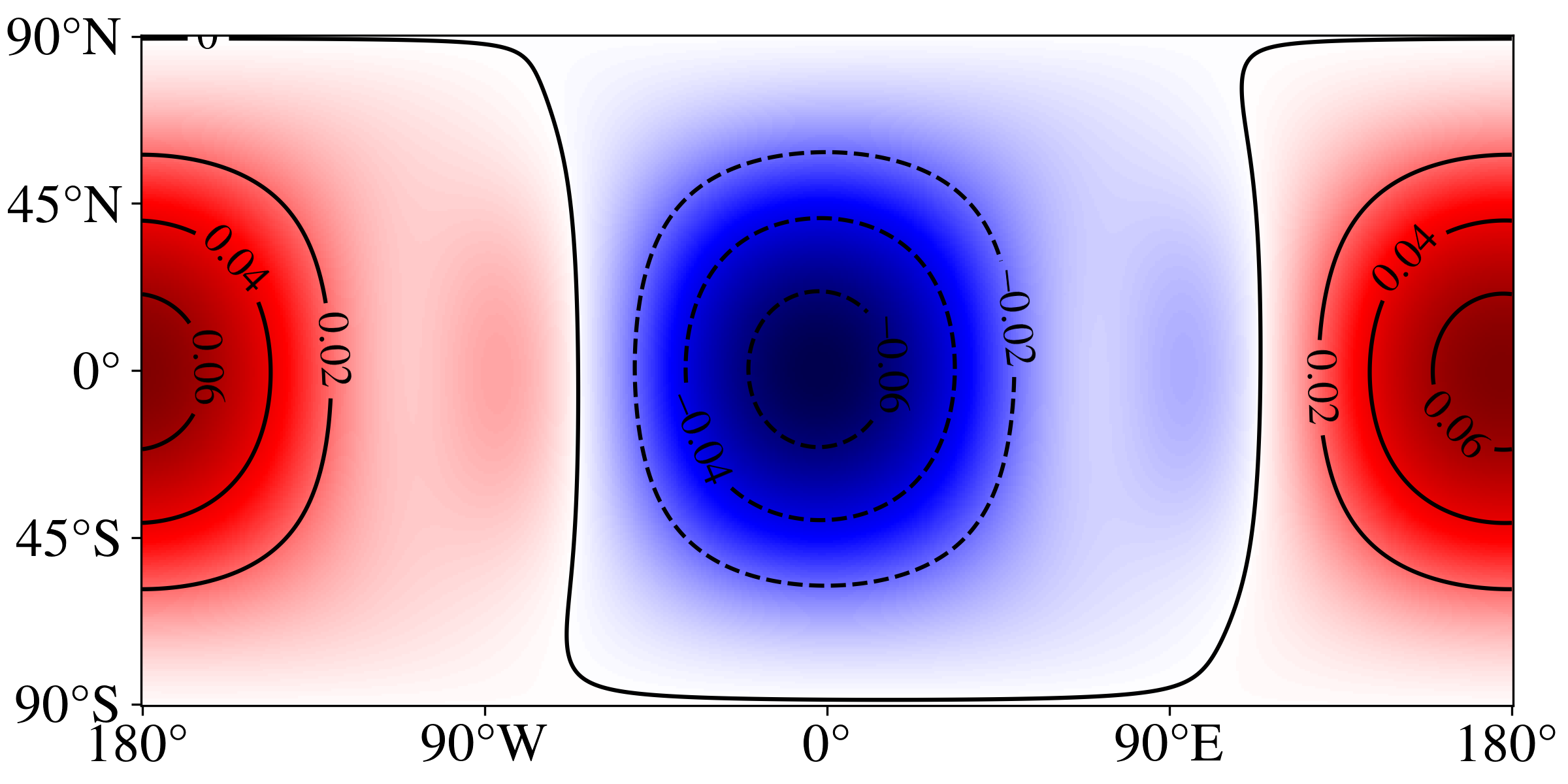}
			\label{fig:suppCallistoMagDiff}
		\end{subfigure}\\
		\begin{subfigure}[b]{0.495\textwidth}
			\caption{$B_y$ difference ($\mathrm{nT}$) vs.~symmetric at $r=2R_C$
			}
			\centering
			\includegraphics[width=\textwidth]{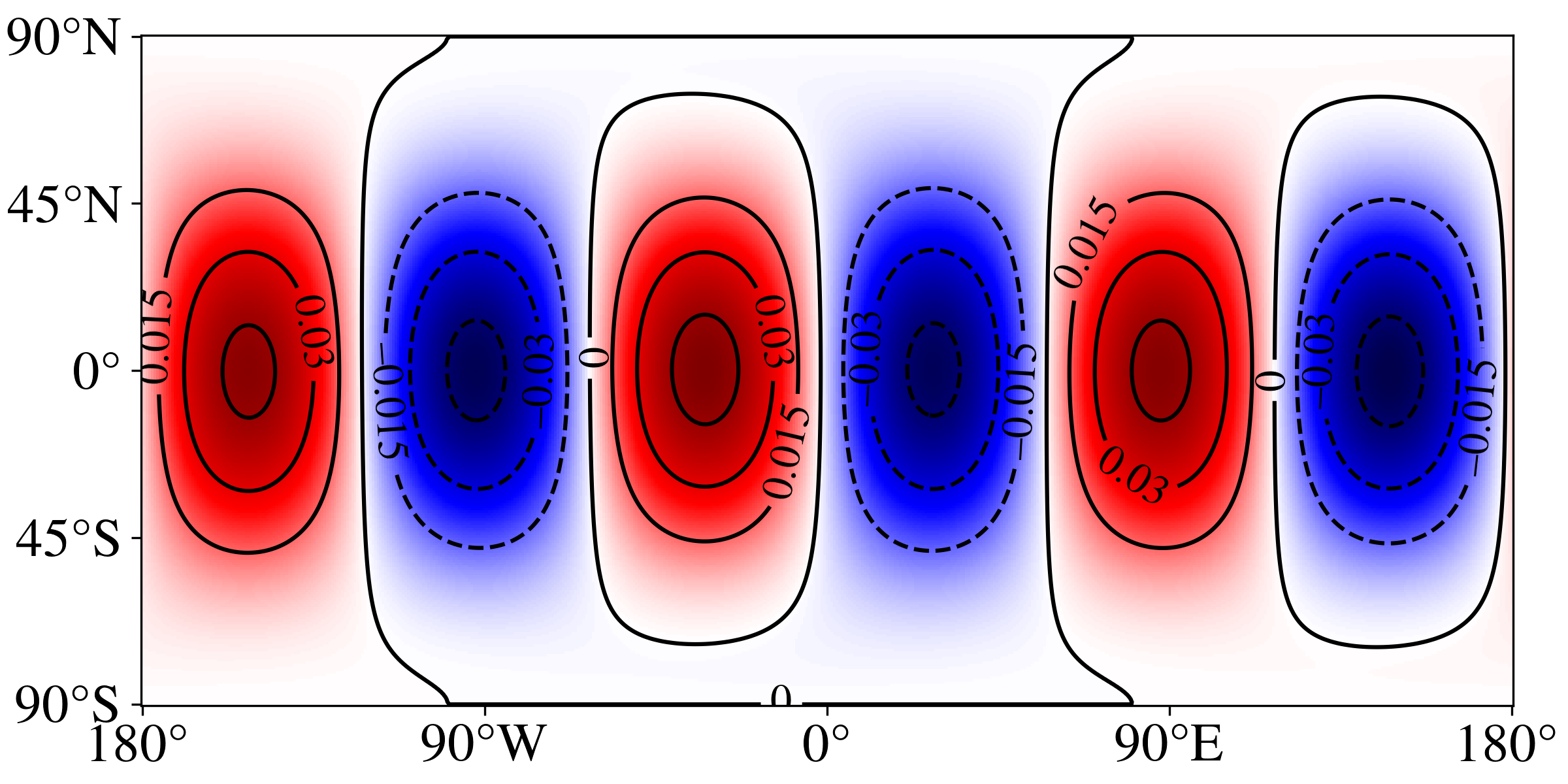}
			\label{fig:suppCallistoYdiff}
		\end{subfigure}
		\hfill
		\begin{subfigure}[b]{0.495\textwidth}
			\caption{$B_z$ difference ($\mathrm{nT}$) vs.~symmetric at $r=2R_C$
			}
			\centering
			\includegraphics[width=\textwidth]{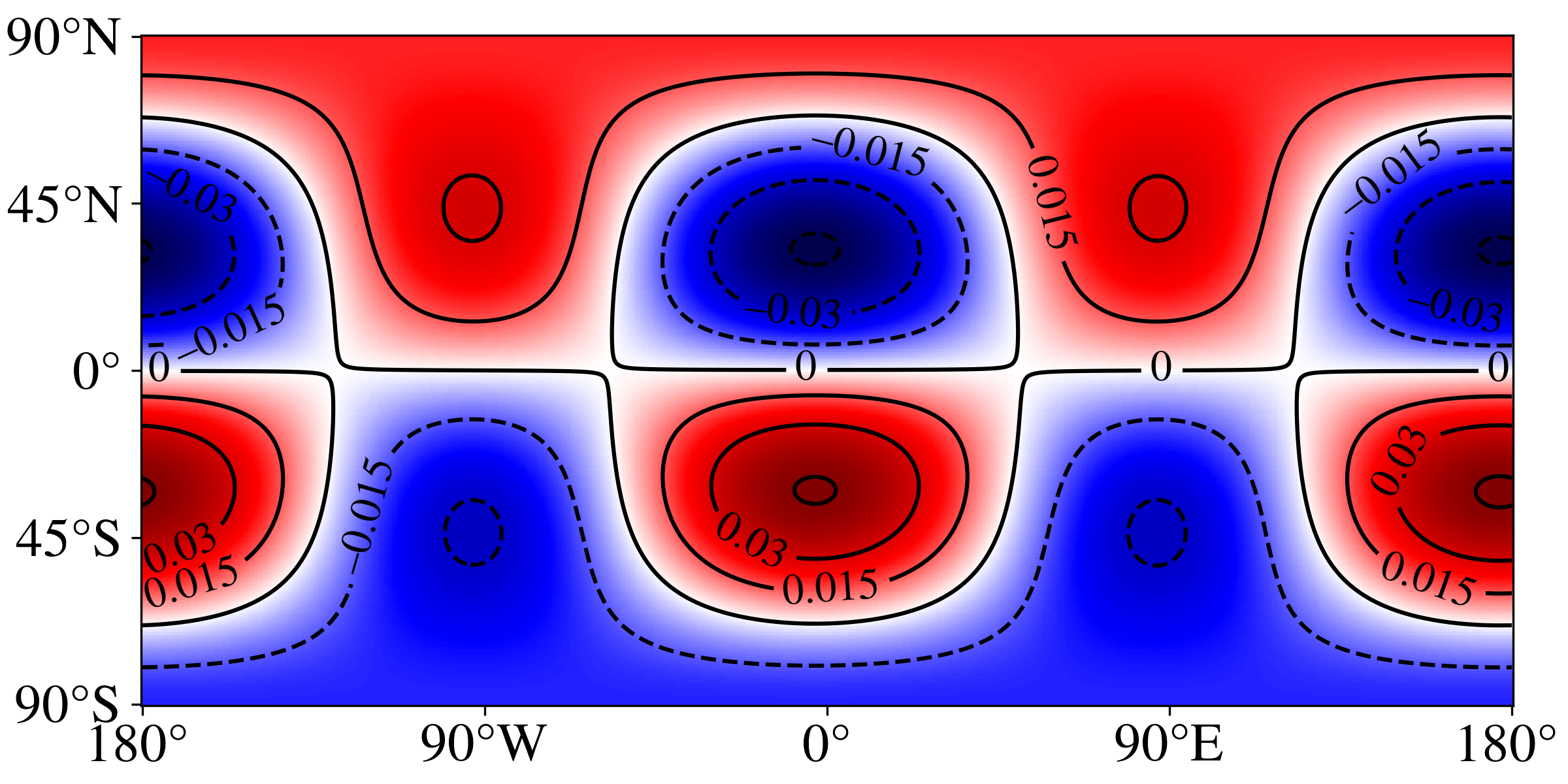}
			\label{fig:suppCallistoZdiff}
		\end{subfigure}
		\caption{
			Callisto model with an asymmetric ionosphere modeled after \citet{hartkorn2017structure}.
			Uniform conductivity, an average ionospheric thickness of $100\siu{km}$, and a total height-integrated conductivity of $800\siu{S}$ are assumed, as a rough approximation of the day--night dichotomy inferred by \citeauthor{hartkorn2017structure}.
			An ocean with dissolved \ch{MgSO4} and a $100\siu{km}$-thick ice shell are assumed.
			The magnetic field is evaluated at the J2000 epoch and at $r=2R_C$, a plausible distance for a spacecraft flyby.
			In this case, the differences due to asymmetry are negligible, owing to the relatively low ionospheric conductivity where the considered asymmetry is present.
			Electrical conductivities in the ocean are calculated using the \textit{PlanetProfile} geophysical modeling framework \citep{vance2021magnetic}.
			For simplicity, only the synodic period is modeled here.
			Animations for the difference in $x$ component and magnitude as they vary throughout the synodic period are included as Supplemental Material.
		}
		\label{fig:suppCallisto}
	\end{figure}

\begin{figure}
		\centering
		\begin{subfigure}[b]{0.495\textwidth}
			\caption{Triton ionosphere thickness ($\mathrm{km}$), $\D=200\siu{km}$
			}
			\centering
			\includegraphics[width=\textwidth]{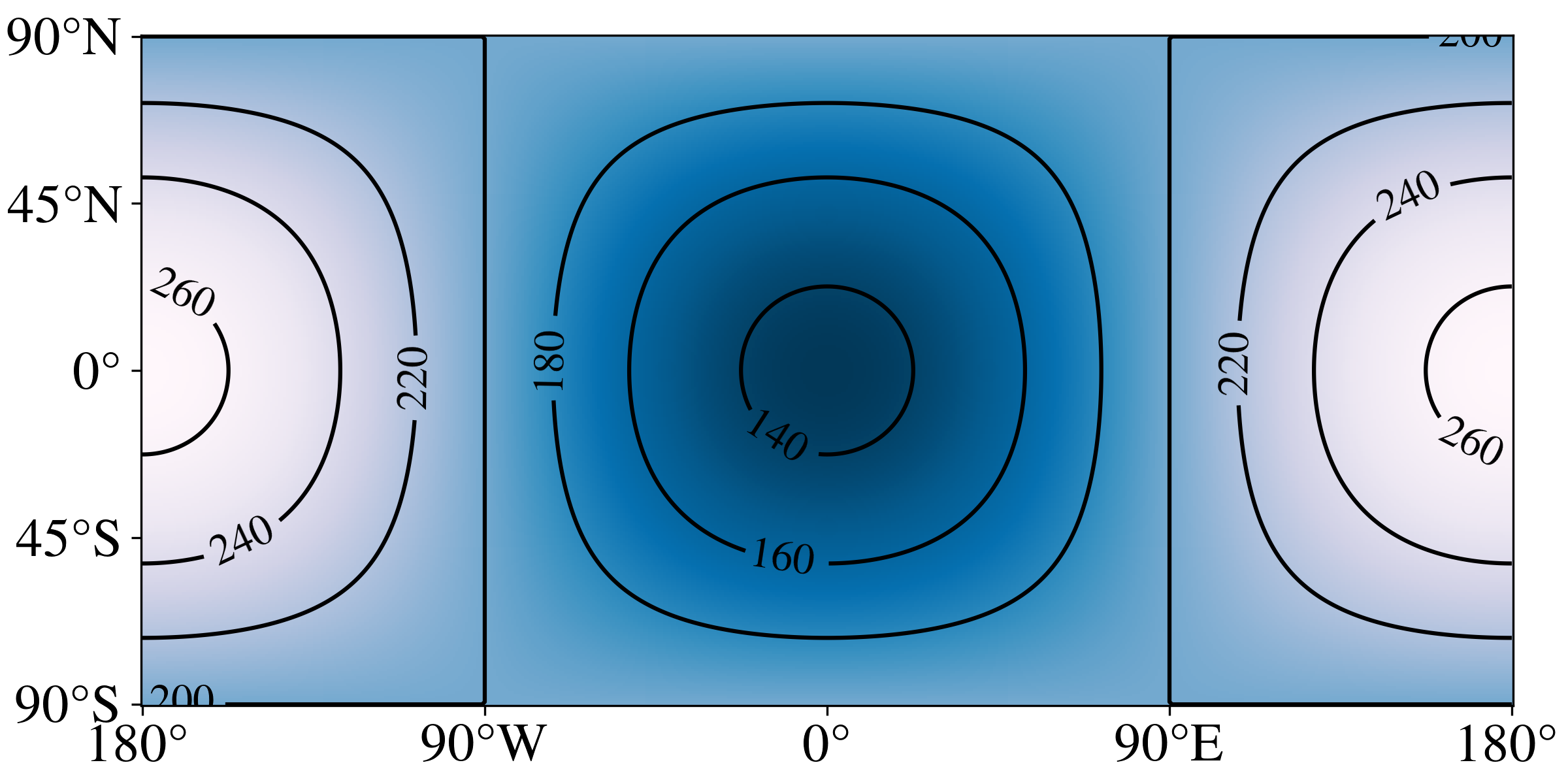}
			\label{fig:suppTritonContour}
		\end{subfigure}
		\hfill
		\begin{subfigure}[b]{0.495\textwidth}
			\caption{Asymmetric model $|\Bind|$ ($\mathrm{nT}$) at $r=2R_T$
			}
			\centering
			\includegraphics[width=\textwidth]{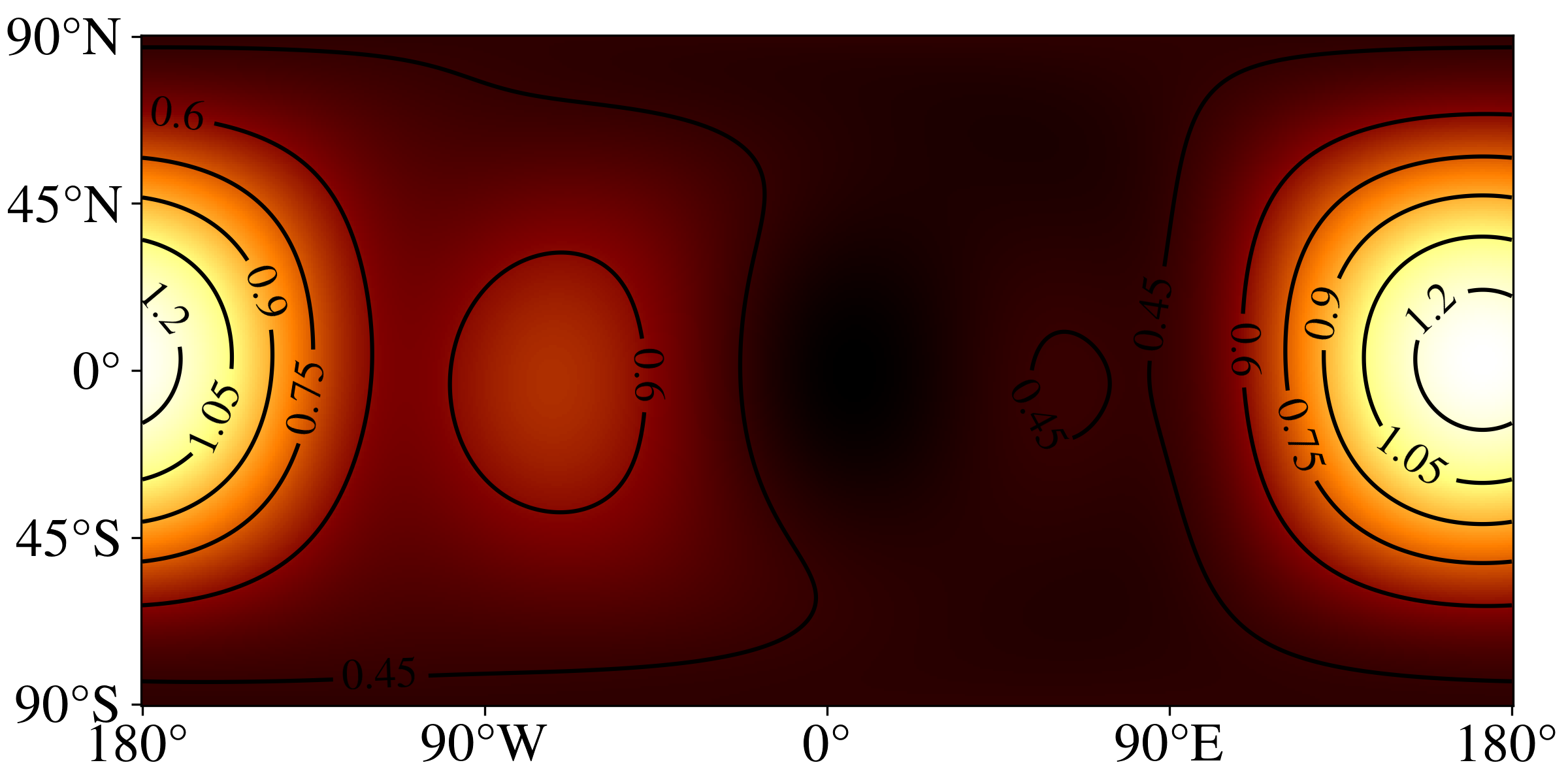}
			\label{fig:suppTritonMag}
		\end{subfigure}\\
		\begin{subfigure}[b]{0.495\textwidth}
			\caption{$B_x$ difference ($\mathrm{nT}$) vs.~symmetric at $r=2R_T$
			}
			\centering
			\includegraphics[width=\textwidth]{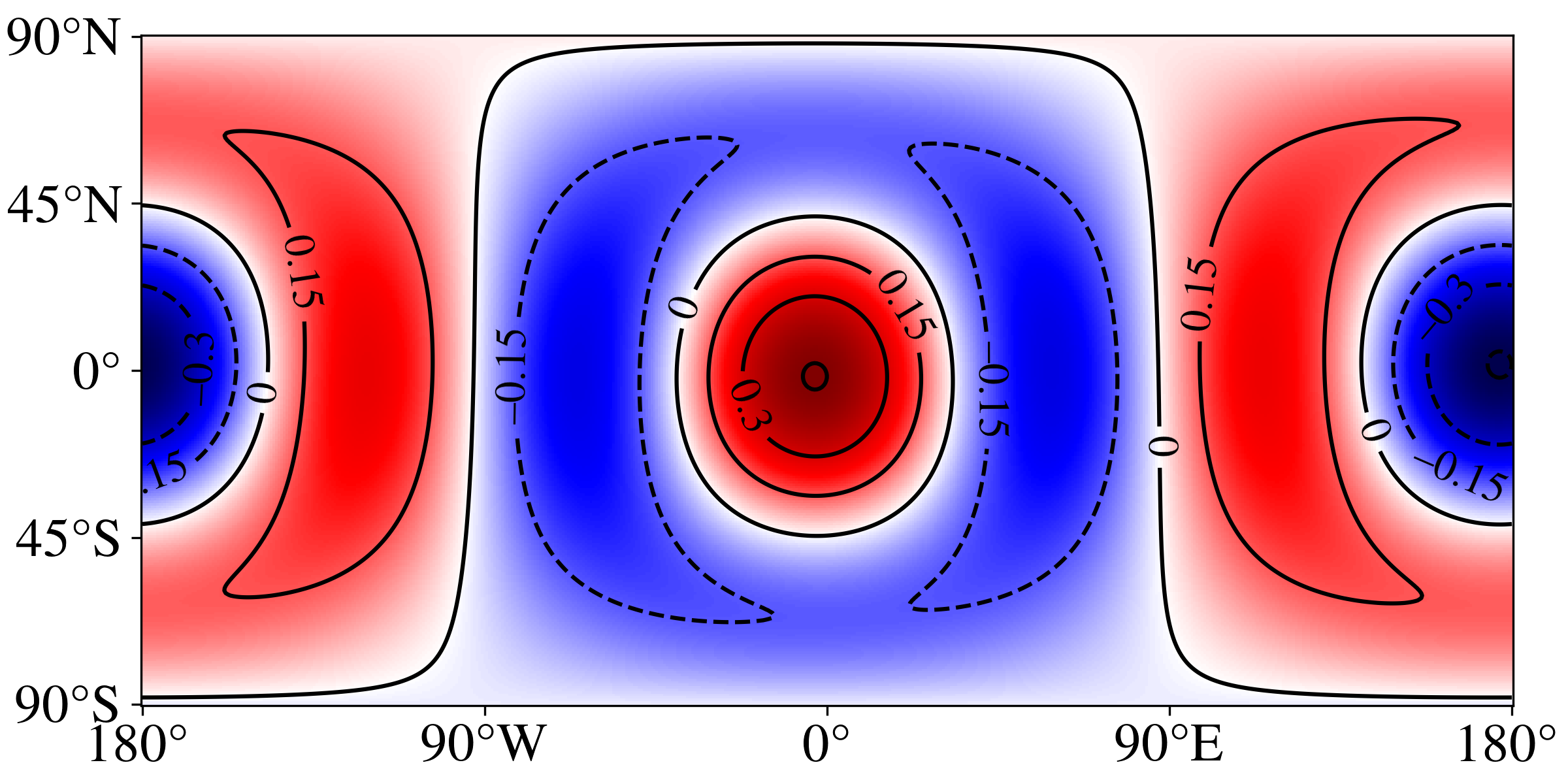}
			\label{fig:suppTritonXdiff}
		\end{subfigure}
		\hfill
		\begin{subfigure}[b]{0.495\textwidth}
			\caption{$|\Bind|$ difference ($\mathrm{nT}$) vs.~symmetric at $r=2R_T$
			}
			\centering
			\includegraphics[width=\textwidth]{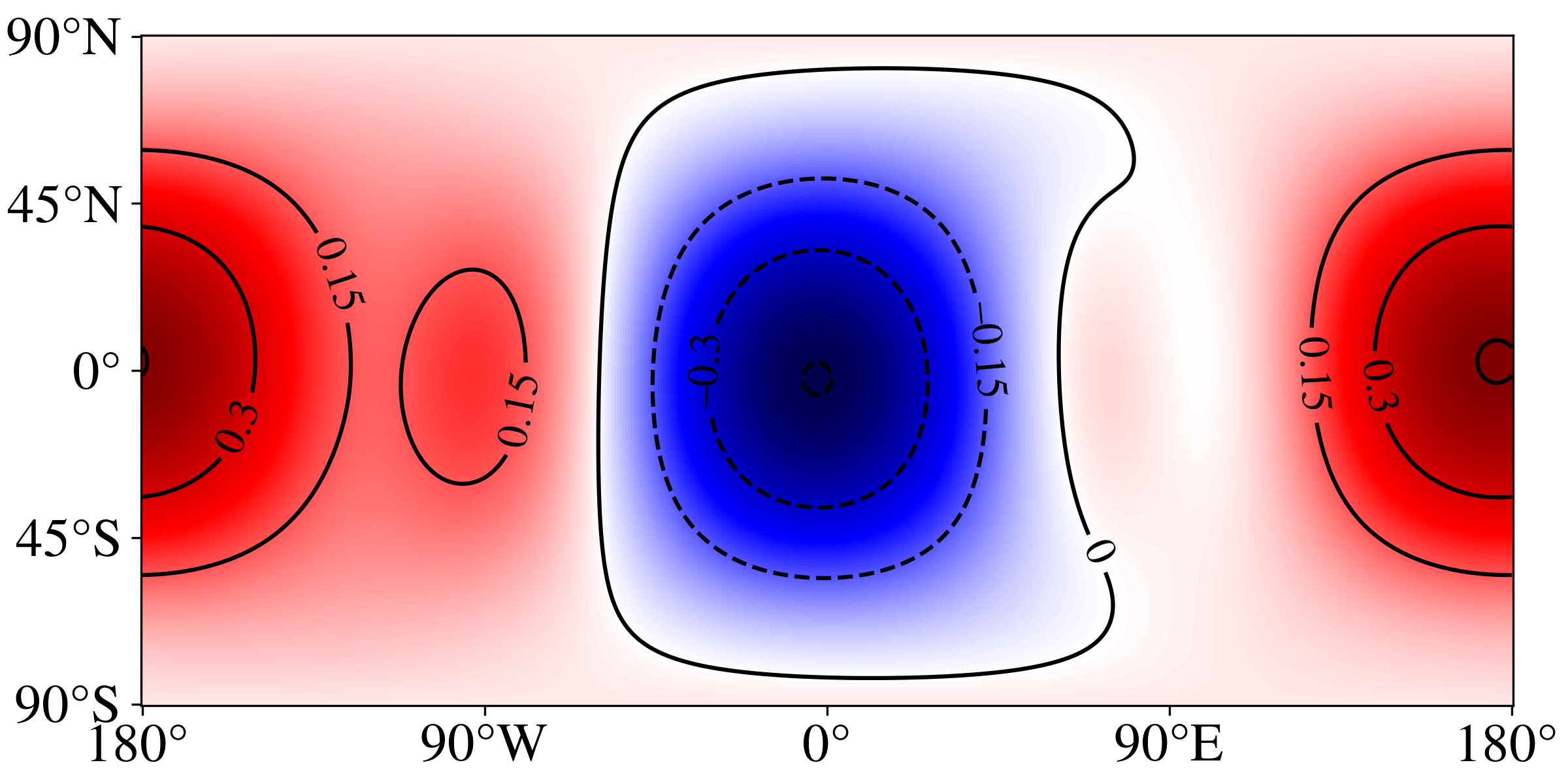}
			\label{fig:suppTritonMagDiff}
		\end{subfigure}\\
		\begin{subfigure}[b]{0.495\textwidth}
			\caption{$B_y$ difference ($\mathrm{nT}$) vs.~symmetric at $r=2R_T$
			}
			\centering
			\includegraphics[width=\textwidth]{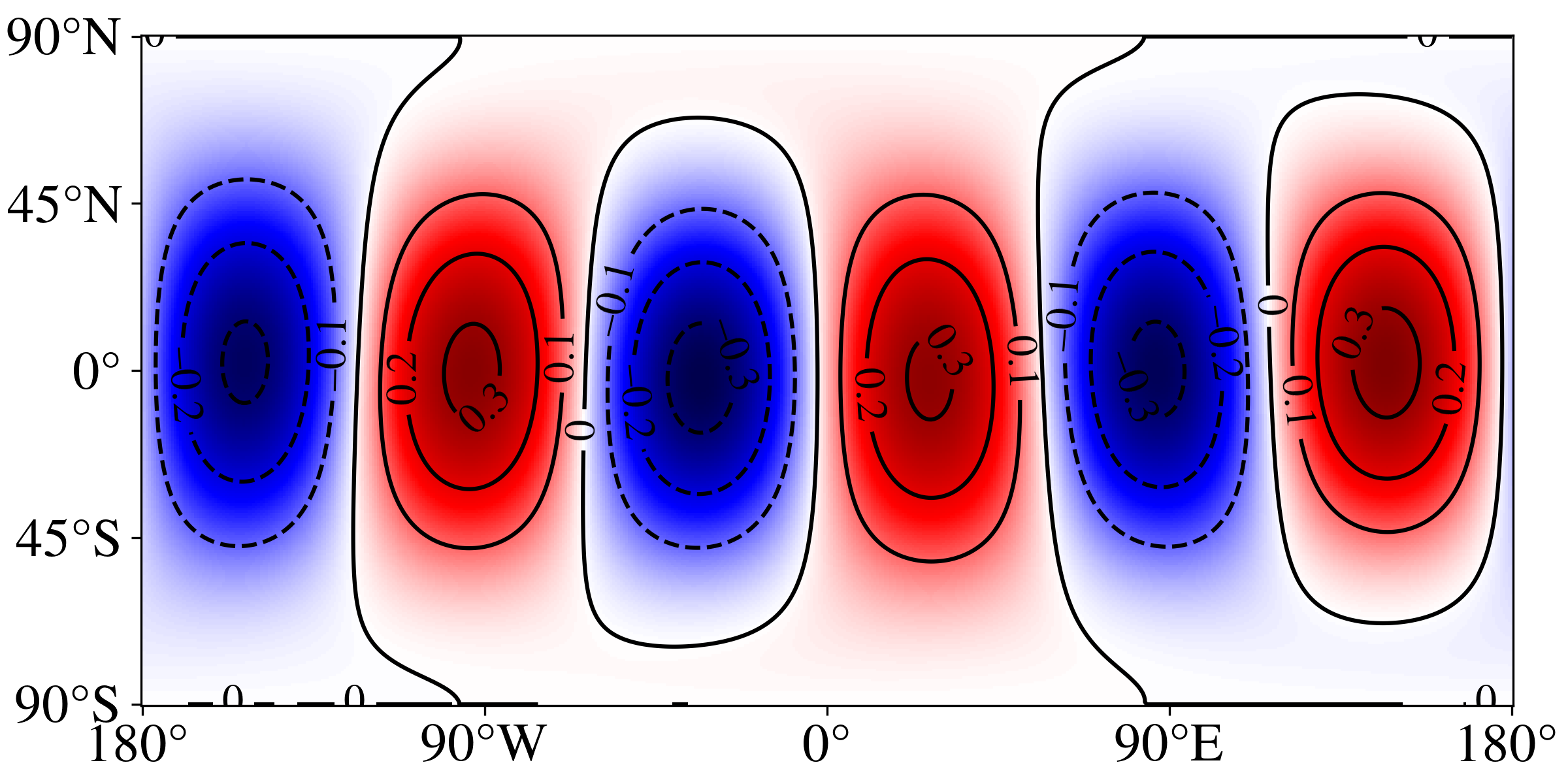}
			\label{fig:suppTritonYdiff}
		\end{subfigure}
		\hfill
		\begin{subfigure}[b]{0.495\textwidth}
			\caption{$B_z$ difference ($\mathrm{nT}$) vs.~symmetric at $r=2R_T$
			}
			\centering
			\includegraphics[width=\textwidth]{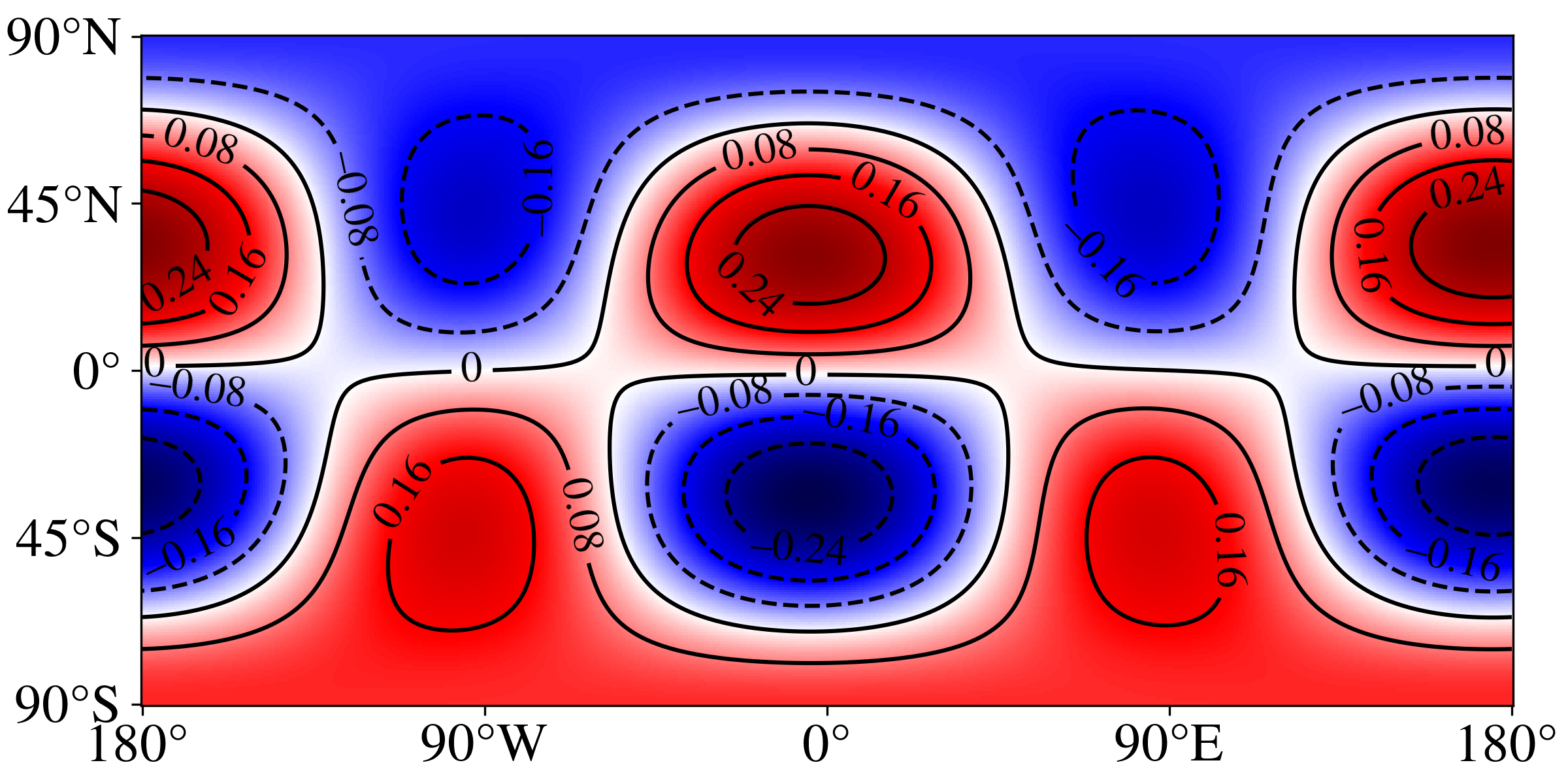}
			\label{fig:suppTritonZdiff}
		\end{subfigure}
		\caption{
			Triton model with an asymmetric ionosphere similar to that supposed for Callisto, an approximation of a day--night dichotomy.
			Uniform conductivity, an average ionospheric thickness of $200\siu{km}$, a lower bound for the ionosphere of $250\siu{km}$ altitude, and a total height-integrated conductivity of $20\siu{kS}$ are assumed, based on the structure inferred by \citet{tyler1989voyager} from \textit{Voyager} measurements.
			The interior structure we suppose for Triton is based on a moment of inertia inferred for Pluto by \citet{hussmann2006subsurface} and geophysical modeling using the \textit{PlanetProfile} framework.
			An ocean with dissolved \ch{MgSO4} and a $112\siu{km}$-thick ice shell are assumed.
			The magnetic field is evaluated at the J2000 epoch and at $r=2R_T$, a plausible distance for a spacecraft flyby.
			Unlike for Callisto, the differences due to asymmetry are a substantial fraction of the overall magnitude of the induced field (compare \Fs\ref{fig:suppTritonXdiff}--\ref{fig:suppTritonZdiff} to \ref{fig:suppTritonMag}), owing to the high conductivity and pronounced asymmetry we suppose for the ionosphere.
			Electrical conductivities in the ocean are determined using the \textit{PlanetProfile} geophysical modeling framework \citep{vance2021magnetic}.
			For simplicity, only the synodic period is modeled here.
			Animations for the difference in $x$ component and magnitude as they vary throughout the synodic period are included as Supplemental Material.
		}
		\label{fig:suppTriton}
	\end{figure}

	In \Fs\ref{fig:suppEuropaHigh}--\ref{fig:suppTriton}, we include a full detailing of analysis products from example applications of our model to several ocean worlds in the solar system (described in \S\ref{sec:results}).
	Blue--white contour maps show the asymmetry models applied for all example cases studied: asymmetric ice--ocean boundaries for Europa and Miranda and asymmetric ionospheres for Triton and Callisto.
	Heat maps show the magnitude of the induced magnetic field predicted for the asymmetric models.
	Some maps are repeated from the main text for completeness.
	Symmetric analog models have the same layer structure as the asymmetric models, and are evaluated using the standard recursion method of \citet{srivastava1966theory}.
	Please refer to the main text (\S\ref{sec:results}) for further details regarding the reasoning behind each model.
	
	Red--blue and other colormaps show the difference in induced magnetic field between our model predictions for the asymmetric cases as compared to spherically symmetric analogs, for all magnetic field vector components and the magnitude.
	Vector components are in IAU coordinates, such that at $0^\circ$ latitude, $0^\circ$ longitude, the parent planet is directly overhead; this is a right-handed coordinate system, with east longitudes positive.
	This coordinate system is rotated approximately $90^\circ$ from the $\phi\Omega$ (``Phi-O,'' {\eg}\ E-Phi-O for Europa) coordinates sometimes used in analysis of spacecraft data.
	All induced fields are evaluated at the J2000 reference epoch, 12:00 pm Jan 1, 2000 TDB and consider only the synodic period.
	Animations of the differences in magnitude and the $B_x$ component throughout a synodic period are included for each model as Supplemental Material.
	
	\subsection{Europa --- effect size with distance}\label{supp:effectSize}
	
\begin{figure}
		\centering
		\includegraphics[width=0.75\textwidth]{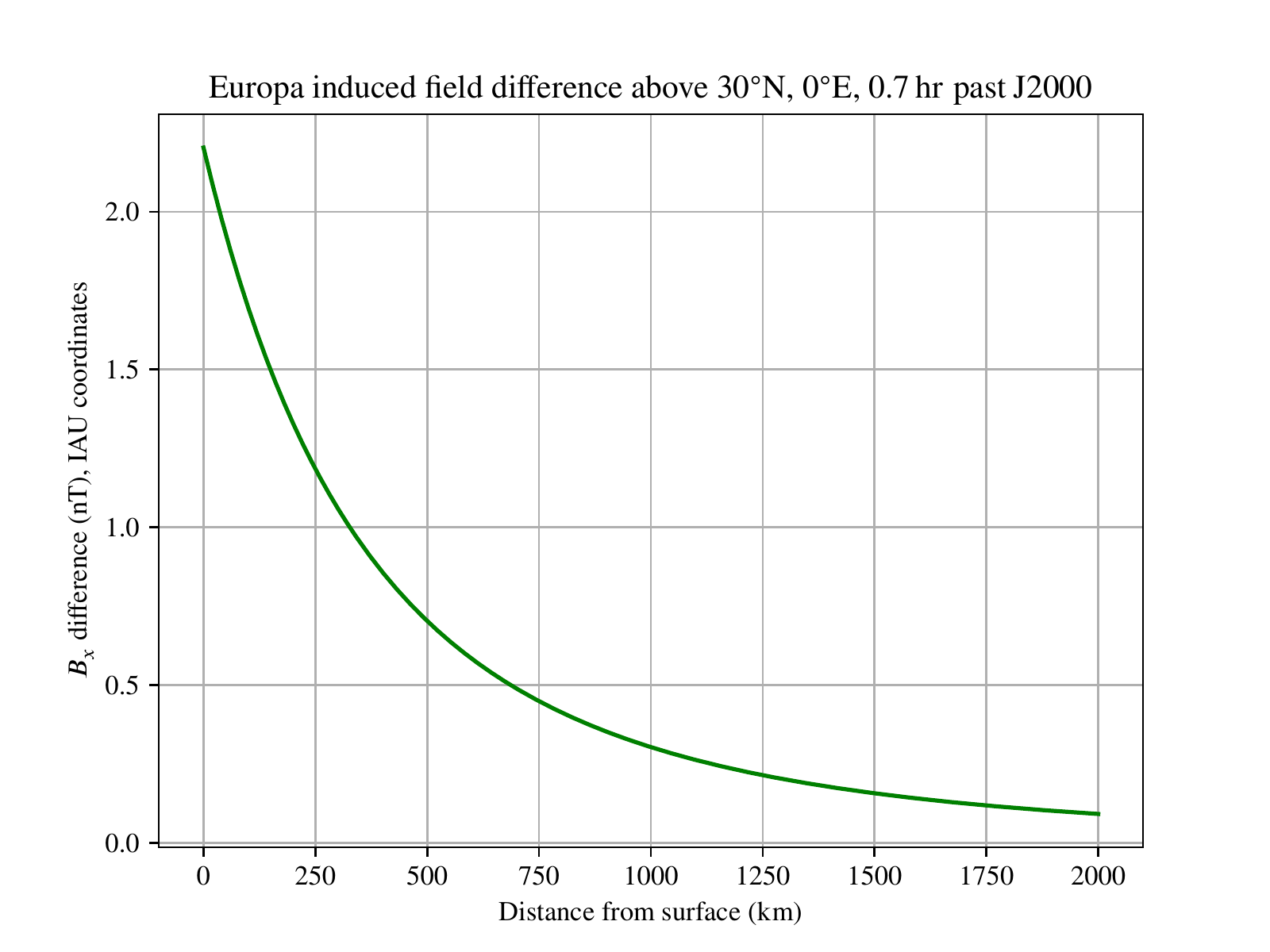}
		\caption{
			Difference in induced field $B_x$ component for our Europa Seawater model as a function of altitude for a fixed point in time.
			The selected surface location $(30^\circ\mathrm{N},0^\circ\mathrm{E})$ and time ($0.7\siu{hr}$ past J2000) maximize the observed difference relative to the spherically symmetric case for this interior model and component---see \F\ref{fig:europaHighXdiff}. 
			Beyond about $1500\siu{km}$ altitude, the difference is around $0.2\siu{nT}$ and likely negligible.
			Repeated from the main text (\F\ref{fig:europaFalloff}).
		}
		\label{fig:suppFalloff}
	\end{figure}

	It may be advantageous for future investigations to pursue modeling efforts that neglect asymmetry, or at least limit studied cases to only include known shapes ({\ie} tidal deformation).
	To facilitate these efforts, we have evaluated the difference in the induced magnetic field resulting from asymmetry as a function of distance from Europa.
	\F\ref{fig:suppFalloff} shows the difference in induced field caused by asymmetry from our Europa Seawater model (\F\ref{fig:suppEuropaHigh}) from the surface upward to $2000\siu{km}$ altitude.
	We selected a time ($0.7\siu{hr}$ after J2000) and surface point (see \F\ref{fig:suppEuropaHighXdiff}) at which the effect size is approximately maximized, with a difference at the surface of about $2.2\siu{nT}$.
	The difference resulting from asymmetry drops to $0.2\siu{nT}$ at about $1\,R_E$ in altitude, around $1500\siu{km}$.
	
\begin{figure}
		\centering
		\includegraphics[width=0.8\textwidth]{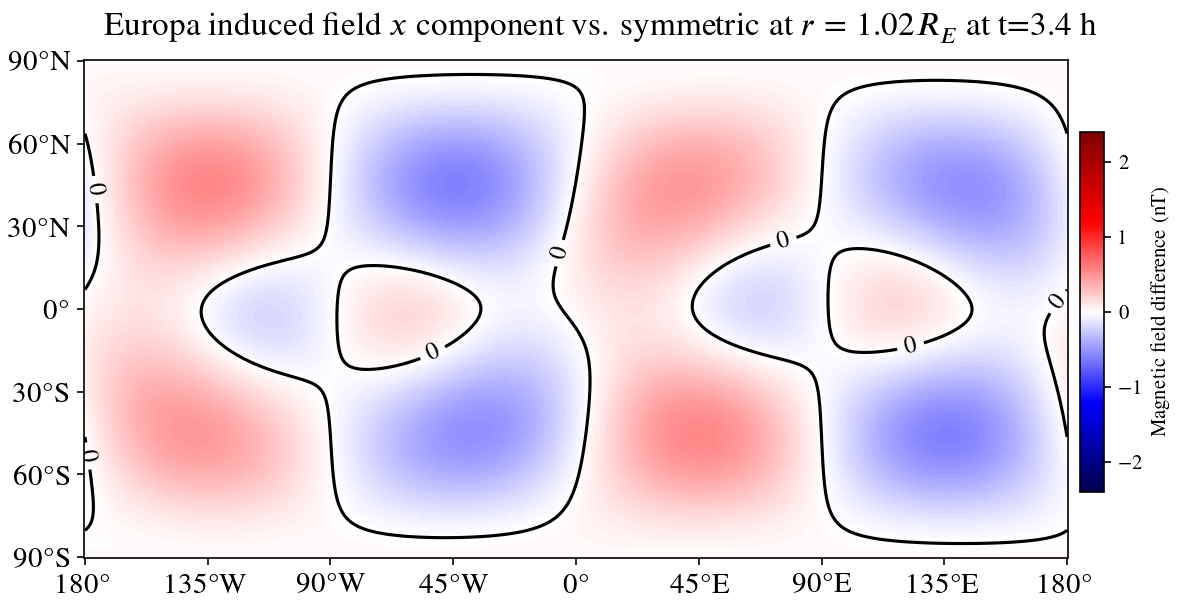}
		\caption{
			Single frame from the animated version of \F\ref{fig:suppEuropaHighXdiff}, at $t=3.4\siu{hr}$ past J2000.
			At this time, the difference resulting from asymmetry is much smaller than at J2000, and is less than $0.6\siu{nT}$ everywhere.
			This state lasts for around $30\siu{min}$.
			A spacecraft reaching its closest approach altitude above $25\siu{km}$ during a flyby at times such as this may be able to ignore the influence of asymmetry under certain conditions.
		}
		\label{fig:suppSmallDiff}
	\end{figure}

	It may also be noted that the effect of asymmetry on the induced field can become small at key points in time during the considered excitation period.
	In \F\ref{fig:suppSmallDiff}, we have extracted a single frame from our animation of the $B_x$ difference for the Europa Seawater model available in the Supplemental Material.
	At the time shown---$3.4\siu{hr}$ past J2000---only the $0\siu{nT}$ contour is plotted, meaning the difference is everywhere less than $0.6\siu{nT}$.
	This condition lasts for about 4\% of the synodic period, or about 30 minutes.
	If a spacecraft reaches its closest approach near this time, at a distance at or above $25\siu{km}$, the difference contributed by the asymmetric ocean will be small.
	This implies that there may be some conditions (including particular asymmetry models) under which the induced field measured by the spacecraft is essentially the same as that predicted by a spherically symmetric model.
	
	\section{Direct expressions for $3j$-symbols and harmonic product coefficients}\label{supp:wignerSY}
	Products of spherical harmonics are often expressed in terms of the Wigner $3j$-symbols.
	The general expression for the $3j$-symbols is presented by several authors \citep[\eg\ ][]{brink1968angular,edmonds1996angular}:
	\begin{align}
		\wigGen = \wigPhas{\ja}{\jb}{\mc}\triCoef{\ja}{\jb}{\jc} \quad\times\phantom{\wig{\ja}{\ma}{\jb}{\mb}{\jc}{\mc}}& \label{eq:wigner}\\
		\wigFact{\ja}{\ma}{\jb}{\mb}{\jc}{\mc} \quad\times\phantom{\wig{\ja}{\ma}{\jb}{\mb}{\jc}{\mc}}& \nonumber\\
		\wigSeri{\ja}{\ma}{\jb}{\mb}{\jc}{\kv} &\nonumber\\
		\mathrm{for} \quad |\ja - \jb| \le \jc \le \ja + \jb \quad\mathrm{and} \quad \mc = -(\ma + \mb), 
		\phantom{\wig{\ja}{\ma}{\jb}{\mb}{\jc}{\mc}\wig{\ja}{\ma}{\jb}{\mb}{\jc}{\mc}}&\nonumber\\
		\wigGen = 0 \quad \mathrm{otherwise.} \phantom{\triCoef{\ja}{\ma}{\jb}{\mb}{\jc}{\mc}}& \nonumber
	\end{align}
	The sum in \Eq\ref{eq:wigner} is over each integer value of $\kv$ for which \textit{all} factorials in the sum are nonnegative.
	For many of the low-degree combinations of spherical harmonics with which we are concerned, the series has only a single term with $\kv=0$.
	
	When $\ma = \mb = \mc = 0$, the $3j$-symbols are non-zero only if $\ja+\jb+\jc$ is even \citep{brink1968angular}.
	Another way to express this is that $\jc$ takes values of $\ja+\jb,~ \ja+\jb-2,~ \ja+\jb-4,~ \dots~ |\ja - \jb|$.
	\Eq\ref{eq:wigner} simplifies under many conditions.
	The results of products of spherical harmonics are always proportional to $\wigZ{\ja}{\jb}{\jc}$.
	In this case, \Eq\ref{eq:wigner} becomes \cite{edmonds1996angular}
	\begin{align}
		\wigZ{\ja}{\jb}{\jc} = (-1)^{(\ja+\jb+\jc)/2}&\triCoef{\ja}{\jb}{\jc} \quad\times \label{eq:wignerzero}\\
		&\qquad\qquad\frac{\left(\frac{1}{2}(\ja+\jb+\jc)\right)!}{\left(\frac{1}{2}(\ja+\jb-\jc)\right)!\left(\frac{1}{2}(\ja-\jb+\jc)\right)!\left(\frac{1}{2}(-\ja+\jb+\jc)\right)!}\nonumber
	\end{align}
	
	Using \Eqs\ref{eq:wigner} and \ref{eq:wignerzero}, we may obtain explicit expressions for the products $\Ynm\Ypq$ and $\dYnm\Ypq$.
	The selection rules for both mixing coefficients are similar, but not identical.
	For $\SY$, the conditions are
	\begin{equation}
		|n - \shapen| \le n' \le n + \shapen, \quad n + \shapen + n' ~\mathrm{is~even,} \quad\mathrm{and} \quad m' = m + \shapem. \label{eq:selectionr}
	\end{equation}
	For $\dSY$, the conditions are instead
	\begin{equation}
		0 < n' \le n + \shapen, \quad n + \shapen + n' ~\mathrm{is~even,} \quad\mathrm{and} \quad m' = m + \shapem. \label{eq:selectiond}
	\end{equation}
	For $\dSY$, the triangular condition is modified because it relates products involving $\dYnm$, which are linear combinations of $Y_{n+1,m}/\sn$ and $Y_{n-1,m}/\sn$, linking each $(n,m)$ to both $(n+2,m)$ and $(n-2,m)$.
	The net result is that either $n' = 1$ or $2$ will have non-zero mixing coefficients for all combinations of $n$ and $p$, and even large $p$ can still impact induced moments of degree $1$ (the dipole moments).
	
	If the selection rules are satisfied, the following expressions apply.
	Otherwise, the coefficient is zero.
	In the radial boundary conditions, the coefficients are
	\begin{align}
		\Ynm\Ypq = \sum_{n'}\SY\Ynmp \phantom{\wigFact{n}{m}{\shapen}{\shapem}{n'}{m'}} \nonumber
		&\\
		\SY = (-1)^\evind \wigNorm{n}{\shapen}{n'}\wigFrac{n}{\shapen}{n'} \quad\times &\label{eq:SYdirect}\\
		\wigFact{n}{m}{\shapen}{\shapem}{n'}{m'} \quad\times &\nonumber\\
		\wigSeriEv{n}{m}{\shapen}{\shapem}{\kv}, &\nonumber\\
		\text{with }~ \nu \equiv \frac{1}{2}(n+\shapen-n'). \phantom{\wigFact{n}{m}{\shapen}{\shapem}{n'}{m'}}\nonumber &
	\end{align}
	The sum over $\kv$ is again limited to those integer values of $\kv$ for which all factorials have nonnegative arguments.
	This limits possible values of $\kv$ to lie between $0$ and $2\evind$, but does not necessarily include those values.
	For example, in the case of $\shapem=\shapen$ with $\evind=1$, $\kv=0$ yields a negative value for $\shapen-\shapem-(2\evind-\kv)$, the final factorial in the series expression.
	This occurs for $n,m=1,-1$, $\shapen,\shapem=2,2$, $n',m'=1,1$, a case of interest for application of our methods because it relates $\shapen=2$ boundary harmonics to the induced dipole moment from a uniform excitation field.
	This condition dictates that the minimum value for $\kv$ is the greatest value among [$0$, $2\evind-(n+m)$, $2\evind-(\shapen-\shapem)$] and the maximum value for $\kv$ is the least value among [$2\evind$, $n-m$, $\shapen+\shapem$].
	In summary:
	\begin{align}
		\kv^- &= \max(0, ~2\evind-(n+m), ~2\evind-(\shapen-\shapem))\\
		\kv^+ &= \min(2\evind, ~n-m, ~\shapen+\shapem).\nonumber
	\end{align}
	
	In the tangential boundary conditions, the products $\dYnm\Ypq$ proportional to a desired $\dYnmp$ may be replaced by $\dSY$:
	\begin{align}
		&\dSY = \frac{ \SYw{} + \sum_{\dg=1}^{\dg^-_\mathrm{max}}\SYw{-2\dg}\prod_{\kv=0}^{\dg-1}\Fcont_{n',2\kv}^- + \sum_{\dg=1}^{\dg^+_\mathrm{max}}\SYw{+2\dg}\prod_{\kv=0}^{\dg-1}\Fcont_{n',2\kv}^+ }{ \left(\mix^-_{n'}\right)^2 +  \left(\mix^+_{n'}\right)^2 - \mix^-_{n'}\mix^+_{n'-2}\Fcont_{n',0}^-  - \mix^+_{n'}\mix^-_{n'+2}\Fcont_{n',0}^+ }, \label{eq:dSYA}\\ 
		&\dg^-_\mathrm{max} \equiv \begin{cases} \fl\left(\frac{1}{2}\left(n'-|m'|\right) \right) &\text{if }n' \ge 3\\
			0 &\text{otherwise}\end{cases}, \quad 
		\dg^+_\mathrm{max} \equiv \fl\left(\frac{1}{2}\left(\nmax+\shapemax-n'\right)\right),\\
		&\Fcont^\pm_{n,\Fsub} \equiv \frac{\mix^\pm_{n\pm\Fsub}\mix^\mp_{n\pm(\Fsub+2)}}{\left(\mix^-_{n\pm(\Fsub+2)}\right)^2 + \left(\mix^+_{n\pm(\Fsub+2)}\right)^2 - \mix^\pm_{n\pm(\Fsub+2)}\mix^\mp_{n\pm(\Fsub+4)}\Fcont^\pm_{n,\Fsub+2}}, \label{eq:FcontA}\\
		&\SYw{} \equiv \mixm\mixmp\SYarg{n-1\,m\shapen\shapem}{n'-1\,m'} +  \mixp\mixpp\SYarg{n+1\,m\shapen\shapem}{n'+1\,m'} -  \mixm\mixpp\SYarg{n-1\,m\shapen\shapem}{n'+1\,m'} -  \mixp\mixmp\SYarg{n+1\,m\shapen\shapem}{n'-1\,m'}, \label{eq:SYwA}\\
		&\mixm = (n+1)\sqrt{\frac{n^2 - m^2}{(2n-1)(2n+1)}}, \quad \mixp = n\sqrt{\frac{(n+1)^2 - m^2}{(2n+1)(2n+3)}}, \label{eq:mixA}
	\end{align}
	where $\kv=0,1,2,\dots$, $\fl(a)$ rounds $a$ down to the nearest integer, and a subscript $m$ is implied on all $\mix$ in all equations where it is omitted, corresponding to the subscript $n$.
	The recursion in $\Fcontnk^+$ continues until $2\kv$ exceeds $\nmax+\shapemax-n$, and the recursion in $\Fcontnk^-$ continues until $2\kv$ exceeds $n-|m|-2$.
	When $2\kv$ exceeds these respective values, $\Fcontnk^\pm=0$.
	$\Fcontnk^-$ is also zero for $n<3$.
	The sums in the numerator of \Eq\ref{eq:dSYA} result from isolating the desired terms of degree $n'$ and order $m'$; the denominator results from solving the isolated term for $\dSY$.
	
	These bounding values result from non-orthogonality of the $\dYnm$.
	For a term of degree $n''$ below the considered value $n'$, non-orthogonality is possible only for $n'' = n'-2$, $n'-4$, etc., but $\dYnmpp$ is defined and non-zero only for $n'' > 0$ and $n'' \ge m''$.
	This constraint limits the non-zero values of $\Fcont^-_{n',0}$ to $n' \ge 3$ and $n' \ge m'+2$.
	The $n''$ above the considered $n'$ that yield non-orthogonal terms are similarly limited by the selection rules for $\SY$, {\ie}\ $n''$ cannot exceed $\nmax+\shapemax$ because no combination of $\dYnm\Ypq \rightarrow \dSY\dYnmp$ results in $n' > \nmax+\shapemax$ and we have assumed the series in $n'$ may be truncated at $n'=\nmax+\shapemax$.

	\section{Sharp transitions in layer conductivities}\label{supp:transitions}
	
	As a matter of practical consideration, here we highlight important challenges for application of the layer method (and thus our results) and strategies for their mitigation.
	The content of this section has also proven useful for validating the functional dependence of the quantities we derive in our solutions.
	In this section, we assume spherical symmetry.
	
	The Bessel functions $\jn$ and $\yn$ always contain complex exponentials with complex arguments $kr$.
	For large $|kr|$, the functions are all asymptotic to a growing exponential divided by $kr$; when $kr$ is close to zero, they are asymptotic to powers of $kr$ \citep{marion1980classical}.
	On planetary scales, $kr$ often takes extreme values in various layers.
	For example, in metallic cores, the conductivity $\sigma$ may be over $10^6\siu{S/m}$ \citep{khurana2002searching}, providing a strong response for all periods of excitation and for any value of $r$.
	$|kr|$ can be large for even moderately conducting oceans on large scales---$|kr|\sim100$ for a spherical Earth-size ocean for a 1-day period of oscillation.
	In contrast, in insulating materials such as ice and rock, conductivity values can be extremely low: $10^{-12} \lesssim \sigma \lesssim 10^{-2}$, depending on hydration state, porosity, and presence of contaminants \citep{glover2015geophysical}.
	Additionally, for all materials, conductivity is pressure- and temperature-dependent.
	Because $|kr|$ can span many orders of magnitude across boundaries, especially at an ice--ocean interface or a core--mantle boundary, application of the recursion relations in \S\ref{supp:intsph} often results in differences of very large terms that are very close together in value, requiring great numerical precision in computation to reliably evaluate.
	With some realistic planetary values, numerical overflow or underflow are assured even for specialized, high-precision libraries such as mpmath\footnote{\url{http://mpmath.org/}} in Python and MPFUN\footnote{MPFUN is available for Unix-based systems on author D.~Bailey's personal website: \url{https://www.davidhbailey.com/dhbsoftware/}} in Fortran.
	
	In handling these challenges, we have found it essential to account for sharp conductivity boundaries by using approximations appropriate to the type of boundary transition.
	Under limiting conditions, the Bessel functions take the following asymptotic forms, derived from \citet{marion1980classical}:
	\begin{align}
		\begin{array}{ll}
			\jnkr \approx \frac{ (-i)^n\eikr - (i)^n\eMikr}{2ikr}, \quad 
			&\ynkr \approx \frac{-(-i)^n\eikr - (i)^n\eMikr}{2kr}, \\[0.5em]
			\jdkr \approx \frac{ (-i)^n\eikr + (i)^n\eMikr}{2}, \quad 
			&\ydkr \approx \frac{ (-i)^n\eikr - (i)^n\eMikr}{2i}
		\end{array}
		\qquad \mathrm{for}~ |kr| \gg n \label{eq:largekr}, \\[1em]
		\begin{array}{ll}
			\jnkr \approx \frac{(kr)^n}{(2n+1)!!}, \quad
			&\ynkr \approx -\frac{(2n-1)!!}{(kr)^{n+1}}, \\[0.5em]
			\jdkr \approx (n+1)\frac{(kr)^n}{(2n+1)!!}, \quad
			&\ydkr \approx n\frac{(2n-1)!!}{(kr)^{n+1}}
		\end{array}
		\qquad \mathrm{for}~ |kr| \ll n. \label{eq:smallkr}
	\end{align}
	These expressions may then be used to identify replacement rules for the Bessel functions as they appear in \Eq\ref{eq:Lambdfull} for extreme layers:
	\begin{align}
		\begin{array}{ll}
			\jdkr \rightarrow -kr\,\ynkr, \quad
			&\ydkr \rightarrow  kr\,\jnkr,\\[0.5em]
			|\jdkr| \gg |\jnkr|, \quad 
			&|\ydkr| \gg |\ynkr|
		\end{array}
		\qquad \mathrm{for}~ |kr| \gg n \label{eq:replhikr}, \\[1em]
		\begin{array}{ll}
			\jdkr \rightarrow (n+1)\,\jnkr, \quad
			&\ydkr \rightarrow -n\,\ynkr,\\[0.5em]
			|\ynkr| \gg |\jnkr|, \quad 
			&|\ydkr| \gg |\jdkr|
		\end{array}
		\qquad \mathrm{for}~ |kr| \ll n \label{eq:repllokr}.
	\end{align}
	Using this information, we can reduce \Eqs\ref{eq:Lambdfull} and \ref{eq:Ae} to account for extreme transitions.
	For the innermost boundary, \Eq\ref{eq:Lambdfull} becomes
	\begin{align}
		\bPnm^2 \rightarrow -\frac{\jn^{2,1} + \frac{\jn^{\star 2,1}}{ik_1r_1}}{\yn^{2,1} + \frac{\yn^{\star 2,1}}{ik_1r_1}} \qquad \mathrm{for}~ |k_1r_1| \gg n, \label{eq:innerhic} \\[1em]
		\bPnm^2 \rightarrow -\frac{\jnP^{2,1}}{\ynP^{2,1}} \qquad \mathrm{for}~ |k_1r_1| \ll n. \label{eq:innerloc}
	\end{align}
	Note that the superscripts in the expressions above are all indices, not exponents---these relate the quantity $\bP$ for the second-innermost layer to $k_1$, $k_2$, and $r_1$ when the innermost layer is a very strong or very poor conductor.
	
	For intermediate layers, when $|kr|$ is large, if we keep only first-order terms in $1/kr$ we obtain the same result as \Eq\ref{eq:innerhic}.
	Consequently, if $|kr|$ is large enough to cause numerical overflows, currents flowing in this layer entirely screen every layer beneath it from oscillations in the magnetic field.
	The behavior is more complicated for an interstitial nonconducting layer, however.
	
	For a $\jl^\mathrm{th}$ (middle) layer with wavenumber $k_\jl$, lower bounding radius $r_\low = r_{\jl-1}$, upper bounding radius $r_\upp = r_\jl$, and wavenumbers $k_\low = k_{\jl-1}$, $k_\upp = k_{\jl+1}$ pertaining to the lower and upper layers respectively, $\bPnm^\upp$ for the upper layer takes the form
	\begin{align}
		\bPnm^\upp \rightarrow -\frac{\jn^{\upp,\upp} + \frac{\jn^{\star \upp,\upp}}{ik_\jl r_\jl}}{\yn^{\upp,\upp} + \frac{\yn^{\star \upp,\upp}}{ik_\jl r_\jl}} \qquad \mathrm{for}~ |k_\jl r_\jl| \gg n, \label{eq:midhic} \\[1em]
		\bPnm^\upp \rightarrow \frac{ \betx + \bPnmL\delx \left(\frac{\rb}{\ru}\right)^{2n+1} }{ \gamx + \bPnmL\epsx \left(\frac{\rb}{\ru}\right)^{2n+1} } = 
		-\frac{\jnP(\krx) - \bPnmL\jnM(\krx)\left(\frac{\rb}{\ru}\right)^{2n+1}}{\ynP(\krx) - \bPnmL\ynM(\krx)\left(\frac{\rb}{\ru}\right)^{2n+1}} \qquad \mathrm{for}~ |k_\jl r_\jl| \ll n. \label{eq:midloc}
	\end{align}
	In \Eq\ref{eq:midloc} we have defined the following quantities based on their similarity to those that appear in \Eqs\ref{eq:ghb}--\ref{eq:ghe} and \ref{eq:Lambd}:
	\begin{align}
		\betx &= \jdju - (n+1)\jnju &= -\krx\jnPju \label{eq:bcross}\\
		\gamx &= \ydju - (n+1)\ynju &= -\krx\ynPju \\
		\delx &= \jdju + n\jnju &= \krx\jnMju \\
		\epsx &= \ydju + n\ynju &= \krx\ynMju  \label{eq:ecross}\\
		\bPnmL &= \frac{\betl + \bPnml\gaml}{\dell + \bPnml\epsl} &= 
		-\frac{\jnPl + \bPnml\ynPl}{\jnMl + \bPnml\ynMl}, \label{eq:LambdL}
	\end{align}
	where $\betl$--$\epsl$ in \Eq\ref{eq:LambdL} are defined as in \Eqs\ref{eq:ghb}--\ref{eq:ghe}---the lower boundary acts as an ``outer'' boundary, as Laplace's equation applies in the space where $k \rightarrow 0$.
	\Eq\ref{eq:midloc} is the only layer recursion relation that contains more than one radius value, because it propagates from the lower boundary to the upper across the ``gap'' nonconducting layer.
	Finally, at the outermost boundary:
	\begin{align}
		\Ampne \rightarrow 1 \qquad |\koB\roB| \gg n, \label{eq:outerhic} \\[1em]
		\Ampne \rightarrow \left(\frac{\roM}{\roB}\right)^{2n+1}\frac{\betoM + \bPn^{\Nl-1}\gamoM}{\deloM + \bPn^{\Nl-1}\epsoM} \qquad |\koB\roB| \ll n. \label{eq:outerloc}
	\end{align}
	For a highly conducting outermost layer, all interior layers are screened and the entire body acts as a perfect conductor.
	For an insulating outer layer, \Eq\ref{eq:outerloc} tells us the net effect is that the outer boundary moves one layer down and the final value for $\Ampne$ is scaled down as if the outer layer were not present.
	
	\Eqs\ref{eq:innerhic}--\ref{eq:outerloc} have been constructed such that numerical overflow is avoided in computation.
	In this formulation, when underflow occurs, a valid approximate result is still successfully computed.
	These expressions are physically valid when the skin depth $s_\jl = 1/\mathrm{Im}(k_\jl)$ is much less or much greater than the layer thickness $D_\jl = r_\jl - r_{\jl-1}$ \citep{styczinski2021induced}.
	
	\bibliographystyle{cas-model2-names}
	\nobibliography{references}